\documentclass[superscriptaddress,aps,pra,reprint,twocolumn,showkeys,letterpaper]{revtex4-1}

\usepackage{amsmath}
\usepackage{amssymb}
\usepackage{graphicx}

\newcommand{\dsum}{\sum}
\newcommand{\Tr}{\text{Tr}}
\newcommand{\cata}{\text{Category~1}}
\newcommand{\catb}{\text{Category~2}}
\newcommand{\catc}{\text{Category~3}}
\newcommand{\catd}{\text{Category~4}}

\begin{document}

\title{Tests for EPR Steering in Two-Mode Systems of Identical Massive Bosons}

\author{B J. Dalton}

\thanks{Corresponding Author}
\email{bdalton@swin.edu.au}

\affiliation{Centre for Quantum and Optical Science, Swinburne University
of Technology, Melbourne, Victoria 3122, Australia}

\affiliation{School of Physics and Astronomy, University of Glasgow, Glasgow
G12 8QQ, United Kingdom}

\author{B. M. Garraway}

\affiliation{Department of Physics and Astronomy, University of Sussex,
Falmer, Brighton BN1 9QH, United Kingdom}

\author{M. D. Reid}

\affiliation{Centre for Quantum and Optical Science, Swinburne University
of Technology, Melbourne, Victoria 3122, Australia}

\date{\today}

\begin{abstract}
In a previous paper tests for entanglement for two mode systems involving
identical massive bosons were obtained. In the present paper we consider
sufficiency tests for EPR steering in such systems. We find that spin
squeezing in any spin component, a Bloch vector test, the Hillery-Zubairy
planar spin variance test and squeezing in two mode quadratures all show
that the quantum state is EPR steerable. We also find a generalisation of
the Hillery-Zubairy planar spin variance test for EPR steering. The relation
to previous correlation tests is discussed. This paper is based on a
detailed classification of quantum states for bipartite systems. States for
bipartite composite systems are categorised in quantum theory as either
separable or entangled, but the states can also be divided differently into
Bell local or Bell non-local states in terms of local hidden variable theory
(LHVT). For the Bell local states there are three cases depending on whether
both, one of or neither of the LHVT probabilities for each sub-system are
also given by a quantum probability involving sub-system density operators.
Cases where one or both are given by a quantum probability are known as
local hidden states (LHS) and such states are non-steerable. The steerable
states are the Bell local states where there is no LHS, or the Bell
non-local states. The relationship between the quantum and hidden variable
theory clasification of states is discussed.
\end{abstract}


\maketitle


\section{Introduction}

\label{Section - Introduction}

Recent papers by Dalton et al.\ \cite{Dalton14a,Dalton16a,Dalton16b}
have dealt with the topic of \emph{bipartite quantum
entanglement} and experimental tests for its demonstration in the
context of two-mode systems of \emph{identical massive bosons}. However,
although the quantum states of composite systems can just be classified
into disjoint sets of \emph{separable} or \emph{entangled} states,
it is also possible to classify them into distinct categories based
on \emph{local hidden variable theory} \cite{Bell65a}, where the
two basic disjoint sub-sets of quantum states are now the \emph{Bell
local} states and the \emph{Bell non-local} states. The latter categorisation
is based on whether or not the probability $P(a,b|A,B,c)$ for measured
outcomes $a,b$ on sub-system observables $A,B$ for state preparation
process $c$, is given by a local hidden variable theory (LHVT) form
$P(a,b|A,B,c)=\sum_{\lambda}P(\lambda|c)P(a|A,c,\lambda)P(b|B,c,\lambda)$
(where preparation $c$ results in a probability distribution $P(\lambda|c)$
for hidden variables $\lambda$, $P(a|A,c,\lambda)$ is the probability
for measured outcome $a$ on sub-system observable$A$ when the hidden
variables are $\lambda$ with $P(b|B,c,\lambda)$ the analogous observable
$B$ probability). Quantum states where $P(a,b|A,B,c)$ is given by
a LHVT form are Bell local, if not they are Bell non-local and associated
with Bell inequality violation experiments.
Hence, in accord with the idea set out in the EPR paper \cite{Einstein35a}
that the predictions based on quantum theory could also be the statistical
outcome of an underlying deterministic theory (involving what we now would
regard as hidden variables), the predictions based on the local hidden
variable theory (the Bell local states) will be regarded as being in
agreement with quantum theory - and the relevant expressions will be
interchangeable. The Bell non-local states will be those quantum states
where the local HVT does not apply, and there is no underlying deterministic
theory that leads to the quantum results. However, within the Bell local
states a further categorisation is possible which is relevant to whether
EPR steering occurs.
Based on the concept of
\emph{local hidden states} introduced by Wiseman et al.\
\cite{Wiseman07a,Jones07a,Cavalcanti09a}, we show that the Bell local
states for \emph{bipartite} systems can be divided into three \emph{disjoint}
sub-categories, with a fourth corresponding to the Bell non-local
states. These four categories of states associated with local hidden
variable theory have differing features regarding entanglement, EPR
steering and Bell non-locality - as will be explained below (see also
\cite{Jevtic15a}).
For readers unfamiliar with the hidden variable theory issue and local
hidden states, a brief overview is presented in the Appendix to this
paper, emphasising the key papers of Einstein, Schrodinger, Bell and
 Werner \cite{Einstein35a,Schrodinger35a,Schrodinger35b,Bell65a,Werner89a} and those of 
Wiseman et al.\ \cite{Wiseman07a,Jones07a,Cavalcanti09a}.

The present paper is one of a series aimed at
developing tests based on experimentally measurable quantities that
are sufficient (though not necessary) for determining which category
applies for specific quantum states of bipartite two-mode systems
of identical massive bosons. The focus of the present paper is on
sufficiency tests for demonstrating \emph{EPR steering }in these systems
- essentially by eliminating two of the four possible categories of
quantum states. 
We find that spin squeezing in any spin component, a Bloch vector test, the Hillery-Zubairy planar spin variance test and squeezing in two mode quadratures are all sufficiency tests to show that the quantum state is EPR steerable. In addition, a generalisation of the Hillery-Zubairy planar spin variance test for EPR steering is also found. Apart from the two planar spin variance tests, the tests depend on applying the local particle number super-selection rule (SSR).

The plan of the paper is as follows. In Section
\ref{Section - Measurement Probabilities in Bipartite Systems}
we begin by first presenting the \emph{quantum theory expressions}
for joint and single measurement probabilities for \emph{bipartite quantum
systems}, and then the possible underlying \emph{local hidden variable} 
\emph{theory} (LHVT) expressions. Only \emph{von Neumann}
measurements will be considered. In accordance with the requirement
that HVT does not give different experimental predictions, the quantum
expressions (\ref{Eq.QuantumJointProb}), (\ref{Eq.QuantumSingleProb})
and (\ref{Eq.ReducedDensOprs}) will be regarded as \emph{always}
applying - irrespective of additional local hidden variable theory
formulae that
apply \emph{as well}. In the present paper,
for quantum theory the preparation process is reflected in the \emph{density
operator }for the system. In HVT the preparation process is reflected
in the \emph{probability function} for the hidden variables. We restrict 
LHVT to a version where the measurement outcomes for the observables
in LHVT are the same as the possible quantum theory outcomes, determined
as the \emph{eigenvalues} of the corresponding quantum \emph{Hermitian
operators}. For simplicity we treat the outcomes as \emph{quantized}
- the generalisation for \emph{continuous} eigenvalues is straightforward.
Important relationships between the probabilities and mean values
for measurements given by quantum theory and by local hidden variable
theory are highlighted. This linkage does not of course apply for
Bell non-local states. The issue of inter-relating the Hermitian operators
and c-number variables that describe the same observable is non-trivial
and is described in Section \ref{Section - Spin Squeezing and Other Tests for EPR Entanglement}
for the specific two mode system of interest. Although LHVT does not
have one unique form, we must choose a version such that its predictions
agree with those from quantum theory.
There would be no point in considering a LHVT that was \emph{not} in
agreement with quantum theory!
A \emph{key point} is that
because LHVT \emph{underlies} quantum theory, \emph{any} result we
establish for mean values, variances of observables using LHVT for
a quantum state that is \emph{also} Bell local, can immediately
be expressed in terms of the equivalent Hermitian \emph{operators}
that describe the same observables, together with the quantum \emph{density
operator} that specifies the \emph{same} state instead of the set
of LHVT \emph{probabilities}.
Obviously, it is also important to consider how 
to inter-relate the Hermitian operators that represent observables in 
quantum theory with the c-number quantities representing the same
observables in LHVT.
General features for joint and single
measurement probabilities are set out in Appendix \ref{Appendix A - Basic Measurement Probabilities}.

In Section \ref{Section - Classes of Quantum States} we then consider
the detailed description of how the quantum states for bipartite systems
may be categorised. We relate our categories of states to the hierarchy
of sub-sets discussed in Refs.\ \cite{Wiseman07a,Jones07a,Cavalcanti09a,He11a}.

In Section \ref{Section - Spin Squeezing and Other Tests for EPR Entanglement}
various tests for \emph{EPR steering} are considered for the case
where each sub-system consists of a \emph{single mode} and the particles 
that occupy it are \emph{massive bosons}, taking into account that 
the local hidden states must comply with the local particle number
\emph{super-selection rule} (see Refs.\
\cite{Dalton14a,Dalton16a,Dalton16b})
since they must be possible quantum states for
the particular sub-system considered on its own. The question of how
to relate the quantum Hermitian operators to the LHVT c-number variables
that describe the same observables is dealt with in this section.
Since mode annihilation and creation operators are not Hermitian we
can replace these by \emph{quadrature operators}, including in expressions
for spin operators and other important quantities. In applying LHVT
the quadrature operators are replaced by c-number quadrature amplitudes.
However, in order to achieve a \emph{reciprocal interconversion} between
the Hermitian operators and the c-number variables that represent
the same observable, it has been necessary to introduces certain additional
\emph{auxiliary observables} and allow the c-number versions of these
to have their own LHVT\ probability distributions. This seems to
be the best version of LHVT to ensure that the quantum theory and
the LHVT are describing the same physical measurements. It turns out
that previous sufficiency tests (see Refs.\
\cite{Dalton14a,Dalton16a,Dalton16b}
for details) for quantum entanglement (\emph{Bloch
vector} test, \emph{spin squeezing} in any spin component $S_{x}$,
$S_{y}$ or $S_{z}$, the Hillery-Zubairy \emph{planar} \emph{spin
variance} test \cite{Hillery06a}, a \emph{two mode quadrature squeezing}
test) can \emph{also} be applied as sufficiency tests for EPR steering
in two mode systems of identical massive bosons. However, in addition
a \emph{different} planar spin variance test for EPR steering involving
 the sum of the variances for spin operators $S_{x}$, $S_{y}$ and
 the mean boson number has been obtained here which also involves the mean
value for $S_{z}$, generalising a result in He et al.\ \cite{He12a}.
This test is a generalisation of the Hillery-Zubairy planar spin variance
test. In addition there are \emph{weak} and \emph{strong correlation}
tests for EPR steering that have been previously obtained by Cavalcanti
et al.\ \cite{Cavalcanti11a}. However, as each of the correlation tests
are equivalent to some of the other tests, we include these in the
Appendices rather than in the main body of the paper. 
The two planar spin variance tests can also be proved without applying the local particle number super-selection rule. However, for convenience we include the proofs for these tests within Section \ref{Section - Spin Squeezing and Other Tests for EPR Entanglement}, as well as covering in Appendices \ref{Appendix- Correlation Tests for EPR Steering} and \ref{Appendix - Correlation Ineq and Spin Operators} the non-SSR dependent proofs based on the correlation tests in Ref.\ \cite{Cavalcanti11a}.
Section \ref{Section - Summary and Conclusion}
provides a summary of the main results. An illustration of applying
the EPR tests is given for the case of the two mode \emph{binomial}
state - which is shown to be EPR steerable.

In Section \ref{Section - Spin Squeezing and Other Tests for EPR Entanglement}
we will identify experiments demonstrating EPR steering in \emph{two
mode} Bose-Einstein condensates according to these tests, such as
in Refs.\ \cite{Gross10a,Riedel10a,Maussang10a,Egorov11a,Gross11a,Peise15a} that have already 
been carried out, though EPR steering was only identified in \cite{Gross11a}
and \cite{Peise15a}. Note also that EPR steering has also recently
been found in three and four mode systems \cite{Kunkel18a,Fadel18a,Lange18a} based on different tests (such as in Ref.\ \cite{Reid09a})
for these multimode cases.
The test in Ref.~\cite{Kunkel18a} for verifying EPR steering involves direct measurement tests on variances of conjugate observables for one sub-system, to see whether the Heisenberg uncertainty principle has been violated after measurements were made on the other sub-system.

Details are set out in Appendices.
Appendix~\ref{Apppendix - Review of Hidden Variable Theory and Local Hidden States}
presents a brief summary of
the \emph{development} of hidden variable theory, and also contains an
overview of the \emph{categorisation} of quantum states both as separable or
entangled on the one hand or as Bell local and Bell non-local on the other,
pointing out that Bell local states may be further sub-categorised in terms
of the presence or otherwise of local hidden states, as introduced by
Wiseman et al. 
Appendix \ref{Appendix A - Basic Measurement Probabilities}
sets out the general relations for measurement probabilities in bipartite
systems. In Appendix \ref{Appendix - Mean Values and Variances} general
properties of mean values and variances are reviewed. Expressions
for classical observables in terms of quadrature amplitudes are given
in Appendix \ref{Appendix - Classical Observables and Quadratures}.
The Werner states are described in Appendix \ref{Appendix - Werner States},
since in various parameter regimes they provide examples of the four
categories of states in the local hidden variable theory model. The
idea behind EPR steering is discussed in Appendix \ref{Appendix - EPR Steering}.
Details for the derivation of the spin squeezing and two mode quadratures
EPR steering tests are presented in Appendices \ref{Appendix - EPR Sterering Othe Approach}
and \ref{Appendix - Variances of Two Mode Quadratures - Cat 2}, The
correlation tests and their forms in terms of spin operators are set
out in Appendices \ref{Appendix- Correlation Tests for EPR Steering}
and \ref{Appendix - Correlation Ineq and Spin Operators}.

\section{Measurement Probabilities in Bipartite Systems}

\label{Section - Measurement Probabilities in Bipartite Systems}

In this Section we set out the expressions for joint and single measurement
probabilities for bipartite systems, both in quantum theory and in
local hidden variable theory. Based on Einstein's view that quantum
theory is under-pinned by LHVT, the relationship between the two approaches
is also pointed out. General results for the probabilities are set
out in Appendix \ref{Appendix A - Basic Measurement Probabilities}.
The same \emph{notation} for observables, their measured outcomes
and the measurement probabilities will be used for both the quantum
theory and LHVT situations.

\subsection{Quantum Theory - Measurement Probabilities}

\label{SubSection - Quantum Theory}

In \emph{quantum theory} the \emph{joint probability} $P(\alpha,\beta|\Omega_{A},\Omega_{B},c)$
for measurement of \emph{any} pair of sub-system \emph{observables}
$\Omega_{A}$ and $\Omega_{B}$ to obtain \emph{any} of their possible
\emph{outcomes} $\alpha$ and $\beta$ when the \emph{preparation}
process is $c$ is given by an expression based on the sub-system
observables $\Omega_{A}$ and $\Omega_{B}$ being represented by quantum
\emph{Hermitian operators} $\widehat{\Omega}_{A}$ and $\widehat{\Omega}_{B}$.
Here simultaneous precise measurement applies because the system operators
involved, $\widehat{\Omega}_{A}\otimes\widehat{1}_{B}$ and $\widehat{1}_{B}\otimes\widehat{\Omega}_{B}$
\emph{commute} and therefore have complete sets of simultaneous
eigenvectors.

We have for the \emph{joint measurement probability }(see Ref.\ \cite{Wiseman07a},
Eq.\ (2)) 
\begin{equation}
P(\alpha,\beta|\Omega_{A},\Omega_{B},c)=\Tr((\widehat{\Pi}_{\alpha}^{A}\otimes\widehat{\Pi}_{\beta}^{B})\widehat{\rho}),\label{Eq.QuantumJointProb}
\end{equation}
where $\widehat{\Pi}_{\alpha}^{A}$ and $\widehat{\Pi}_{\beta}^{B}$
are \emph{projectors} onto the \emph{eigenvector spaces} for $\widehat{\Omega}_{A}$
and $\widehat{\Omega}_{B}$ associated with the real \emph{eigenvalues}
$\alpha$ and $\beta$ that in quantum theory are the \emph{possible}
measurement outcomes. We have $\widehat{\Omega}_{A}\widehat{\Pi}_{\alpha}^{A}
=\alpha\widehat{\Pi}_{\alpha}^{A}=\widehat{\Pi}_{\alpha}^{A}\widehat{\Omega}_{a}$,
and similar expressions for $\widehat{\Pi}_{\beta}^{B}$. Clearly
the quantum expression for the joint probability satisfies the general
probability requirement (\ref{Eq.SumJointProballOutcomes}) that the
sum over all possible outcomes is unity - the sum rules over $\alpha$
and $\beta$ being implemented via the \emph{projector properties}
$\sum_{\alpha}\widehat{\Pi}_{\alpha}^{A}=\widehat{1}^{A}$
and $\sum_{\beta}\widehat{\Pi}_{\beta}^{B}=\widehat{1}^{B}$
involving the sub-system \emph{unit operators} and $\Tr\widehat{\rho}=1$.

The quantum theory expressions for the \emph{single measurement probabilities}
\begin{eqnarray}
P(\alpha|\Omega_{A},c) & = & \Tr((\widehat{\Pi}_{\alpha}^{A}\otimes\widehat{1}^{B})\widehat{\rho}),\nonumber \\
P(\beta|\Omega_{B},c) & = & \Tr((\widehat{1}^{A}\otimes\widehat{\Pi}_{\beta}^{B})\widehat{\rho}),\label{Eq.QuantumSingleProb}
\end{eqnarray}
for (respectively) measuring $\Omega_{A}$ to have outcome $\alpha$
irrespective of $\Omega_{B}$ and $\beta$ or for measuring $\Omega_{B}$
to have outcome $\beta$ irrespective of $\Omega_{A}$ and $\alpha$
both follow from (\ref{Eq.SingleProbMeasureB}) or (\ref{Eq.SingleProbMeastA})
and the projector properties. The single measurement probabilities
can be expressed in terms of \emph{reduced density operators }$\widehat{\rho}^{A}$
and $\widehat{\rho}^{B}$ for the sub-systems 
\begin{align}
\widehat{\rho}^{A} & = \Tr_{B}(\widehat{\rho}),&
P(\alpha|\Omega_{A},c) & =  \Tr_{A}(\widehat{\Pi}_{\alpha}^{A}  \widehat{\rho}^{A}),\nonumber \\
\widehat{\rho}^{B}&=\Tr_{A}(\widehat{\rho}),&
P(\beta|\Omega_{B},c)&=\Tr_{B}(\widehat{\Pi}_{\beta}^{B}\,\widehat{\rho}^{B}). 
\label{Eq.ReducedDensOprs}
\end{align}
The proof of the results (\ref{Eq.ReducedDensOprs}) for $P(\alpha|\Omega_{A},c)$
and $P(\beta|\Omega_{B},c)$ is straight-forward. Note that in general
the reduced density operators require first knowing the \emph{overall}
system density operator $\widehat{\rho}$. The joint and single measurement
probabilities are related via (\ref{Eq.SingleProbMeastA}) and (\ref{Eq.SingleProbMeasureB}),
as easily shown using $\sum_{\alpha}\widehat{\Pi}_{\alpha}^{A}=\widehat{1}^{A}$
and $\sum_{\beta}\widehat{\Pi}_{\beta}^{B}=\widehat{1}^{B}$.
Using similar considerations and $\Tr\widehat{\rho}=1$, the single measurement
probabilities also satisfy the sum rules (\ref{Eq.SingleProbSumRules}).

The \emph{conditional probabilities }are given by the general expressions
(\ref{Eq.LHVTConditionalProb}) that apply for both quantum and LHVT
cases.

The \emph{mean value} for joint measurement outcomes of the observables
$\widehat{\Omega}_{A}$ and $\widehat{\Omega}_{B}$\ will be given
by 
\begin{eqnarray}
\left\langle \widehat{\Omega}_{A}\otimes\widehat{\Omega}_{B}\right\rangle  & = & \sum\limits _{\alpha,\beta}\alpha\,\beta\,P(\alpha,\beta|\Omega_{A},\Omega_{B},c)\nonumber \\
 & = & \Tr(\widehat{\Omega}_{A}\otimes\widehat{\Omega}_{B})\widehat{\rho},\label{Eq.MeanQThy}
\end{eqnarray}
where the results $\sum_{\alpha}\alpha\,\widehat{\Pi}_{\alpha}^{A}=\widehat{\Omega}_{A}$\ and
$\sum_{\beta}\beta\,\widehat{\Pi}_{\beta}^{B}=\widehat{\Omega}_{B}$\ and
(\ref{Eq.QuantumJointProb}) have been used.

The mean value for the measurement of a single observable $\widehat{\Omega}_{A}$
is
\begin{multline}
\left\langle \widehat{\Omega}_{A}\right\rangle =\sum\limits _{\alpha}\alpha\,P(\alpha|\Omega_{A},c)=\Tr(\widehat{\Omega}_{A}\otimes\widehat{1}_{B})\widehat{\rho}\\=\Tr_{A}(\widehat{\Omega}_{A}\,\widehat{\rho}^{A}),\label{Eq.MeanQThySingle}
\end{multline}
as can be derived from (\ref{Eq.QuantumJointProb}) and (\ref{Eq.ReducedDensOprs}).

It is worth noting that for systems of identical massive bosons \emph{super-selection
rules} (SSR) require the overall density operator $\widehat{\rho}$
to commute with the \emph{total }number operator $N$ (\emph{global}
particle number SSR - see for example Refs.\ \cite{Dalton16a,Dalton16b}
and references therein for discussions on SSR). Consequently the density
operator for a two mode system 
\begin{multline}
\widehat{\rho}  =  \sum\limits _{n_{A},n_{B}}\sum\limits _{m_{A},m_{B}}\rho(n_{A},n_{B};m_{A},m_{B})\\
\times(\left\vert n_{A}\right\rangle \otimes\left\vert n_{B}\right\rangle )(\left\langle m_{A}\right\vert \otimes\left\langle m_{B}\right\vert )\label{Eq.GenDensOpr}
\end{multline}
is such that $\rho(n_{A},n_{B};m_{A},m_{B})=0$ unless $n_{A}+n_{B}=m_{A}+m_{B}$.
It is then straightforward to show that the reduced density operator
$\widehat{\rho}^{A}$ for mode $A$ is given by
\begin{equation}
\widehat{\rho}^{A}=\sum\limits _{n_{A}}(\sum\limits _{n_{B}}\rho(n_{A},n_{B};n_{A},n_{B}))(\left\vert n_{A}\right\rangle \left\langle n_{A}\right\vert ),\label{Eq.GenRDO}
\end{equation}
which is SSR compliant for the \emph{sub-system} particle number $N_{A}$
(\emph{local} particle number SSR). This feature will turn out to
be relevant for evaluating terms associated with the EPR steering
tests.
Note that in general the reduced density operator
$\widehat{\rho }^{A}$ depends on the full density matrix for
\emph{both} sub-systems, unlike that for a local hidden state.

\subsection{Local Hidden Variable Theory - Measurement Probabilities}

\label{SubSection - Local Hidden Variable Theory}

A \emph{hidden variable} \emph{theory} (HVT) is based on hidden variables
$\lambda$ which describe the \emph{real} or \emph{underlying} state
of the system, and which are determined with a \emph{probability}
$P(\lambda|c)$ for a preparation process $c$. The probability $P(\lambda|c)$
is real, positive and its sum over all possible hidden variables
is also unity. Thus 
\begin{equation}
\sum\limits _{\lambda}P(\lambda|c)=1.\label{Eq.LHVProbHiddenVar}
\end{equation}
The preparation process is thus reflected in the \emph{probability
function} for the hidden variables $c\rightarrow P(\lambda|c)$. In
order to maintain generality, the nature of the hidden variables and
what fundamental equations determine them is best left unspecified.
We are also ignoring any time delay between preparation of the state
and measurements on it, so dynamical evolution of hidden variables
during this interval is irrelevant. Discussion of successive measurements
is not considered here, so whether the hidden variables change as
a result of measurement is also beyond the scope of this paper. The
key feature is that having been determined in the preparation process,
the hidden variables still determine the outcome probabilities in
separated sub-systems.

In \emph{local hidden variable theory} the \emph{joint probability}
$P(\alpha,\beta|\Omega_{A},\Omega_{B},c)$ for measurement of \emph{any}
pair of sub-system \emph{observables} $\Omega_{A}$ and $\Omega_{B}$
to obtain \emph{any} of their possible \emph{outcomes} $\alpha$ and
$\beta$ when the \emph{preparation} process is $c$ is given by an
expression involving measurement probabilities $P(\alpha|\Omega_{A},c,\lambda)$
and $P(\beta|\Omega_{B},c,\lambda)$ for the \emph{separate} sub-systems,
and which depend on the hidden variables $\lambda$. The sub-system
observables $\Omega_{A}$ and $\Omega_{B}$ are represented by \emph{c-numbers}
rather than Hermitian operators. Here $P(\alpha|\Omega_{A},c,\lambda)$
is the probability that measurement of the \emph{observable} $\Omega_{A}$
of sub-system $A$ results in \emph{outcome} $\alpha$ when the \emph{hidden
variable} are $\lambda$, with a similar definition for $P(\beta|\Omega_{B},c,\lambda)$.

For a \emph{LHVT} the \emph{joint probability} $P(\alpha,\beta|\Omega_{A},\Omega_{B},c)$
for measurement of \emph{any} pair of sub-system \emph{observables}
$\Omega_{A}$ and $\Omega_{B}$ to obtain \emph{any} of their possible
\emph{outcomes} $\alpha$ and $\beta$ when the \emph{preparation}
process is $c$ is given by (see Ref.\ \cite{Wiseman07a}, Eq.~(3), Ref.\ \cite{Cavalcanti09a}, Eq.~(15)) 
\begin{multline}
P(\alpha,\beta|\Omega_{A},\Omega_{B},c)  =  \sum\limits _{\lambda}P(\alpha|\Omega_{A},c,\lambda)
P(\beta|\Omega_{B},c,\lambda)
\\ \times
P(\lambda|c) . \label{Eq.GeneralLocalHVTJointProb}
\end{multline}
In \emph{LHVT} the \emph{hidden variables} $\lambda$ are \emph{global}
and first determined (probabilistically) via the \emph{preparation}
process, but then act \emph{locally} to determine the \emph{sub-system}
measurement \emph{probabilities\,}$P(\alpha|\Omega_{A},c,\lambda)\,$and
$P(\beta|\Omega_{B},c,\lambda)$- even in the situation where the
sub-systems are localised in \emph{well-separated} spatial regions
and the two sub-system measurements occur \emph{simultaneously}. The
probabilities are then finally combined in accordance with \emph{classical
probability theory} to determine the joint measurement probability.
States for which the joint probability is given by the local hidden
variable theory Eq.\ (\ref{Eq.GeneralLocalHVTJointProb}) are referred
to as \emph{Bell local}. State where this does not apply are the \emph{Bell
non-local} states.

In a \emph{non-local hidden variable theory} we would just have $P(\alpha,\beta|\Omega_{A},\Omega_{B},c)
= \sum_{\lambda}P(\alpha,\beta|\Omega_{A},\Omega_{B},c,\lambda)\,P(\lambda|c)$,
with no local sub-system probabilities involved. Here $P(\alpha,\beta|\Omega_{A},\Omega_{B},c,\lambda)$
is the joint probability that measurement of the \emph{observables}
$\Omega_{A}$, $\Omega_{B}$, of sub-systems $A$, $B$ results in
\emph{outcomes} $\alpha$, $\beta$ when the \emph{hidden variables}
are represented by $\lambda$, and
$P(\alpha,\beta|\Omega_{A},\Omega_{B},c)$
is not factorisable.

For LHVT the sub-system probabilities $P(\alpha|\Omega_{A},c,\lambda)$ and
$P(\beta|\Omega_{B},c,\lambda)$ are \emph{not necessarily} given
by quantum expressions such as (\ref{Eq.QuantumSingleProb}) though
they \emph{may} be. Following the approach of Refs.\
\cite{Wiseman07a,Jones07a} we will introduce a \emph{more specific} notation
(subscript $Q$) to distinguish cases where $P(\alpha|\Omega_{A},c,\lambda)\,$and/or
$P(\beta|\Omega_{B},c,\lambda)$ \emph{are} given by quantum expressions
from those where they are not. When the $P_{Q}(\gamma|\Omega_{C},c,\lambda)$
for sub-system $C$ $(C=A,B)$ are determined from a quantum expression
which involves a \emph{density operator} $\widehat{\rho}^{C}(c,\lambda)$
for sub-system $C$ determined from the hidden variables $\lambda$,
then $\widehat{\rho}^{C}(c,\lambda)$ specifies a so-called \emph{local
hidden state} (LHS).

The \emph{single measurement} probabilities $P(\alpha|\Omega_{A},c,\lambda)$ and
$P(\beta|\Omega_{B},c,\lambda)$ must of course satisfy the general
requirements of being \emph{real}, \emph{positive} and such that their
\emph{sum} over all possible outcomes is \emph{unity} for each value
$\lambda$ of the LHV in accordance with the general requirements
(\ref{Eq.SingleProbSumRules}). Thus 
\begin{equation}
\sum\limits _{\alpha}P(\alpha|\Omega_{A},c,\lambda)=1,\qquad\sum\limits _{\beta}P(\beta|\Omega_{B},c,\lambda)=1 . \label{Eq.LHVTheorySingleProb}
\end{equation}
By combining (\ref{Eq.LHVProbHiddenVar}) and (\ref{Eq.LHVTheorySingleProb})
it is straightforward to show that the joint probability $P(\alpha,\beta|\Omega_{A},\Omega_{B},c)$
satisfies the standard probability sum rule (\ref{Eq.SumJointProballOutcomes}).
Again, using (\ref{Eq.LHVTheorySingleProb}) and (\ref{Eq.LHVTSingleProb})
the general relationships (\ref{Eq.SingleProbMeastA}) and (\ref{Eq.SingleProbMeasureB})
between the joint and single measurement probabilities occur.

The overall probability $P(\alpha|\Omega_{A},c)$ that measurement
of the \emph{observable} $\Omega_{A}$ of sub-system $A$ results
in \emph{outcome} $\alpha$ when the \emph{preparation }process is
$c$ irrespective of the outcome for measurement of the \emph{observable}
$\Omega_{B}$ of sub-system $B$ is obtained by summing $P(\alpha,\beta|\Omega_{A},\Omega_{B},c)$
over $\beta$ (see (\ref{Eq.SingleProbMeastA})), so it is given by
the sum over the possible values $\lambda$ of the hidden variables
of the $P(\alpha|\Omega_{A},c,\lambda)$ times the preparation probability
$P(\lambda|c)$. A similar expression applies for $P(\beta|\Omega_{B},c)$.
Thus using (\ref{Eq.GeneralLocalHVTJointProb}) and (\ref{Eq.LHVTheorySingleProb})
\begin{eqnarray}
P(\alpha|\Omega_{A},c) & = & \sum\limits _{\lambda}P(\alpha|\Omega_{A},c,\lambda)\,P(\lambda|c),\nonumber \\
P(\beta|\Omega_{B},c) & = & \sum\limits _{\lambda}P(\beta|\Omega_{B},c,\lambda)\,P(\lambda|c).\label{Eq.LHVTSingleProb}
\end{eqnarray}
Under the condition of Bell locality, the results (\ref{Eq.LHVTSingleProb})
show that in a LHVT the measurement probability for an observable
$\Omega_{A}$ of sub-system $A$ is independent of the results for
measuring an observable $\Omega_{B}$ of sub-system $B$, and do not
even depend on which observable $\Omega_{B}$ is being measured. The
same applies if the sub-systems are reversed. This important result
for LHVT is called the \emph{no-signalling theorem }and shows that
a choice of observable to be measured in one sub-system cannot affect
the result of measurements in the other sub-system.

The \emph{conditional probabilities }are given by the general expressions
(\ref{Eq.LHVTConditionalProb}) that apply for both quantum and LHVT
cases.

We can use (\ref{Eq.GeneralLocalHVTJointProb}) to obtain an expression
for the \emph{mean value} of the \emph{joint measurement} of observables
$\Omega_{A}$ and $\Omega_{B}$ when the preparation process is $c$.
This will be given by 
\begin{eqnarray}
\left\langle \Omega_{A}\otimes\Omega_{B}\right\rangle  & = & \sum\limits _{\alpha,\beta}\alpha\,\beta\,P(\alpha,\beta|\Omega_{A},\Omega_{B},c)\nonumber \\
 & = & \sum\limits _{\lambda}\left\langle \Omega_{A}(c,\lambda)\right\rangle \,\left\langle \Omega_{B}(c,\lambda)\right\rangle \,P(\lambda|c),\label{Eq.ExpnValueJointMeastLHVT}
\end{eqnarray}
where $\left\langle \Omega_{A}(c,\lambda)\right\rangle \equiv\left\langle \Omega_{A}(\lambda)\right\rangle $
is the expectation value of observable $\Omega_{A}$ when the preparation
process $c$ leads to hidden variables $\lambda$, with $\left\langle \Omega_{B}(c,\lambda)\right\rangle \equiv\left\langle \Omega_{B}(\lambda)\right\rangle $
the corresponding expectation value for observable $\Omega_{B}$.
These are given by
\begin{eqnarray}
\left\langle \Omega_{A}(c,\lambda)\right\rangle  & = & \sum\limits _{\alpha}\alpha\,P(\alpha|\Omega_{A},c,\lambda)\nonumber \\
\left\langle \Omega_{B}(c,\lambda)\right\rangle  & = & \sum\limits _{\beta}\beta\,P(\beta|\Omega_{B},c,\lambda).\label{Eq.MeanSubSysObservHVs}
\end{eqnarray}

The \emph{mean value} for the measurement of a \emph{single observable}
$\Omega_{A}$ is
\begin{equation}
\left\langle \Omega_{A}\right\rangle =\sum\limits _{\alpha}\alpha\,P(\alpha|\Omega_{A},c)=\sum\limits _{\lambda}\left\langle \Omega_{A}(c,\lambda)\right\rangle \,P(\lambda|c)\label{Eq.MeanValueSingleLHVT}
\end{equation}
as can be derived from (\ref{Eq.LHVTSingleProb}) and (\ref{Eq.MeanSubSysObservHVs}).
A similar result applies for $\left\langle \Omega_{B}\right\rangle $.

In a \emph{deterministic} (or non-fuzzy) version of LHVT
$\left\langle \Omega_{A}(c,\lambda)\right\rangle =\alpha(c,\lambda)$
and $\left\langle \Omega_{B}(c,\lambda)\right\rangle =\beta(c,\lambda)$,
where $\alpha(c,\lambda)$ and $\beta(c,\lambda)$ are \emph{specific}
allowed outcomes for measurement of the observables when the preparation
process $c$ leads to hidden variables $\lambda$. Here the hidden
variables $\lambda$ determine \emph{unique} measurement outcomes
$\alpha(c,\lambda)$ and $\beta(c,\lambda)$. In the deterministic case
\begin{equation}
\left\langle \Omega_{A}\otimes\Omega_{B}\right\rangle =\sum\limits _{\lambda}\alpha(c,\lambda)\,\beta(c,\lambda)\,P(\lambda|c),\label{Eq.ExpnValueJointMeastNonFuzzyLHVT}
\end{equation}
which is a form originally used for $\left\langle \Omega_{A}\otimes\Omega_{B}\right\rangle $
by Bell (see Ref.\ \cite{Bell65a}). Thus in a non-fuzzy version
of LHVT the hidden variables \emph{uniquely} specify the measurement
outcomes, and it is only because the hidden variables are \emph{not
known} that they must be averaged over.

\subsection{Links between Quantum and Local Hidden Variable Theory}

\label{SubSection - Links between Quantum and LHVT}

In accordance with Einstein's basic idea that quantum theory predictions
for $P(\alpha,\beta|\Omega_{A},\Omega_{B},c)$ and $P(\alpha|\Omega_{A},c)$,
$P(\beta|\Omega_{B},c)$ are \emph{correct}, but can be \emph{interpreted}
in terms of an underlying \emph{reality} represented by a hidden variable
theory, it follows that the \emph{same} joint probability in (\ref{Eq.GeneralLocalHVTJointProb})
can \emph{also} be determined from the quantum theory expression (\ref{Eq.QuantumJointProb}).
Similarly for the single measurement probabilities $P(\alpha|\Omega_{A},c)$,
$P(\beta|\Omega_{B},c)$. Note that this \emph{assumes} that the particular
quantum state for the composite system \emph{can} be interpreted via
local hidden variable theory, which by definition excludes the \emph{Bell
non-local} states. As we have already noted, there are \emph{actual}
Bell non-local states where the quantum results are \emph{not} accountable
via LHVT - either \emph{theoretically} or \emph{experimentally}.
So it is only when we are considering \emph{Bell local} states that
these inter-relationships can be applied.

As indicated in Section \ref{Section - Introduction}, a key issue
is how to inter-relate the Hermitian operators that describe the observables
in quantum theory to the c-number variables describing the same observables
in LHVT, in order that valid comparisons between the predictions of
quantum and LHVT can be made.The approach that will be used is to
express all the quantum theory observables of interest in terms of
Hermitian operators associated with observables (such as position
and momentum) that have a classical counterpart, and then choose the
equivalent LHVT observables to have the same form as those in quantum
theory, except that the Hermitian operators will be replaced by c-number
variables. As indicated in Section \ref{Section - Introduction} it
will be necessary to introduce auxiliary observables whose c-number
versions have separate probability distributions. The procedure will
be discussed in more detail in Section \ref{Section - Spin Squeezing and Other Tests for EPR Entanglement}.

For \emph{Bell local} states, equating the LHVT (\ref{Eq.LHVTSingleProb})
and quantum theory (\ref{Eq.ReducedDensOprs})\ expressions for the
single measurement \emph{probability} $P(\alpha|\Omega_{A},c)$ we
obtain a LHVT - quantum theory relationship
\begin{eqnarray}
P(\alpha|\Omega_{A},c) & = & \sum\limits _{\lambda}P(\alpha|\Omega_{A},c,\lambda)\,P(\lambda|c),\qquad\qquad \text{LHVT}\nonumber \\
 & = & \Tr((\widehat{\Pi}_{\alpha}^{A}\otimes\widehat{1}^{B})\widehat{\rho}).\qquad\qquad\qquad\qquad \text{QT}\nonumber \\
&&
\label{Eq.LHVTQThyReln}
\end{eqnarray}
As $\Tr((\widehat{\Pi}_{\alpha}^{A}\otimes\widehat{1}^{B})\widehat{\rho})=\Tr_{A}(\widehat{\Pi}_{\alpha}^{A}\,\widehat{\rho}^{A})$
this shows that the \emph{hidden variable theory probability} $P(\alpha|\Omega_{A},c,\lambda)$
associated with single sub-system $A$ measurements and the \emph{reduced
density operator} $\widehat{\rho}^{A}$ for sub-system $A$ are inter-related.
A similar result applies for $P(\beta|\Omega_{B},c)$. However, this
relationship does \emph{not} mean that $P(\alpha|\Omega_{A},c,\lambda)$
can always be \emph{determined} from a sub-system density operator
which is \emph{not} dependent on the overall quantum state $\widehat{\rho}$
describing \emph{both} sub\textendash systems together - in general
the reduced density operator for each sub\textendash system is determined
from the \emph{full} density operator $\widehat{\rho}$. However,
when there is a local hidden state, the reduced density operator $\widehat{\rho}^{A}$
may be replaced by the form $\widehat{\rho}^{A}(c,\lambda)$ - which
is determined specifically for sub-system $A$ for preparation process
$c$ via the hidden variables $\lambda$.

Similar considerations apply for \emph{Bell local} states to the joint
measurement \emph{probability} $P(\alpha,\beta|\Omega_{A},\Omega_{B},c)$.
We have a second LHVT - quantum theory relationship:
\begin{align}
P(\alpha ,\beta |\Omega _{A},\Omega _{B},c) &= \sum\limits_{\lambda}
P(\alpha |\Omega _{A},c,\lambda )\,P(\beta |\Omega_{B},c,\lambda )  
\nonumber\\ & \qquad \qquad \times 
P(\lambda |c) ,
 \hspace{0.25\columnwidth} \text{LHVT}  \nonumber \\
\nonumber \\
&= \Tr((\widehat{\Pi }_{\alpha }^{A}\otimes \widehat{\Pi }_{\beta }^{B})%
\widehat{\rho })\,.  \hspace{0.25\columnwidth} \text{QT}  
\nonumber\\ &
\label{Eq.LHVTQThyRelnJointProb}
\end{align}
Also, for \emph{Bell local} states we can inter-relate the quantum
and LHVT \emph{mean values} of the\ \emph{joint} measurement of observables
$\Omega_{A}$\ and $\Omega_{B}$\ when the preparation process is
$c$. Using (\ref{Eq.MeanQThy}) and (\ref{Eq.ExpnValueJointMeastLHVT})
we have 
\begin{eqnarray}
\left\langle \Omega_{A}\otimes\Omega_{B}\right\rangle  & = & \sum\limits _{\lambda}\left\langle \Omega_{A}(c,\lambda)\right\rangle \,\left\langle \Omega_{B}(c,\lambda)\right\rangle \,P(\lambda|c),
\quad \text{LHVT}\nonumber \\
 & = & \Tr(\widehat{\Omega}_{A}\otimes\widehat{\Omega}_{B})\widehat{\rho}=\left\langle \widehat{\Omega}_{A}\otimes\widehat{\Omega}_{B}\right\rangle \qquad\quad \text{QT}\nonumber \\
\label{Eq.MeanValueJointMeastsQThyLHVT}
\end{eqnarray}
in cases where the LHVT can be applied.

In the case of \emph{mean values} for a \emph{single} observable,
we have similarly
\begin{eqnarray}
\left\langle \Omega_{A}\right\rangle  & = & \left\langle \Omega_{A}\otimes1_{B}\right\rangle =\sum\limits _{\lambda}\left\langle \Omega_{A}(c,\lambda)\right\rangle \,P(\lambda|c),\qquad \text{LHVT}\nonumber \\
 & = & \Tr[(\widehat{\Omega}_{A}\otimes\widehat{1}_{B})\widehat{\rho}]=\left\langle \widehat{\Omega}_{A}\otimes\widehat{1}_{B}\right\rangle =\left\langle \widehat{\Omega}_{A}\right\rangle \qquad \text{QT}\nonumber \\
\label{Eq.MeanValueSingleMeastQThyLHVT}
\end{eqnarray}
for \emph{Bell local} states. A similar result applies for $\left\langle \Omega_{B}\right\rangle $.
These results are all useful for \emph{inter-converting} LHVT and
quantum theory expressions, for the Bell local states.

The above results assume that there is a well-defined \emph{relationship}
for the c-numbers that represent the observables $\Omega_{A}$, $\Omega_{B}$
in LHVT and the Hermitian operators $\widehat{\Omega}_{A}$, $\widehat{\Omega}_{B}$
that represent the \emph{same} observables in quantum theory. It is
also required that the LHVT involves the same measurement \emph{outcomes}
$\alpha$, $\beta$ apply as for quantum theory. Other constraints
on the LHVT\ probability distributions would need to be imposed if
the LHVT is required to be consistent with quantum theory features
such as the \emph{Heisenberg uncertainty principle} for observables
with non-commuting quantum operators. This issue is not addressed
here.

As previously emphasised, a key point is that because LHVT \emph{underlies}
quantum theory, \emph{any} result we establish mean values, variances
of observables $\Omega_{A}$, $\Omega_{B}$ using LHVT for a quantum
state that is \emph{also} Bell local, can immediately be expressed
in terms of the equivalent Hermitian \emph{operators} observables
$\widehat{\Omega}_{A}$, $\widehat{\Omega}_{B}$ that describe the
same observables, together with the quantum \emph{density operator}
$\widehat{\rho}$ that specifies the \emph{same} state instead of
the set of LHVT \emph{probabilities} $P(\alpha|\Omega_{A},c,\lambda)\,$,
$P(\beta|\Omega_{B},c,\lambda)\,$and$\,P(\lambda|c)$.
Except in the case of a LHS there are no quantum
expressions for quantities such as $P(\alpha |\Omega _{A},c,\lambda )$, $%
\left\langle \Omega _{A}(c,\lambda )\right\rangle $, so no attempt will be
made to replace these by quantum expressions. Also, both the Bell
inequalities and the tests for EPR\ steering only involve mean values of
various observables, a primary emphasis will be on the two expressions
(\ref{Eq.MeanValueSingleMeastQThyLHVT}) and (\ref{Eq.MeanValueJointMeastsQThyLHVT})
involving mean values of either single sub-system observables or pairs of
such observables.

We will also need to consider the mean values for observables which
in quantum theory are given by the \emph{sum} of \emph{products} of
sub-system Hermitian operators, where the operators for each sub-system
do not necessarily commute - $[\widehat{\Omega}_{A1},\widehat{\Omega}_{A2}]\neq0$
etc.. The links between quantum theory and LHVT for these cases are
set out in Appendix \ref{Appendix - Mean Values and Variances}.

\section{Categories of Quantum States for Bipartite Systems}

\label{Section - Classes of Quantum States}

\subsection{Two Hierarchies of Bipartite Quantum States}

\label{SubSection - Two Hierarchies of Bipartite Q States}

As indicated in Section \ref{Section - Introduction} there are various
ways the quantum states for bipartite systems can be \emph{categorised},
and quantum states falling into a particular category in one scheme
\emph{may not} all end up in the same category in a different scheme.
Jones et al.\ \cite{Jones07a} (as elaborated by Cavalcanti et al.\ \cite{Cavalcanti09a}),
established a hierarchy of \emph{bipartite quantum states} can be
established based on \emph{LHVT models} for the \emph{joint probability}
$P(\alpha,\beta|\Omega_{A},\Omega_{B},c)$ for measurement of \emph{any}
pair of sub-system \emph{observables} $\Omega_{A}$ and $\Omega_{B}$
to obtain \emph{any} of their possible \emph{outcomes} $\alpha$ and
$\beta$ when the \emph{preparation} process is $c$. However before
considering this hierarchy we first identify a \emph{classification}
based purely on \emph{quantum state models}.

\subsection{Separable and Entangled States}

\label{SubSection - Separable and Entangled}

The quantum states for bipartite composite systems may be divided
into \emph{two classes} - the \emph{separable} and the \emph{entangled}
states. We will refer to this scheme as the \emph{Quantum Theory Classification
Scheme} (QTCS).

The \emph{separable} states are those whose preparation is described
by the density operator
\begin{equation}
\widehat{\rho}_{sep}=\sum\limits _{R}P_{R}\,\widehat{\rho}_{R}^{A}\otimes\widehat{\rho}_{R}^{B},\label{Eq.QuantSepState}
\end{equation}
where $\widehat{\rho}_{R}^{A}$ and $\widehat{\rho}_{R}^{B}$ are
\emph{possible} quantum states for sub-systems $A$ and $B$ respectively
and $P_{R}$ is the probability that this \emph{particular pair} of
sub-system states is prepared. Each distinct pair is listed by $R$.
This follows the preparation process for separable states described
by Werner \cite{Werner89a}. Such quantum states are of the same form
as what Werner \cite{Werner89a} referred to as \emph{uncorrelated
states}, but which nowadays would be referred to as separable or \emph{non-entangled}
states. The \emph{entangled} states are simply the quantum states
that are \emph{not }separable. A detailed discussion of the significance
of separable and entangled states, and tests for distinguishing these
is given in many articles and textbooks (see for example \cite{Dalton16a,Dalton16b}). Clearly for each choice of sub-systems a given
quantum state is \emph{either} separable \emph{or} entangled - it
cannot be both.

For the present we note that \emph{if} the quantum state is \emph{separable}
then from (\ref{Eq.QuantumJointProb}) and (\ref{Eq.QuantSepState})
the joint probability $P(\alpha,\beta|\Omega_{A},\Omega_{B},c)$ is
given by
\begin{eqnarray}
P(\alpha ,\beta |\Omega _{A},\Omega _{B},c) &=&\sum\limits_{R}P_{R}\,\Tr_{A}(%
\widehat{\Pi }_{\alpha }^{A}\,\widehat{\rho }_{R}^{A})\,\Tr_{B}(\widehat{\Pi }%
_{\beta }^{B}\,\widehat{\rho }_{R}^{B}) , \nonumber \\
&&  \label{Eq.SepStateJointProb} \\
&=&\sum\limits_{R}P_{R}\,P(\alpha |\Omega _{A},c(A,R))\,  \nonumber \\
&&\qquad\times P(\beta |\Omega _{B},c(B,R)),  \label{Eq.SepStateJointProbForm}
\end{eqnarray}%
where 
\begin{eqnarray}
P(\alpha |\Omega _{A},c(A,R)) &=&\Tr_{A}(\widehat{\Pi }_{\alpha }^{A}\,%
\widehat{\rho }_{R}^{A}) \,, \nonumber \\
P(\beta |\Omega _{B},c(B,R)) &=&\Tr_{B}(\widehat{\Pi }_{\beta }^{B}\,\widehat{%
\rho }_{R}^{B})  \label{Eq.SubSystemSingleProb}
\end{eqnarray}%
are the \emph{sub-system probabilities} for outcomes $\alpha$, $\beta$
for measurements of observables $\Omega_{A}$, $\Omega_{B}$ when
the sub-system preparations specify density operators as $c(A,R)\rightarrow\widehat{\rho}_{R}^{A}$,
$c(B,R)\rightarrow\widehat{\rho}_{R}^{B}$.

Alternatively, \emph{if} the joint probability is given by (\ref{Eq.SepStateJointProb})
for \emph{all} observables and outcomes then we can show that $P(\alpha,\beta|\Omega_{A},\Omega_{B},c)=\Tr((\widehat{\Pi}_{\alpha}^{A}\otimes\widehat{\Pi}_{\beta}^{B})\widehat{\rho})$,
where $\widehat{\rho}=\sum_{R}P_{R}\,\widehat{\rho}_{R}^{A}\otimes\widehat{\rho}_{R}^{B}$
- so the state is separable. Thus the density operator definition
and the joint probability expression for a separable state are \emph{equivalent}.

\subsection{Bell Local and Non-Local States}

\label{SubSection - Bell Local and Non-Local}

Based on LHVT the quantum states for bipartite composite systems may
\emph{also} be \emph{differently} divided into \emph{two other classes}
- the \emph{Bell local} and the \emph{Bell-non-local} states. We
will refer to this scheme as the \emph{Local Hidden Variable Theory
Classification Scheme} (LHVTCS). As we will see, there is \emph{no
simple} relationship between the entangled states on the one hand
and the Bell non-local states on the other, (nor between the separable
states on the one hand and the Bell local states on the other). The
\emph{Bell local} states are those for which the joint probability
$P(\alpha,\beta|\Omega_{A},\Omega_{B},c)$ is given by the LHVT expression
(\ref{Eq.GeneralLocalHVTJointProb}) \emph{as well as} the quantum
theory expression (\ref{Eq.QuantumJointProb}). In contrast, the \emph{Bell
non-local} states are those for which there is \emph{no} LHVT expression
(\ref{Eq.GeneralLocalHVTJointProb}) for the joint probability - this
is \emph{only} given by the quantum theory expression (\ref{Eq.QuantumJointProb}).

Before looking at \emph{further classes} of quantum states defined
in terms of LHVT we first present an important result, namely that
\emph{all} \emph{separable} states are \emph{Bell local}. The \emph{formal
similarity} between the hidden variable theory expression for the
joint probability (\ref{Eq.GeneralLocalHVTJointProb}) and the quantum
expression (\ref{Eq.SepStateJointProbForm}) for a \emph{separable}
state is noticeable. We can then identify the probabilistic choice
$R$ for the preparation of the \emph{particular pair} of sub-system
states $\widehat{\rho}_{R}^{A}$ and $\widehat{\rho}_{R}^{B}$ with
a \emph{particular choice} of hidden variables $\lambda$, thus $R\rightarrow\lambda$.
The $\widehat{\rho}_{R}^{A}$ and $\widehat{\rho}_{R}^{B}$ thus specify
\emph{local hidden states}. Then the probability $P_{R}$ for this particular
pair of sub-system states $\widehat{\rho}_{R}^{A}$ and $\widehat{\rho}_{R}^{B}$
can be identified with the hidden variable probability $P(\lambda|c)$,
thus $P_{R}\rightarrow P(\lambda|c)$. Next, the probabilities $P(\alpha|\Omega_{A},c(A,R))\,$and
$P(\beta|\Omega_{B},c(B,R))$ for the single sub-system probabilities
can be identified with the hidden variable probabilities $P(\alpha|\Omega_{A},c,\lambda)\,$and
$P(\beta|\Omega_{B},c,\lambda)$, thus $P(\alpha|\Omega_{A},c(A,R))\rightarrow P(\alpha|\Omega_{A},c,\lambda)$
and $P(\beta|\Omega_{B},c,\lambda)\rightarrow P(\beta|\Omega_{B},c,\lambda)$.
With these identifications the joint probability $P(\alpha,\beta|\Omega_{A},\Omega_{B},c)$
for a separable state (\ref{Eq.SepStateJointProbForm}) \emph{is}
of the general form for the joint probability $P(\alpha,\beta|\Omega_{A},\Omega_{B},c)$
for a Bell local state (\ref{Eq.GeneralLocalHVTJointProb}). Hence
the separable states are Bell local.

Thus, for the quantum \emph{separable} states the joint probability
can be written as 
\begin{multline}
P(\alpha,\beta|\Omega_{A},\Omega_{B},c)=\sum\limits _{\lambda}P_{Q}(\alpha|\Omega_{A},c,\lambda)\,P_{Q}(\beta|\Omega_{B},c,\lambda)
\\ \times
P(\lambda|c)\label{Eq.LHVJointProbCategory1}
\end{multline}
where the single probabilities are given by \emph{quantum
theory} expressions 
\begin{align}
P(\alpha|\Omega_{A},c,\lambda) & =  \Tr_{A}(\widehat{\Pi}_{\alpha}^{A}\,\widehat{\rho}_{R}^{A})=P_{Q}(\alpha|\Omega_{A},c,\lambda),\nonumber \\
P(\beta|\Omega_{B},c,\lambda) & =  \Tr_{B}(\widehat{\Pi}_{\beta}^{B}\,\widehat{\rho}_{R}^{B})=P_{Q}(\beta|\Omega_{B},c,\lambda),\label{Eq.SingleProbSepStates}
\end{align}
where the subscript $Q$ indicates that a quantum theory expression
applies.

It therefore follows that \emph{all} Bell \emph{non-local} states
are \emph{quantum entangled}. After all, if the quantum state is Bell
non-local and is also separable, then the separable state expression
(\ref{Eq.SepStateJointProbForm}) applies for the joint measurement
probability, which being of the required form for LHVT leads to the
contradictory result that the state was Bell local. Thus, \emph{all}
quantum separable states are \emph{Bell local} and \emph{all} Bell
non-local states are quantum \emph{entangled}. Note however that the
converses are \emph{not} true. As we will see, \emph{some} Bell
local states are \emph{not} quantum separable, that is they are quantum
entangled. Similarly, \emph{some} quantum entangled states are \emph{not}
Bell non-local, that is they are Bell local. This last result was
established by Werner \cite{Werner89a}. 

\subsection{Categories of Bell Local States}

\label{SubSection - Categories of Bell Local States}

This situation for separable states suggests that the \emph{Bell local
states} for \emph{bipartite} systems may be divided up into \emph{three}
classes depending on the \emph{number} of single sub-system probabilities
that are \emph{definitely} described by quantum expressions involving
the \emph{density operator} $\widehat{\rho}^{C}(c,\lambda)$ for a
\emph{local hidden state} (LHS) and a \emph{projector} $\widehat{\Pi}_{\omega}^{C}$
associated with measurement outcome $\omega$ for observable $\widehat{\Omega}_{C}$.
For bipartite systems there are three possibilities: firstly, \emph{Category
1} states where \emph{both} $P(\alpha|\Omega_{A},c,\lambda)$ and $P(\beta|\Omega_{B},c,\lambda)$
are given by quantum expressions as in (\ref{Eq.SingleProbSepStates});
secondly, \emph{Category 2} states where only \emph{one} is given by a quantum
expression; and thirdly, \emph{Category 3} states where \emph{neither} is
given by a quantum expression. The three classes or categories are
mutually exclusive - a given Bell local state can only be in \emph{one}
of the three classes. We now introduce a \emph{different} notation
in which (as in Eq.\ (\ref{Eq.SingleProbSepStates})) the \emph{presence}
of the sub-script $Q$ on a sub-system LHV probability indicates that
it \emph{can} be obtained from a quantum expression involving a sub-system
density operator for a local hidden state, and the \emph{absence}
of the sub-script $Q$ indicates that it is \emph{not} determined
from a quantum expression. Note that our notation \emph{differs} from
that in Refs.\ \cite{Wiseman07a,Jones07a,Cavalcanti09a}
where the $P(\alpha|\Omega_{A},c,\lambda)$ could be \emph{either}
$P(\alpha|\Omega_{A},c,\lambda)\,$(non-quantum) \emph{or} $P_{Q}(\alpha|\Omega_{A},c,\lambda)$
(quantum) in our notation. Hence in the present notation the joint
probabilities for the \emph{Bell local} states in \emph{Categories
1, 2 and 3} are given by
\begin{align}
  P(\alpha,\beta|\Omega_{A},\Omega_{B},c)  =  \sum\limits _{\lambda}&P_{Q}(\alpha|\Omega_{A},c,\lambda)\,P_{Q}(\beta|\Omega_{B},c,\lambda)
\nonumber\\ & \times
P(\lambda|c), \qquad\cata\label{Eq.CategoryOneStates}\\
P(\alpha,\beta|\Omega_{A},\Omega_{B},c)  =  \sum\limits _{\lambda}&P(\alpha|\Omega_{A},c,\lambda)\,P_{Q}(\beta|\Omega_{B},c,\lambda)
\nonumber\\ & \times
P(\lambda|c),
\qquad\catb\label{Eq.CategoryTwoStates}\\
P(\alpha,\beta|\Omega_{A},\Omega_{B},c)  =  \sum\limits _{\lambda}&P(\alpha|\Omega_{A},c,\lambda)\,P(\beta|\Omega_{B},c,\lambda)
\nonumber\\ & \times
P(\lambda|c).\qquad\catc\label{Eq.CategoryThreeStates}
\end{align}
When a quantum expression applies:
\begin{eqnarray}
\,P_{Q}(\alpha|\Omega_{A},c,\lambda) & = & \Tr_{A}(\widehat{\Pi}_{\alpha}^{A}\,\widehat{\rho}^{A}(c,\lambda)),\nonumber \\
\,P_{Q}(\beta|\Omega_{B},c,\lambda) & = & \Tr_{B}(\widehat{\Pi}_{\beta}^{B}\,\widehat{\rho}^{B}(c,\lambda)),\label{Eq.SingleProbLHVQuantumSubSys}
\end{eqnarray}
where $\widehat{\rho}^{A}(c,\lambda)$ and $\widehat{\rho}^{B}(c,\lambda)$
are the sub-system density operators for the local hidden states associated
with hidden variables $\lambda$ for preparation $c$. By \emph{convention}
for Category 2 states we choose $B$ to be the sub-system where the
single probability is given by a quantum expression.

We also list as \emph{Category 4} states those for which the joint
probability is \emph{not} given by \emph{any} of Eqs.\
(\ref{Eq.CategoryOneStates}), (\ref{Eq.CategoryTwoStates})
and (\ref{Eq.CategoryThreeStates}):
\begin{multline}
P(\alpha,\beta|\Omega_{A},\Omega_{B},c) \neq 
\text{Eqs.~(\ref{Eq.CategoryOneStates}), (\ref{Eq.CategoryTwoStates}) or (\ref{Eq.CategoryThreeStates})}.
\\
\catd \label{Eq.CategoryFourStates}
\end{multline}
For these states the joint probability is \emph{only} given by the
quantum theory expression (\ref{Eq.QuantumJointProb}). The Category
4 states are of course the \emph{Bell non-local} states, and such
states \emph{do} occur. If Einstein's realist approach applied there
would be \emph{no} Category 4 states.

To avoid confusion we note that Wiseman et al.\ \cite{Wiseman07a} also
introduced the term \emph{local hidden state model} to refer to the
situation when \emph{at least one} sub-system is associated with a
local hidden state. Thus the \emph{LHS model} applies to Category
1 and Category 2 states, but not to Category 3 and Category 4 states.

Clearly, all separable states are Category 1 states, and all Category
1 states are separable. The \emph{Category 1} states may also be just
referred to as \emph{separable} states. However, \emph{Category 2},
\emph{Category 3} and \emph{Category 4} states must be quantum \emph{entangled}
states. The four different categories of bipartite states have differing
features in regard to \emph{entanglement} based on their distinction
via the number of sub-systems associated with a local hidden state.

The feature of \emph{EPR steering} of sub-system $B$ from sub-system
$A$ is fully discussed in Refs.\ \cite{Wiseman07a,Jones07a,Cavalcanti09a}, and requires there must be \emph{no} local
hidden state $\widehat{\rho}^{B}(c,\lambda)$ for sub-system $B$.
For such states the sub-system $B$ said to be \emph{non-steerable}
from sub-system $A$. For completeness, a brief presentation of the
physical argument involved based on a consideration of states that
are conditional on the outcomes of measurements on sub-system $A$,
is set out in Appendix \ref{Appendix - EPR Steering}. Thus EPR steering
requires the \emph{failure} of the \emph{LHS model}. Hence Category
1 and Category 2 states are \emph{non-steerable}, whereas Category
3 and Category 4 states are \emph{steerable} since no local hidden
state for sub-system $B$ is involved. The Category 3 states, which
are Bell local, entangled, non LHS and steerable are sometimes referred
to as \emph{EPR entangled} states. Thus, based on their distinction
via the number of sub-systems associated with a local hidden state,
the four different categories of bipartite states also have differing
features in regard to \emph{EPR steering}.

As we have now seen, the Bell local states for \emph{bipartite} systems
can be divided up into three \emph{non-overlapping} subsets, each
of which has different features for the sub-system LHV probabilities
$P(\alpha|\Omega_{A},c,\lambda)\,$and $P(\beta|\Omega_{B},c,\lambda)$.
This distinctiveness between the sub-sets is of particular convenience
when we consider tests for various categories of states. However,
it should again be emphasised that other researchers
(\cite{Wiseman07a,Jones07a} and \cite{Cavalcanti09a}) have used a \emph{hierarchy}
of non disjoint sub-sets. This is because in certain of their definitions
the sub-system probabilities can be either given by quantum or non-quantum
expressions. In their scheme the sub-sets overlap, with each set being
a sub-set of a larger set. In their scheme Category 1 states (the
separable states) would be a sub-set of a set (the LHS states) consisting
of Category 1 and Category 2 states, where at least one sub-system
is in a local hidden state. In their scheme the Category 1 and Category
2 states would be combined and be a sub-set of a combined set (the
Bell local states) consisting of Category 1, Category 2 and Category
3 states. Thus the present scheme and that in Refs.\
\cite{Wiseman07a,Jones07a} and \cite{Cavalcanti09a} are \emph{not} the same
though they are \emph{related}, and this needs to be taken into
account when discussing tests. The \emph{overall scheme} used \emph{here}
is shown in Fig.~1, where the features for all the different sets
of states for bipartite composite systems are set out.
\begin{figure}[htb]
  \includegraphics[width=\columnwidth]{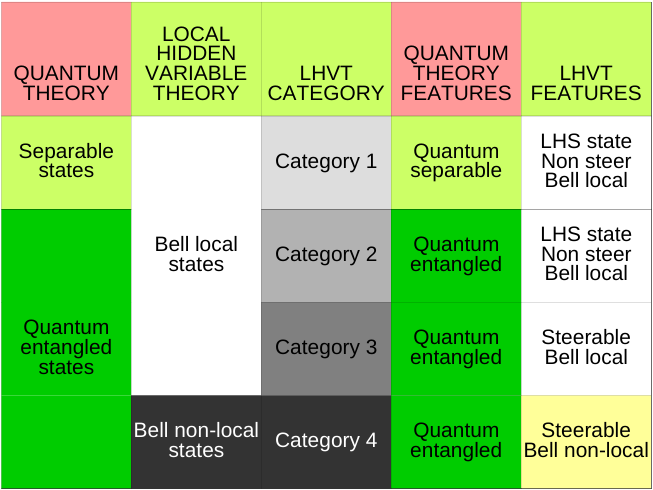}
  \caption[]{%
The Quantum Theory and the Local Hidden Variable Theory
Classification Schemes (QTCS and LHVCS). The two categories of quantum
states in the QTCS are shown in the left column and the two basic categories
of quantum states in the LHVCS are shown in the second left column. The four
more detailed categories of quantum states in the LHVCS are shown in the
third left column, whilst the right two columns lists the features of the
four categories of LHVCS states in both the QTCS and LHVCS schemes.}
\end{figure}

The mixed states introduced by Werner \cite{Werner89a} provide examples
of the three categories of Bell local states and of the Bell non-local
states. These are certain $U\otimes U$ invariant states ($(\widehat{U}\otimes\widehat{U})\,\widehat{\rho}_{W}\,(\widehat{U}^{\dag}\otimes\widehat{U}^{\dag})=\widehat{\rho}_{W}$,
where $\widehat{U}$ is any \emph{unitary} operator) for two $d$
dimensional sub-systems. Depending on the parameter $\eta$ (or $\phi$)
the Werner states (see Eq, (\ref{Eq.WernerStates})) may be separable
or entangled. They may also be Bell local and in one of the three
categories described above, or they may be Bell non-local. For completeness
the Werner states are described in Appendix \ref{Appendix - Werner States}.
The GHZ (or maximally entangled) pure state for two sub-systems, each
consisting of a spin $1/2$ particle considered by He et al.\ \cite{He11a},
and given by $(\left\vert \frac{1}{2},+\frac{1}{2}\right\rangle _{A}\left\vert \frac{1}{2},+\frac{1}{2}\right\rangle _{B}+\left\vert \frac{1}{2},-\frac{1}{2}\right\rangle _{A}\left\vert \frac{1}{2}-\frac{1}{2}\right\rangle _{B})/\sqrt{2}$
is an example of a Category 3 state, since it is entangled and steerable,
but is still Bell-local. As mentioned previously, the singlet state
\cite{Clauser69a} for the same system - given by $(\left\vert \frac{1}{2},+\frac{1}{2}\right\rangle _{A}\left\vert \frac{1}{2},+\frac{1}{2}\right\rangle _{B}-\left\vert \frac{1}{2},-\frac{1}{2}\right\rangle _{A}\left\vert \frac{1}{2}-\frac{1}{2}\right\rangle _{B})/\sqrt{2}$
- is an example of a Category 4 state, since it is entangled, steerable,
and is Bell non-local as it violates a Bell inequality. 

\section{Tests for EPR\ Steering in Bipartite Systems}

\label{Section - Spin Squeezing and Other Tests for EPR Entanglement}

\subsection{General Considerations}

\label{SubSection - General Tests for EPR Steering}

In a number of papers (see the review papers \cite{Dalton16a,Dalton16b}
and references therein) various tests for quantum \emph{entanglement}
have been formulated, recently in the particular context of bipartite
systems of identical massive bosons \cite{Dalton14a}. The focus was
on the situation of single mode sub-systems. These include spin and
two mode quadrature squeezing, Bloch vector and correlation tests.
An important issue then is: Are these tests also valid for detecting
\emph{EPR steering} or do some of them fail? As for the entanglement
tests, for the EPR steering tests we also focus on \emph{single
mode} sub-systems. Of course any test that detects EPR steering must
of necessity also detect entanglement, but a test that demonstrates
entanglement does not necessarily demonstrate EPR steering. In this
situation we are looking for conditions where there is no local hidden
state for sub-system $B$ - or in other words, the quantum state does
\emph{not }have a joint measurement probability as in Eqs.\ (\ref{Eq.CategoryOneStates})
and (\ref{Eq.CategoryTwoStates}) for Category 1 or Category 2 states.
Thus EPR steering requires the failure of the LHS model. As the tests
for \emph{quantum entanglement} previously obtained have already found
the conditions under which Category 1 probabilities fail, we \emph{then}
know that the quantum state must be in Category 2, Category 3 or Category
4. If we can then show that it is \emph{not} in Category 2 because
the joint measurement probability (\ref{Eq.CategoryTwoStates}) also
fails, then the state \emph{must} be in Category 3 or Category 4 -
in other words it is an \emph{EPR steerable} state. We would then
have found a \emph{test} for \emph{EPR steering}. Note that for the
Category 2 states the sub-system $A$ probabilities $P(\alpha|\Omega_{A},c,\lambda)$
in LHVT\ are \emph{not} given by a quantum expression involving a
\emph{sub-system} density operator. \emph{This feature} must be taken
into account when considering the tests for EPR\ steering. However,
the issue of how to treat mean values and variances in the context
of LHVT in general requires some consideration, so we have set this
out in Appendix \ref{Appendix - Mean Values and Variances}.

Note however that a test that demonstrates EPR steering only shows
that the quantum state is either Category 3 or Category 4, both of
which are entangled states. To demonstrate \emph{Bell non-locality}
(Category 4 states) will require different tests - notably those involving
violations of a \emph{Bell inequality}. This will be the subject of
a later paper. As has been emphasised in Section \ref{Section - Introduction},
showing that a Bell inequality is violated demonstrates that the state
cannot be in Categories 1, 2 or 3, so it must be a Bell non-local
state (Category 4). However, we emphasise again the point that the
tests presented here show what category (or categories) the quantum
state \emph{cannot} belong to - which does \emph{not always} determine
what category of quantum state \emph{must} apply. The tests are
those of \emph{sufficiency} not \emph{necessity}.

In the present paper, as in previous work in
Refs.\ \cite{Dalton14a,Dalton16a,Dalton16b}, we focus on tests for bipartite
systems involving \emph{identical massive bosons}. Consequently, when
quantum states either for the overall system or for a sub-system are
involved these must comply with the \emph{symmetrization} principle
and \emph{super-selection rules} involving the total boson number
for either the overall system or for the sub-system. In particular,
for Category 2 states (as well as Category 1 states) the local hidden
state $\widehat{\rho}^{B}(c,\lambda)$ for the sub-system $B$ that
is treated quantum mechanically must have zero coherences between
Fock states with differing sub-system boson number $N_{B}$. The LHS
must be a possible quantum state for sub-system $B$. The issue of
super-selection rules is discussed fully in \cite{Dalton16a}.

Also, as in these papers both the overall system and the two sub-systems
will be specified in terms of \emph{modes} (or single particle states
that the particles may occupy) based on a second quantization treatment,
rather than in terms of labeled identical \emph{particles -} as might
be thought appropriate in a first quantization method. Cases with
differing \emph{numbers} of particles are just different \emph{states}
of the (multi) modal system, not different systems, as in first quantization.

In addition, since the mean values of various observables are involved
in the tests for showing the state is not Category 2, we can use Eqs.
(\ref{Eq.MeanValueSingleMeastQThyLHVT}) and (\ref{Eq.MeanValueJointMeastsQThyLHVT})
for overall system mean values to replace LHVT theory expressions
by quantum theory expressions at suitable stages in the derivations
- both when a sub-system $B$ LHS $\widehat{\rho}^{B}(c,\lambda)$
occurs or when we wish to evaluate the mean value of a sub-system
$A$ observable $\Omega_{A}$ allowing for all values of the hidden
variables $\lambda$. However, there will be situations for Category
2 states where we need to consider the mean value of a sub-system
$A$ observable $\Omega_{A}$ when the hidden variables have particular
values. In this case some general properties of classical probabilities
$P(\alpha|\Omega_{A},c,\lambda)$ are useful. These are not dependent
on $P(\alpha|\Omega_{A},c,\lambda)$ being obtained from a hidden
state density operator $\widehat{\rho}^{A}(c,\lambda)$. One is that
the mean of the square of a real observable is never less than the
square of the mean for the observable, that is
\begin{equation}
\left\langle \Omega_{A}^{2}(c,\lambda)\right\rangle \geq(\left\langle \Omega_{A}(c,\lambda)\right\rangle )^{2} . \label{Eq.ClassMeanSquareResult}
\end{equation}
Another, is a Cauchy inequality
\begin{equation}
\sum\limits _{\lambda}C(\lambda)\,P(\lambda|c)\geq\left(\sum\limits
  _{\lambda}\sqrt{C(\lambda)}\,P(\lambda|c)\right)^{2} \label{Eq.CauchyInequality}
\end{equation}
for $C(\lambda)\geq0$, such as the case $\ C(\lambda)=\left\langle \Omega_{A}^{2}(c,\lambda)\right\rangle $.
The proof of the first is elementary, the second is proved in Ref.\ \cite{Dalton16a}.
These results are only used to derive correlation tests (see Appendix
\ref{Appendix- Correlation Tests for EPR Steering}).

Finally, since LHVT deals with physical quantities that are \emph{classical}
observables we need to express various non-Hermitian quantum mechanical
operators that we need to consider - such as mode annihilation\ and
creation operators - in terms of quantum operators that are Hermitian.
Any non-Hermitian operator $\widehat{\Omega}$ can always be expressed
in terms of Hermitian operators $\widehat{\Omega}_{1}$ and $\widehat{\Omega}_{2}$
as $\widehat{\Omega}=$ $\widehat{\Omega}_{1}+i$ $\widehat{\Omega}_{2}$
and the latter operators would be equivalent to classical observables
$\Omega_{1}$\ and $\Omega_{2}$, so the corresponding classical
observable will be $\Omega=$ $\Omega_{1}+i$ $\Omega_{2}$. The mean
value $\langle \widehat{\Omega}\rangle $ will then be
equal to $\langle \widehat{\Omega}_{1}\rangle
+i\langle \widehat{\Omega}_{2}\rangle $.
Note that two independent sets of measurements for the generally
incompatible $\widehat{\Omega}_{1}$ and $\widehat{\Omega}_{2}$ would
be needed to separately determine $\langle \widehat{\Omega}_{1}\rangle $
and $\langle \widehat{\Omega}_{2}\rangle $. For the corresponding
classical observable we take $\langle \Omega\rangle =
\langle \Omega_{1}\rangle +i\langle \Omega_{2}\rangle $
- see Eq.\ (\ref{Eq.MeanComplexCombHVTDefn}) in Appendix \ref{Appendix - Mean Values and Variances}.
The bosonic annihilation and creation operators for each of the single
mode sub-systems are not Hermitian, so we replace these by pairs of
\emph{quadrature operators }$\widehat{x},\widehat{p}$, which are
then associated with classical \emph{quadrature observables} $x,p$
when LHVT\ is being considered. As we will see, we also need new
\emph{auxiliary} Hermitian operators $\widehat{U},\widehat{V}$ as
well, which are sums of products of quadrature operators and these
will also be associated with classical observables $U,V$ in the LHVT.
All the physical observables that we need to consider have quantum
operators that can be written as linear combinations of products $\widehat{\Omega}_{A}\otimes\widehat{\Omega}_{B}$,
where both $\widehat{\Omega}_{A}$ and $\widehat{\Omega}_{B}$ are
Hermitian - including cases where $\widehat{\Omega}_{A}=\widehat{1}_{A}$
or $\widehat{\Omega}_{B}=\widehat{1}_{B}$. Such products can then
be replaced by $\Omega_{A}\otimes\Omega_{B}$, where $\Omega_{A}$
and $\Omega_{B}$ are the corresponding classical observables. Using
this procedure both quantum and hidden variable theory expressions
can be used for the joint measurement probabilities and mean values.

\subsection{Spin and Quadrature Tests for EPR Steering}

\label{SubSection - Tests for EPR Steeing}

We now obtain a number of inequalities for spin and quadrature observables
that apply for Category 2 (and Category 1) states and apply these
to obtain tests for EPR steering. First, we consider whether tests
that have been shown to be sufficient to demonstrate quantum entanglement
(violation of Category 1) (see Ref.\ \cite{Dalton16b} for details)
are also valid for demonstrating EPR steering. Obviously a test that
demonstrates EPR steering must also demonstrate quantum entanglement,
but a test that demonstrates entanglement does not necessarily demonstrate
EPR steering. We first consider the Bloch vector tests, then spin
squeezing tests for $S_{z\text{ }}$and for the other spin components,
followed by planar spin variance tests (such as the Hillery-Zubairy
test) which involve the sum of the variances for $S_{x\text{ }}$and
$S_{y\text{ }}$and finally two mode quadrature squeezing tests .
Of these possible tests, the Bloch vector test, spin squeezing in
any spin component, the Hillery-Zubairy spin variance test and squeezing
in any two mode quadrature are valid for demonstrating EPR steering.
We also consider a generalised version of the Hillery- Zubairy spin
variance test, which also shows that EPR steering occurs. Finally,
we consider for completeness weak and strong correlation tests in
Appendix \ref{Appendix- Correlation Tests for EPR Steering}, though
these are equivalent to certain of the tests involving spin operators
already set out in this Section. 

\subsection{Quadrature Amplitudes}

\label{SubSection - Quadrature Amplitudes}

The non-Hermitian quantum mode annihilation or creation operators
can be replaced by their Hermitian components,which are the \emph{quadrature
operators}. In quantum theory these are given by 
\begin{align}
\widehat{x}_{A} & =  \frac{1}{\sqrt{2}}(\widehat{a}+\widehat{a}^{\dag}),&\widehat{p}_{A}&=\frac{1}{\sqrt{2}i}(\widehat{a}-\widehat{a}^{\dag}),\nonumber \\
\widehat{x}_{B} & =  \frac{1}{\sqrt{2}}(\widehat{b}+\widehat{b}^{\dag}),&\widehat{p}_{B}&=\frac{1}{\sqrt{2}i}(\widehat{b}-\widehat{b}^{\dag}),\label{Eq.PositionMtmOprs}
\end{align}
which have the same commutation rules as the position and momentum
operators for distinguishable particles in units where $\hbar=1$.
Thus $[\widehat{x}_{A},\widehat{p}_{A}]=[\widehat{x}_{B},\widehat{p}_{B}]=i$
as for cases where $A$, $B$ were distinguishable particles. It is
then reasonable to assume that there are equivalent classical observables
$x_{A},p_{A},x_{B},p_{B}$ and that their measurement outcomes would
be real numbers, and further more for sub-systems not being treated
quantum mechanically (such as sub-system $A$ in the context of Category
2 states) these outcomes can \emph{actually} be measured in \emph{experiment}
and probabilities and mean values such as $P(\alpha|\Omega_{A},c,\lambda)$
and $\left\langle \Omega_{A}(\lambda)\right\rangle $ can be assigned
as in a hidden variable treatment of sub-system $A$. However, in
considering Category 2 states the probabilities and mean values such
as $P(\beta|\Omega_{B},c,\lambda)$\ and $\left\langle \Omega_{B}(\lambda)\right\rangle $\
for the sub-system $B$\ are also given by quantum expressions involving
sub-system density operators $\widehat{\rho}^{B}(\lambda)$.

We can write the mode annihilation and creation operators in terms
of the quadrature operators as $\widehat{a}=(\widehat{x}_{A}+i\widehat{p}_{A})/\sqrt{2}$,
$\widehat{a}^{\dag}=(\widehat{x}_{A}-i\widehat{p}_{A})/\sqrt{2}$,
$\widehat{b}=(\widehat{x}_{B}+i\widehat{p}_{B})/\sqrt{2}$, $\widehat{b}^{\dag}=(\widehat{x}_{B}-i\widehat{p}_{B})/\sqrt{2}$
and then show that important observables can be expressed in terms
of the quadrature operators. In the case of the \emph{spin
operators} (defined as $\widehat{S}_{x}=(\widehat{b}^{\dag}\widehat{a}+\widehat{a}^{\dag}\widehat{b})/2$,
$\widehat{S}_{y}=(\widehat{b}^{\dag}\widehat{a}-\widehat{a}^{\dag}\widehat{b})/2i$,
$\widehat{S}_{z}=(\widehat{b}^{\dag}\widehat{b}-\widehat{a}^{\dag}\widehat{a})/2$)
and the\ \emph{number operators} (defined as $\widehat{N}=$\ $\widehat{N}_{A}+\widehat{N}_{B}$
with $\widehat{N}_{A}=\widehat{a}^{\dag}\widehat{a}$, $\widehat{N}_{B}=\widehat{b}^{\dag}\widehat{b}$
being the separate mode number operators - note that $\widehat{S}_{x}^{2}+\widehat{S}_{y}^{2}+\widehat{S}_{z}^{2}=\frac{\widehat{N}}{2}(\frac{\widehat{N}}{2}+1)$),
all these quantities can be expressed in terms of the quadrature operators
as follows 
\begin{eqnarray}
\widehat{S}_{x} & = & \frac{1}{2}(\widehat{x}_{A}\widehat{x}_{B}+\widehat{p}_{A}\widehat{p}_{B}),\qquad\widehat{S}_{y}=\frac{1}{2}(\widehat{p}_{A}\widehat{x}_{B}-\widehat{x}_{A}\widehat{p}_{B}),\nonumber \\
\widehat{S}_{z} & = & \frac{1}{4}(\widehat{x}_{B}^{2}-\widehat{x}_{A}^{2}+\widehat{p}_{B}^{2}-\widehat{p}_{A}^{2})-\frac{1}{2}\widehat{V}_{B}+\frac{1}{2}\widehat{V}_{A},\nonumber \\
\widehat{N} & = & \frac{1}{2}(\widehat{x}_{B}^{2}+\widehat{x}_{A}^{2}+\widehat{p}_{B}^{2}+\widehat{p}_{A}^{2})-\widehat{V}_{B}-\widehat{V}_{A},\nonumber \\
\label{Eq.SpinOprsQuadOprs}
\end{eqnarray}
which are all linear combinations of products of two quadrature operators.
Here we have introduced the \emph{auxiliary} Hermitian operators 
\begin{eqnarray}
\widehat{V}_{A} & = & \frac{1}{2i}(\widehat{x}_{A}\widehat{p}_{A}-\widehat{p}_{A}\widehat{x}_{A})=\frac{1}{2}\widehat{1}_{A\text{ }},\nonumber \\
\widehat{V}_{B} & = & \frac{1}{2i}(\widehat{x}_{B}\widehat{p}_{B}-\widehat{p}_{B}\widehat{x}_{B})=\frac{1}{2}\widehat{1}_{B\text{ }},\label{Eq.AuxOprsV}
\end{eqnarray}
using the commutation rules. These operators could represent observables
in quantum theory, albeit rather useless ones since all eigenstates
have the same eigenvalue of $1/2$. In terms of the quadrature and
auxiliary operators the \emph{mode number} and
mode \emph{number difference} operators are:
\begin{eqnarray}
\widehat{N}_{A} &=&\frac{1}{2}(\widehat{x}_{A}^{2}+\widehat{p}_{A}^{2})-%
\widehat{V}_{A},  \nonumber \\
\widehat{N}_{B} &=&\frac{1}{2}(\widehat{x}_{B}^{2}+\widehat{p}_{B}^{2})-%
\widehat{V}_{B} , \label{Eq.ModeNumbOprs} \\
\widehat{N}_{-} &=&\widehat{N}_{B}-\widehat{N}_{A}=2\widehat{S}_{z} ,
\nonumber \\
&=&\frac{1}{2}(\widehat{x}_{B}^{2}+\widehat{p}_{B}^{2}-\widehat{x}_{A}^{2}-%
\widehat{p}_{A}^{2})-\widehat{V}_{B}+\widehat{V}_{A} . 
\label{Eq.ModeNumberDiff}
\end{eqnarray}

As \emph{spin squeezing} was a test for \emph{entanglement} \cite{Dalton16b},
spin squeezing expressions for $\widehat{S}_{x}^{2}$, $\widehat{S}_{y}^{2}$\ and
$\widehat{S}_{z}^{2}$ will be required. We find that for $\widehat{S}_{x}^{2}$
and $\widehat{S}_{y}^{2}$ 
\begin{eqnarray}
\widehat{S}_{x}^{2} & = & \frac{1}{4}(\widehat{x}_{A}^{2}\widehat{x}_{B}^{2}+\widehat{p}_{A}^{2}\widehat{p}_{B}^{2})+\frac{1}{2}(\widehat{U}_{A}\widehat{U}_{B}-\widehat{V}_{A}\widehat{V}_{B}),\label{Eq.SxSquare}\\
\widehat{S}_{y}^{2} & = & \frac{1}{4}(\widehat{p}_{A}^{2}\widehat{x}_{B}^{2}+\widehat{x}_{A}^{2}\widehat{p}_{B}^{2})-\frac{1}{2}(\widehat{U}_{A}\widehat{U}_{B}+\widehat{V}_{A}\widehat{V}_{B}).\label{Eq.SySquare}
\end{eqnarray}
The spin operators thus involve the quadrature operators for both
modes. Here we have introduced two \emph{further} distinct auxiliary
Hermitian combinations of the quadrature operators for each mode:
\begin{eqnarray}
\widehat{U}_{A} & = & \frac{1}{2}(\widehat{x}_{A}\widehat{p}_{A}+\widehat{p}_{A}\widehat{x}_{A})=\frac{1}{2i}((\widehat{a})^{2}-(\widehat{a}^{\dag})^{2}),\nonumber \\
\widehat{U}_{B} & = & \frac{1}{2}(\widehat{x}_{B}\widehat{p}_{B}+\widehat{p}_{B}\widehat{x}_{B})=\frac{1}{2i}((\widehat{b})^{2}-(\widehat{b}^{\dag})^{2}),\label{Eq.UOprs}
\end{eqnarray}
where using the commutation rules the operators $\widehat{U}_{A}$\ and
$\widehat{U}_{B}$ can also be expressed in terms of mode annihilation
and creation operators.

In addition to the spin operators we can also define \emph{two mode
quadrature operators} in terms of the quadrature operators for both
modes \cite{Dalton16b}. These depend on a \emph{phase parameter}
$\theta$. There are two sets given by 
\begin{align}
\widehat{X}_{\theta}(\pm) & = 
 \frac{1}{2}\left(\widehat{a}\,e^{-i\theta}\pm\widehat{b}\,e^{+i\theta}+\widehat{a}^{\dag}\,e^{+i\theta}\pm\widehat{b}^{\dag}\,e^{-i\theta}\right),\nonumber \\
\widehat{P}_{\theta}(\pm) & = 
 \frac{1}{2i}\left(\widehat{a}\,e^{-i\theta}\mp\widehat{b}\,e^{+i\theta}-\widehat{a}^{\dag}\,e^{+i\theta}\pm\widehat{b}^{\dag}\,e^{-i\theta}\right).\label{Eq.TwoModeQuadOprs}
\end{align}
It is easy to see that $\widehat{P}_{\theta}(\pm)=\widehat{X}_{\theta+\pi/2}(\pm)$
and that $[\widehat{X}_{\theta}(+),\widehat{P}_{\theta}(+)]=[\widehat{X}_{\theta}(-),\widehat{P}_{\theta}(-)]=i$.
The Heisenberg uncertainty principle is given by $\left\langle \Delta X_{\theta}^{2}(\pm)\right\rangle \left\langle \Delta P_{\theta}^{2}(\pm)\right\rangle \geq1/4$
and a state is two mode quadrature \emph{squeezed} if one of $\left\langle \Delta X_{\theta}^{2}(\pm)\right\rangle $
or $\left\langle \Delta P_{\theta}^{2}(\pm)\right\rangle $ is less
than $1/2$. In Reference \cite{Dalton16b} we showed that \emph{two
mode quadrature squeezing} was a sufficiency test for \emph{entanglement}.
We can write the two mode quadrature operators in terms of the single
mode quadrature operators as:
\begin{eqnarray}
\widehat{X}_{\theta}(\pm) & = & \frac{1}{\sqrt{2}}\left(\widehat{x}_{A}\cos\theta+\widehat{p}_{A}\sin\theta\pm\widehat{x}_{B}\cos\theta\pm\widehat{p}_{B}\sin\theta\right),\nonumber \\
\widehat{P}_{\theta}(\pm) & = & \frac{1}{\sqrt{2}}\left(-\widehat{x}_{A}\sin\theta+\widehat{p}_{A}\cos\theta\mp\widehat{x}_{B}\sin\theta\pm\widehat{p}_{B}\cos\theta\right) . 
\nonumber \\ &&
\label{Eq.TwoModeQuadOprsB}
\end{eqnarray}
The square of the two mode quadrature operators $\widehat{X}_{\theta}(\pm)$
are given by 
\begin{eqnarray}
\widehat{X}_{\theta}(\pm)^{2} & = & \frac{1}{2}\left\{ \widehat{x}_{A}^{2}\cos^{2}\theta+\widehat{p}_{A}^{2}\sin^{2}\theta+2\widehat{U}_{A}\sin\theta\cos\theta\right\} \nonumber \\
 &  & +\frac{1}{2}\left\{ \widehat{x}_{B}^{2}\cos^{2}\theta+\widehat{p}_{B}^{2}\sin^{2}\theta+2\widehat{U}_{B}\sin\theta\cos\theta\right\} \nonumber \\
 &  & \pm\left\{ \widehat{x}_{A}\widehat{x}_{B}\cos^{2}\theta+\widehat{p}_{A}\widehat{p}_{B}\sin^{2}\theta  \right. \nonumber\\
&& \left. \qquad
+\widehat{x}_{A}\widehat{p}_{B}\sin\theta\cos\theta+\widehat{p}_{A}\widehat{x}_{B}\sin\theta\cos\theta\right\} . \nonumber \\
&&
\label{Eq.TwoModeQuadSquare}
\end{eqnarray}
The expression for $\widehat{P}_{\theta}(\pm)^{2}$ can be obtained
using $\widehat{P}_{\theta}(\pm)=\widehat{X}_{\theta+\pi/2}(\pm)$.

The fundamental quantum Hermitian operators $\widehat{x}_{A}$, $\widehat{p}_{A}$, $\widehat{x}_{B}$, $\widehat{p}_{B}$
for the two mode system plus the auxiliary Hermitian operators $\widehat{U}_{A}$, $\widehat{V}_{A}$, $\widehat{U}_{B}$, $\widehat{V}_{B}$
all correspond to physical quantities that could be measured, with
real eigenvalues as the outcomes. Following the general approach described
in Section \ref{Section - Introduction}, for local hidden variable
theory these quantities correspond to classical observables $x_{A}$, $p_{A}$, $x_{B}$, $p_{B}$
and $U_{A}, V_{A}, U_{B}, V_{B}$, for which single observable hidden
variable probabilities $P(\alpha|\Omega_{A},c,\lambda)$ and $P(\beta|\Omega_{B},c,\lambda)$
apply - from which joint probabilities $P(\alpha,\beta|\Omega_{A},\Omega_{B},c)$
can be obtained via (\ref{Eq.GeneralLocalHVTJointProb}). The physical
observables involved in the tests such as the spin operators, their
squares and the number operators can all be expressed in terms of
the quadrature and auxiliary operators as sums of products of the
form $\widehat{\Omega}_{A}\otimes\widehat{\Omega}_{B}$. For the local
hidden variable theory treatment the corresponding classical observables
will be the \emph{same} as the quantum expressions, but now with the
quantum Hermitian operators \emph{replaced} by the corresponding classical
observable. For the classical spin components $S_{x}$, $S_{y}$\ and
$S_{z}$\ and the number observable $N$ the expressions in terms
of quadrature amplitudes $x$, $p$ and auxiliary observables $U$,
$V$ are 
\begin{eqnarray}
S_{x} & = & \frac{1}{2}(x_{A}x_{B}+p_{A}p_{B}),\qquad S_{y}=\frac{1}{2}(p_{A}x_{B}-x_{A}p_{B}),\nonumber \\
S_{z} & = & \frac{1}{4}(x_{B}^{2}-x_{A}^{2}+p_{B}^{2}-p_{A}^{2})-\frac{1}{2}V_{B}+\frac{1}{2}V_{A},\nonumber \\
N & = & \frac{1}{2}(x_{B}^{2}+x_{A}^{2}+p_{B}^{2}+p_{A}^{2})-V_{B}-V_{A}\,.\nonumber \\
\label{Eq.SpinCompts}
\end{eqnarray}
The expressions in terms of quadrature amplitudes $x$, $p$ and auxiliary
observables $U$, $V$ for the sub-system particle numbers and their
difference are 
\begin{eqnarray}
N_{A} & = & \frac{1}{2}(x_{A}^{2}+p_{A}^{2})-V_{A},\qquad N_{B}=\frac{1}{2}(x_{B}^{2}+p_{B}^{2})-V_{B},\nonumber \\
N & = & N_{A}+N_{B},\nonumber \\
N_{-} & = & N_{B}-N_{A}=2S_{z},\nonumber \\
 & = & \frac{1}{2}(x_{B}^{2}+p_{B}^{2}-x_{A}^{2}-p_{A}^{2})-V_{B}+V_{A}\,.\label{Eq.ModeNumbersObserv}
\end{eqnarray}
The two mode quadrature observables are given by 
\begin{align}
X_{\theta}(\pm) &=  \frac{1}{\sqrt{2}}\left(x_{A}\cos\theta+p_{A}\sin\theta\pm x_{B}\cos\theta\pm p_{B}\sin\theta\right),\nonumber \\
P_{\theta}(\pm) &=  \frac{1}{\sqrt{2}}\left(
-x_{A}\sin\theta
+p_{A}\cos\theta
\mp x_{B}\sin\theta
\pm p_{B}\cos\theta\right).\nonumber\\
&\label{Eq.TwoModeQuadObserv}
\end{align}
For completeness we set out expressions for other observables in Appendix
\ref{Appendix - Classical Observables and Quadratures}. The reverse
process for the replacement of the classical observables $x_{A}$,
$x_{B}$, $p_{A}$, $p_{B}$ by $\widehat{x}_{A}$, $\widehat{x}_{B}$,
$\widehat{p}_{A}$, $\widehat{p}_{B}$ and $U_{A}$, $U_{B}$, $V_{A}$,
$V_{B}$ by $\widehat{U}_{A}$, $\widehat{U}_{B}$, $\widehat{V}_{A}$,
$\widehat{V}_{B}$ requires using (\ref{Eq.PositionMtmOprs}), (\ref{Eq.UOprs})
and (\ref{Eq.AuxOprsV}) to give the correct quantum Hermitian operators.
This requires writing $V_{A}=(x_{A}p_{A}-p_{A}x_{A})/2i$ and $U_{A}=(x_{A}p_{A}+p_{A}x_{A})/2$
etc.\ before substituting $x_{A}$ by $\widehat{x}_{A}$, $p_{A}$
by $\widehat{p}_{A}$ etc., rather than $V_{A}=0$ and $U_{A}=2x_{A}p_{A}$
etc., but this is not surprising as c-number variables are not mathematically
identical to Hermitian operators. Carrying out this replacement in
the \emph{classical} spin components $S_{x}$, $S_{y}$\ and $S_{z}$\ and
the number observable $N$ also gives the correct \emph{quantum} operators,
as also occurs for the squares of these observables as well. Once
again we emphasise that we only need single measurement LHVT\ probabilities
$P(\alpha|\Omega_{A},c,\lambda)$ with $\Omega_{A}=x_{A}$, $p_{A}$,
$U_{A}$ or $V_{A}$ and $P(\beta|\Omega_{B},c,\lambda)$ with $\Omega_{B}=x_{B}$,
$p_{B}$, $U_{B}$ or $V_{B}$ to treat the classical observables
such as $S_{x}$, $S_{y}$\ and $S_{z}$\ and $N$ or $X_{\theta}(\pm)$,
$P_{\theta}(\pm)$ via hidden variable theory.

The local hidden variable theory for these new observables is defined by
measurement probability functions for each sub-system. For sub-system $A$
this will be $P(\alpha _{A},\beta _{A},\xi _{A},\eta
_{A}|x_{A},p_{A},U_{A},V_{A},c,\lambda )$ for the measurement outcomes 
$\alpha _{A},\beta _{A},\xi _{A},\eta _{A}$ for $x_{A}$, $p_{A}$, $U_{A}$ and 
$V_{A}$ respectively, with an analogous probability for $x_{B}$, $p_{B}$, 
$U_{B}$ and $V_{B}$. Note that as the measurement outcomes for $V_{A}$ and 
$V_{B}$ are required to be the same as in quantum theory for \emph{any}
choice of preparation probability $P(\lambda |c)$, we must have 
\begin{multline}
P(\alpha _{A},\beta _{A},\xi _{A},\eta
_{A}|x_{A},p_{A},U_{A},V_{A},c,\lambda ) \\=\delta _{\eta _{A},1/2}Q(\alpha
_{A},\beta _{A},\xi _{A}|x_{A},p_{A},U_{A},c,\lambda ) , \\
\shoveleft
P(\alpha _{B},\beta _{B},\xi _{B},\eta
_{B}|x_{B},p_{B},U_{B},V_{B},c,\lambda ) \\=\delta _{\eta _{B},1/2}Q(\alpha
_{B},\beta _{B},\xi _{B}|x_{B},p_{B},U_{B},c,\lambda )  
  .
  \label{Eq.MeastProbsVaVb2}
\end{multline}
These requirements have implications for the mean values $\left\langle V_{A}(\lambda)\right\rangle $,
though only the final mean value $\left\langle V_{A}\right\rangle$ is
required for the EPR steering tests.

\subsection{Bloch Vector Test for EPR Steering}

\label{SubSection - Bloch Vector Test EPR Steering}

\subsubsection{Mean Values of Spin Components S$_{x}$and S$_{y}$ - Category 2
States\ }

We now consider the mean value for spin components for the Category
2 states. For example, in the case of the spin component $S_{x}$
\begin{align}
\left\langle S_{x}\right\rangle &=\dsum\limits_{\lambda }P(\lambda
|c)\left\langle S_{x}(\lambda )\right\rangle  \nonumber \\
&=\frac{1}{2}\dsum\limits_{\lambda }
\bigl(\left\langle x_{A}(\lambda )\right\rangle \left\langle x_{B}(\lambda )\right\rangle
_{Q}+\left\langle p_{A}(\lambda )\right\rangle \left\langle p_{B}(\lambda
)\right\rangle _{Q}\bigr)
  \nonumber \\
&
\qquad\qquad
\times
P(\lambda |c)
  \label{Eq.MeanSxLHSModel}
\end{align}%
using (\ref{Eq.SpinCompts}) and (\ref{Eq.ExpnValueJointMeastLHVT}).
This expression involves the hidden variable mean values for the (classical)
observables $x_{A}$ and $p_{A}$ of sub-system $A$ and the local
hidden state mean values for the quantum quadrature operators $\widehat{x}_{B}$
and $\widehat{p}_{B}$. The latter must also correspond to quantum
mean values, for a physically realisable quantum state for sub-system
$B$. Thus $\left\langle x_{B}(\lambda)\right\rangle _{Q}=\Tr(\widehat{x}_{B}\widehat{\rho}^{B}(\lambda))$
and $\left\langle p_{B}(\lambda)\right\rangle _{Q}=\Tr(\widehat{p}_{B}\widehat{\rho}^{B}(\lambda))$.
Since sub-system $B$ is to be treated quantum mechanically then the
density operator $\widehat{\rho}^{B}(\lambda)$ would be required
to both satisfy the \emph{symmetrisation principle} and be \emph{local
particle number SSR} compliant. Hence there is a constraint based
on the local hidden state $\widehat{\rho}^{B}(\lambda)$ being a \emph{possible
state} for sub-system $B$ that requires the state to be local particle
number SSR compliant.

In this case then since both $\widehat{x}_{B}$ and $\widehat{p}_{B}$
are just linear combinations of $\widehat{b}$ and $\widehat{b}^{\dag}$
we have 
\begin{eqnarray}
\left\langle x_{B}(\lambda)\right\rangle _{Q} & = & \Tr\frac{1}{\sqrt{2}}(\widehat{b}+\widehat{b}^{\dag})\widehat{\rho}^{B}(\lambda)=0,\nonumber \\
\left\langle p_{B}(\lambda)\right\rangle _{Q} & = & \Tr\frac{1}{\sqrt{2}i}(\widehat{b}-\widehat{b}^{\dag})\widehat{\rho}^{B}(\lambda)=0,\label{Eq.MeanSubSystB}\\
\left\langle S_{x}(\lambda)\right\rangle  & = & 0,\qquad\left\langle S_{y}(\lambda)\right\rangle =0,\label{Eq.MeanSxLambda}
\end{eqnarray}
and thus for Category 2 states 
\begin{equation}
\left\langle S_{x}\right\rangle =0,\qquad\left\langle S_{y}\right\rangle =0 \,.
\end{equation}
We do not need to know the outcome for $\left\langle x_{A}(\lambda)\right\rangle $
or $\left\langle p_{A}(\lambda)\right\rangle $.

So that if LHVT is to give the same prediction as quantum theory then
on reverting to quantum operators and using (\ref{Eq.MeanValueJointMeastsQThyLHVT})
we have for Category 2 states 
\begin{equation}
\left\langle \widehat{S}_{x}\right\rangle =0,\qquad \text{and}\qquad\left\langle \widehat{S}_{y}\right\rangle =0.\label{Eq.MeanSxSy}
\end{equation}
These two results are the same as for a quantum separable (Category
1) state. 

\subsubsection{Bloch Vector Test}

From (\ref{Eq.MeanSxSy}) for Category 2 (or Category 1) states we
immediately see that if 
\begin{equation}
   \left\langle \widehat{S}_{x}\right\rangle \neq0\qquad \text{or}\qquad\left\langle \widehat{S}_{y}\right\rangle \neq0,\label{Eq.BlochVectorTestLHS}
\end{equation}
then the quantum state cannot be in Category 2 (or Category 1). The
\emph{Bloch vector test} $\langle \widehat{S}_{x}\rangle \neq0$
or $\langle \widehat{S}_{y}\rangle \neq0$ now also shows
that the state is \emph{EPR steered} as well as just being entangled.

Experiments in two mode BEC by \cite{Gross10a,Egorov11a}
have found non-zero behaviour for $\langle \widehat{S}_{x}\rangle $,
$\langle \widehat{S}_{y}\rangle $. These experiments therefore
demonstrate EPR steering, though only entanglement was claimed to
have been shown \cite{Gross10a}. The application of the Bloch vector
test for EPR steering to the experiment in \cite{Egorov11a} is discussed
more fully elsewhere \cite{Opanchuk17a}.

\subsection{Spin Squeezing Tests for EPR Steering}

\label{SubSection - Spin Squeezing Test for EPR Steering}

\subsubsection{Mean Values of Spin Component S$_{z}$and Number N - Category 2 States\ }

For the other spin component $S_{z}$ we find using (\ref{Eq.ModeNumbersObserv}) that
for the Category 2 states 
\begin{equation}
\left\langle S_{z}\right\rangle =\frac{1}{2}\left\langle 1_{A}\otimes N_{B}\right\rangle -\frac{1}{2}\left\langle N_{A}\otimes1_{B}\right\rangle .\label{Eq.SzLHSModel}
\end{equation}
As in the quantum separable state case $\left\langle S_{z}\right\rangle $
is not necessarily zero.

\subsubsection{Variances of Spin Components S$_{x}$and S$_{y}$ - Category 2 States}

As $\langle S_{x}(\lambda)\rangle =\langle S_{y}(\lambda)\rangle =0$
from (\ref{Eq.MeanSxLambda}) we see that $\langle \Delta S_{x}^{2}(\lambda)\rangle =\langle S_{x}^{2}(\lambda)\rangle $
and $\langle \Delta S_{y}^{2}(\lambda)\rangle =\langle S_{y}^{2}(\lambda)\rangle $.
Using (\ref{Eq.MeanValueJointMeastsQThyLHVT}), the LHVT expression
for $S_{x}^{2}$ obtained from the classical form of (\ref{Eq.SxSquare})
and after applying the inequality (\ref{Eq.VarianceInequalityHVT})
we then have the following inequalities for Category 2 states 
\begin{eqnarray}
\left\langle \Delta S_{x}^{2}\right\rangle  
\geq &&  \sum\limits_{\lambda }P(\lambda |c)  \nonumber \\
&&\times \left[
\frac{1}{4}(\left\langle x_{A}^{2}(\lambda )\right\rangle \left\langle
x_{B}^{2}(\lambda )\right\rangle _{Q}+\left\langle p_{A}^{2}(\lambda
)\right\rangle \left\langle p_{B}^{2}(\lambda )\right\rangle _{Q}) \right.\nonumber\\ 
&& \left. \quad
+\frac{1}{2}(\left\langle U_{A}(\lambda )\right\rangle \left\langle
U_{B}(\lambda )\right\rangle _{Q}-\left\langle V_{A}(\lambda )\right\rangle
\left\langle V_{B}(\lambda )\right\rangle _{Q})%
\right]  \nonumber \\
\left\langle \Delta S_{y}^{2}\right\rangle 
\geq && \sum\limits_{\lambda }P(\lambda |c)  \nonumber \\
&&\times \left[
\frac{1}{4}(\left\langle p_{A}^{2}(\lambda )\right\rangle \left\langle
x_{B}^{2}(\lambda )\right\rangle _{Q}+\left\langle x_{A}^{2}(\lambda
)\right\rangle \left\langle p_{B}^{2}(\lambda )\right\rangle _{Q}) \right.\nonumber\\ 
&& \left. \quad 
-\frac{1}{2}(\left\langle U_{A}(\lambda )\right\rangle \left\langle
U_{B}(\lambda )\right\rangle _{Q}+\left\langle V_{A}(\lambda )\right\rangle
\left\langle V_{B}(\lambda )\right\rangle _{Q})%
\right] . \nonumber \\
&&  \label{Eq.VarSxSyIneqLHSModel}
\end{eqnarray}

\subsubsection{Evaluation of Expressions Needed - Category 2 States}
\label{SuSubSection- Evaluation of Expressions Cat 2}

To consider spin squeezing, spin variance and correlation tests for
EPR\ steering based on the Category 2 states we will need to consider
the following additional quantum theory based expressions: $\left\langle x_{B}^{2}(\lambda)\right\rangle _{Q}$,
$\left\langle p_{B}^{2}(\lambda)\right\rangle _{Q}$, $\left\langle V_{B}(\lambda)\right\rangle _{Q}$,
$\left\langle U_{B}(\lambda)\right\rangle _{Q}$ and the following
non-quantum expressions $\left\langle x_{A}^{2}(\lambda)\right\rangle $,
$\left\langle p_{A}^{2}(\lambda)\right\rangle $, $\left\langle V_{A}(\lambda)\right\rangle $.

Starting with the \emph{quantum} theory expressions (\ref{Eq.PositionMtmOprs})
we find that 
\begin{eqnarray}
\left\langle x_{B}^{2}(\lambda)\right\rangle _{Q} & = & \Tr((\widehat{b}^{\dag}\widehat{b})\widehat{\rho}^{B}(\lambda))+\frac{1}{2},\nonumber \\
 & = & \left\langle N_{B}(\lambda)\right\rangle _{Q}+\frac{1}{2},\label{Eq.ResultXBSquareQuant}\\
\left\langle p_{B}^{2}(\lambda)\right\rangle _{Q} & = & \left\langle N_{B}(\lambda)\right\rangle _{Q}+\frac{1}{2},\label{Eq.ResultPBSquareQuant}
\end{eqnarray}
where the commutation rules have been used and the SSR constraints
eliminate the $\Tr((\widehat{b})^{2}\widehat{\rho}^{B}(\lambda))$
and $\Tr((\widehat{b}^{\dag})^{2}\widehat{\rho}^{B}(\lambda))$ terms.
Note that $\left\langle N_{B}(\lambda)\right\rangle _{Q}\geq0$.

Then using (\ref{Eq.UOprs}) we find that 
\begin{eqnarray}
\left\langle U_{B}(\lambda)\right\rangle _{Q} & = & \frac{1}{2i}\Tr((\widehat{b})^{2}-(\widehat{b}^{\dag})^{2})\widehat{\rho}^{B}(\lambda)\nonumber \\
 & = & 0,\label{Eq.MeanUBLHSModel}
\end{eqnarray}
again due to the SSR constraints on the hidden state $\widehat{\rho}^{B}(\lambda)$.

Also, using (\ref{Eq.AuxOprsV}) 
\begin{eqnarray}
\left\langle V_{B}(\lambda)\right\rangle _{Q} & = & \frac{1}{2}\Tr_{B}(\widehat{1}_{B}\widehat{\rho}^{B}(\lambda))\nonumber \\
 & = & \frac{1}{2}\label{Eq.MeanVBLHSModel}
\end{eqnarray}
since the trace of a density operator is unity. Using (\ref{Eq.ResultXBSquareQuant}),
(\ref{Eq.ResultPBSquareQuant}) and (\ref{Eq.MeanVBLHSModel}) we
confirm the result that $\left\langle N_{B}(\lambda)\right\rangle _{Q}=\frac{1}{2}\left\langle x_{B}^{2}(\lambda)\right\rangle _{Q}+\frac{1}{2}\left\langle p_{B}^{2}(\lambda)\right\rangle _{Q}-\left\langle V_{B}(\lambda)\right\rangle _{Q}$
consistent with (\ref{Eq.ModeNumbersObserv}). Result (\ref{Eq.MeanVBLHSModel})
also follows directly from (\ref{Eq.MeastProbsVaVb2}).

For the \emph{local hidden variable theory} expressions involving sub-system 
$A$ we have using (\ref{Eq.ModeNumbersObserv}) 
\begin{equation}
\left\langle x_{A}^{2}(\lambda )\right\rangle +\left\langle
p_{A}^{2}(\lambda )\right\rangle =2\left\langle N_{A}(\lambda )\right\rangle
+2\left\langle V_{A}(\lambda )\right\rangle
\,.  \label{Eq.MeanNA}
\end{equation}
Note the analogous result for sub-system $B$.

Using the results
(\ref{Eq.AuxOprsV}), (\ref{Eq.VarSxSyIneqLHSModel})--(\ref{Eq.MeanNA})
we now have for Category 2 states
\begin{equation}
  \begin{split}
&\left\langle \Delta S_{x}^{2}\right\rangle  \\
&\geq \sum\limits_{\lambda }P(\lambda |c)
  \left( \frac{1}{2}\left(\left\langle N_{B}(\lambda )\right\rangle _{Q}
  +\frac{1}{2}\right)(\left\langle
N_{A}(\lambda )\right\rangle +\left\langle V_{A}(\lambda )\right\rangle )
\right.\\ &\left.
\qquad\qquad\qquad - \frac{1}{4}\left\langle V_{A}(\lambda )\right\rangle \right)  \,,
\\
&\geq \frac{1}{2}\left\langle N_{A}\otimes N_{B}\right\rangle +\frac{1}{2}%
\left\langle V_{A}\otimes N_{B}\right\rangle +\frac{1}{4}\left\langle
N_{A}\otimes 1_{B}\right\rangle 
\\&\quad+\frac{1}{4}\left\langle V_{A}\otimes
1_{B}\right\rangle -\frac{1}{4}\left\langle V_{A}\otimes 1_{B}\right\rangle \,,\\
&\geq \frac{1}{2}\left\langle \widehat{N}_{A}\otimes \widehat{N}%
_{B}\right\rangle +\frac{1}{2}\left\langle \widehat{V}_{A}\otimes \widehat{N}%
_{B}\right\rangle +\frac{1}{4}\left\langle \widehat{N}_{A}\otimes \widehat{1}%
_{B}\right\rangle  \,,\\
&\geq \frac{1}{2}\left\langle \widehat{N}_{A}\otimes \widehat{N}%
_{B}\right\rangle +\frac{1}{4}\left\langle \widehat{1}_{A}\otimes \widehat{N}%
_{B}\right\rangle +\frac{1}{4}\left\langle \widehat{N}_{A}\otimes \widehat{1}%
_{B}\right\rangle  \,,\\ &\\
& \left\langle \Delta S_{y}^{2}\right\rangle  \\
&\geq \frac{1}{2}\left\langle \widehat{N}_{A}\otimes \widehat{N}%
_{B}\right\rangle +\frac{1}{4}\left\langle \widehat{1}_{A}\otimes \widehat{N}%
_{B}\right\rangle +\frac{1}{4}\left\langle \widehat{N}_{A}\otimes \widehat{1}%
_{B}\right\rangle \,.
\end{split}
\label{Eq.ResultVarSxSyLHS}
\end{equation}
Note that moving from line one to line two only involves LHVT expressions,
whereas moving from line two to line three involves replacing the LHVT
overall mean values by the equivalent quantum expressions, and in the next
line the quantum operator $\widehat{V}_{A}$ is replaced by
$\widehat{1}_{A}/2$. These inequalities are the same as those for Category 1 states
(see \cite{Dalton16b}). Note that the SSR for the LHS have been used in
deriving these last results.
Also from (\ref{Eq.SzLHSModel})
\begin{equation}
\frac{1}{2}|\left\langle S_{z}\right\rangle |\;\leq\frac{1}{4}\left\langle 1_{A}\otimes N_{B}\right\rangle +\frac{1}{4}\left\langle N_{A}\otimes1_{B}\right\rangle 
. \label{Eq.ResultSzLHS}
\end{equation}
The last line follows from the LHVT expression $\langle 1_{A}\otimes N_{B}\rangle $
giving the mean number of bosons in mode $B$ and for this to be the
same as the quantum theory expression $\langle \widehat{1}_{A}\otimes\widehat{N}_{B}\rangle $.
As the eigenvalues of the number operator $\widehat{N}_{B}=\widehat{b}^{\dag}\widehat{b}$
are never negative $\langle \widehat{1}_{A}\otimes\widehat{N}_{B}\rangle $
and hence $\langle 1_{A}\otimes N_{B}\rangle $ is never
negative, so $|\langle 1_{A} \otimes N_{B}\rangle | = \langle 1_{A} \otimes N_{B} \rangle $.
Similarly, $\langle N_{A} \otimes1_{B} \rangle $ is never
negative. This result is the same as that for Category 1 states (see
Ref.\ \cite{Dalton16b}).

Combining (\ref{Eq.ResultVarSxSyLHS}) and (\ref{Eq.ResultSzLHS})
we find using LHVT that for Category 2 states 
\begin{eqnarray}
\left\langle \Delta S_{x}^{2}\right\rangle -\frac{1}{2}|\left\langle S_{z}\right\rangle | & ~\geq~ & \frac{1}{2}\left\langle N_{A}\otimes N_{B}\right\rangle ,\nonumber \\
\left\langle \Delta S_{y}^{2}\right\rangle -\frac{1}{2}|\left\langle S_{z}\right\rangle | & \geq & \frac{1}{2}\left\langle N_{A}\otimes N_{B}\right\rangle ,
\end{eqnarray}
so as the LHVT is required to predict the same results as for quantum
theory we have for Category 2 states 
\begin{eqnarray}
\left\langle \Delta\widehat{S}_{x}^{2}\right\rangle -\frac{1}{2}\left|\left\langle \widehat{S}_{z}\right\rangle \right| & ~\geq~ & \frac{1}{2}\left\langle \widehat{N}_{A}\otimes\widehat{N}_{B}\right\rangle \geq0 , \label{Eq.SpinSqSxLHS}\\
\left\langle \Delta\widehat{S}_{y}^{2}\right\rangle -\frac{1}{2}\left|\left\langle \widehat{S}_{z}\right\rangle \right| & \geq & \frac{1}{2}\left\langle \widehat{N}_{A}\otimes\widehat{N}_{B}\right\rangle \geq0 . \label{Eq.SpinSqSyLHS}
\end{eqnarray}
The expression $\frac{1}{2}\langle \widehat{N}_{A}\otimes\widehat{N}_{B}\rangle $
is never negative because the eigenvalues of $\widehat{N}_{A}$ and
$\widehat{N}_{B}$ are never negative.

\subsubsection{Spin Squeezing Tests}

From Eq.\ (\ref{Eq.MeanSxSy}) we immediately see that for a quantum
state where the observable $\widehat{S}_{z}$ is squeezed with respect
to $\widehat{S}_{x}$ or with respect to $\widehat{S}_{y}$, then
it cannot be a Category 2 state, because spin squeezing in $\widehat{S}_{z}$
requires $\langle \Delta\widehat{S}_{z}^{2}\rangle $ to
be less than either $|\langle \widehat{S}_{x}\rangle |/2$
or $|\langle \widehat{S}_{y}\rangle |/2$ and this is impossible
for both Category 1 (see Ref.\ \cite{Dalton16b}) and Category 2 states
- where $\langle \widehat{S}_{x}\rangle =\langle \widehat{S}_{y}\rangle =0$.
This condition also rules out $\widehat{S}_{x}$ or $\widehat{S}_{y}$
being squeezed with respect to $\widehat{S}_{z}$, or $\widehat{S}_{z}$
being squeezed with respect to $\widehat{S}_{x}$ or $\widehat{S}_{y}$.
In Ref.\ \cite{Dalton16b} it was shown that spin squeezing involving
$\widehat{S}_{z}$ provided a test for entanglement. Here we see that
spin squeezing involving the observable $\widehat{S}_{z}$ shows the
state is \emph{EPR steered} as well as merely being \emph{entangled}.

From Eqs.\ (\ref{Eq.SpinSqSxLHS}) and (\ref{Eq.SpinSqSyLHS}) we see
that for Category 2 states$\ (\langle \Delta\widehat{S}_{x}^{2}\rangle -\frac{1}{2}|\langle \widehat{S}_{z}\rangle |)\geq0$
and $(\langle \Delta\widehat{S}_{y}^{2}\rangle -\frac{1}{2}|\langle \widehat{S}_{z}\rangle |)\geq0$.
Hence we find that for Category 2 states there is no spin squeezing
in $\widehat{S}_{x}$ compared to $\widehat{S}_{y}$ (or vice versa).
For Category 1 states we also find that $(\langle \Delta\widehat{S}_{x}^{2}\rangle -\frac{1}{2}|\langle \widehat{S}_{z}\rangle |)\geq\frac{1}{2}\langle \widehat{N}_{A}\otimes\widehat{N}_{B}\rangle \geq0$
and $(\langle \Delta\widehat{S}_{y}^{2}\rangle -\frac{1}{2}|\langle \widehat{S}_{z}\rangle |)\geq\frac{1}{2}\langle \widehat{N}_{A}\otimes\widehat{N}_{B}\rangle \geq0$
(see Eq.\ (31) in Ref.\ \cite{Dalton16b}). Hence spin squeezing in
$\widehat{S}_{x}$ versus $\widehat{S}_{y}$ (or vice versa) is a
test for entanglement, so the state is \emph{not} in Category 1. Thus
spin squeezing in $\widehat{S}_{x}$ versus $\widehat{S}_{y}$ (or
vice versa) is therefore also a test for \emph{EPR steering}.

Overall then we now see that \emph{spin squeezing} in \emph{any} spin
component $\widehat{S}_{\alpha}$ with respect to another component
$\widehat{S}_{\beta}$ 
\begin{equation}
\left\langle \Delta\widehat{S}_{\alpha}^{2}\right\rangle <\frac{1}{2}\left|\left\langle \widehat{S}_{\gamma}\right\rangle \right|\qquad \text{and} \qquad\left\langle \Delta\widehat{S}_{\beta}^{2}\right\rangle >\frac{1}{2}\left|\left\langle \widehat{S}_{\gamma}\right\rangle \right|\label{Eq.SpinSqueezingTests}
\end{equation}
(where $\alpha,\beta,\gamma$ are $x,y,z$ in cyclic order) is a sufficiency
test for \emph{EPR steering}. Hence \emph{spin squeezing} in \emph{any}
spin component $\widehat{S}_{\alpha}$ with respect to another component
$\widehat{S}_{\beta}$ shows that the state is \emph{EPR steered}
as well as just being entangled.

Experiments in two mode BEC by \cite{Gross10a,Riedel10a,Maussang10a}
have found spin squeezing in $S_{z}$. These experiments
therefore demonstrate EPR steering, though only entanglement was claimed
to have been shown in \cite{Gross10a,Riedel10a}.

\subsection{Planar Spin Variance Tests for EPR Steering}

\label{SubSection - Planar Spin Variance Tests for EPR Steering}

\subsubsection{Mean Values of Total Boson Number N - Category 2 States\ }

For the number observable $N$ we have from (\ref{Eq.ModeNumbersObserv})
\begin{equation}
\left\langle N\right\rangle =\left\langle 1_{A}\otimes N_{B}\right\rangle +\left\langle N_{A}\otimes1_{B}\right\rangle . \label{Eq.ResultNLHS}
\end{equation}
This result is the same as that for Category 1 states (see Ref.~\cite{Dalton16b}).

\subsubsection{Hillary-Zubairy Planar Spin Variance Test}

The Hillery-Zubairy spin variance test \cite{Hillery06a} for quantum
entanglement is $\langle \Delta\widehat{S}_{x}^{2}\rangle +\langle \Delta\widehat{S}_{y}^{2}\rangle -\frac{1}{2}\langle \widehat{N}\rangle <0$.
We now consider the quantity $\langle \Delta S_{x}^{2}\rangle +\langle \Delta S_{y}^{2}\rangle 
- \frac{1}{2}\langle N\rangle $ for
Category 2 states using the results based on LHVT in Eqs.\ (\ref{Eq.ResultVarSxSyLHS})
and (\ref{Eq.ResultNLHS}). We find that 
\begin{equation}
   \left\langle \Delta S_{x}^{2}\right\rangle +\left\langle
   \Delta S_{y}^{2}\right\rangle - \frac{1}{2}\left\langle N\right\rangle 
   \geq  \left\langle N_{A}\otimes N_{B}\right\rangle 
   \geq  0   .
\label{Eq.HilleryFormLHSB}
\end{equation}
Thus if LHVT is to predict the same result as quantum theory it follows
that for Category 2 states that 
\begin{equation}
\left\langle \Delta\widehat{S}_{x}^{2}\right\rangle +\left\langle \Delta\widehat{S}_{y}^{2}\right\rangle - \frac{1}{2}\left\langle \widehat{N}\right\rangle \geq0 . 
\label{Eq.PlanarSpinVarCat2States}
\end{equation}
This result also applies for Category 1 states (see Eqs.\ (82,83)
in Ref.\ \cite{Dalton16b} for details, or directly from Eq.\ (\ref{Eq.IneqSpinOprsCat1})).

Hence we can say that if
\begin{equation}
\left\langle \Delta\widehat{S}_{x}^{2}\right\rangle +\left\langle \Delta\widehat{S}_{y}^{2}\right\rangle \mathbf{\ }-\frac{1}{2}\left\langle \widehat{N}\right\rangle <0 
\,, \label{Eq.HiilarySpinVarianceTest}
\end{equation}
then the state is not in Category 2. It also shows that it is not
in Category 1 (separable states), this being the Hillery-Zubairy planar
spin variance test \cite{Hillery06a} for entanglement. This condition
can also be written as 
\begin{equation}
E_{HZ}=\frac{\left\langle \Delta\widehat{S}_{x}^{2}\right\rangle +\left\langle \Delta\widehat{S}_{y}^{2}\right\rangle }{\frac{1}{2}\left\langle \widehat{N}\right\rangle }<1 \,, \label{Eq.HZTestB}
\end{equation}
which is the form given in Ref.\ \cite{He12a}.

Hence the \emph{Hillary-Zubairy planar spin variance} inequality
is a sufficiency test for \emph{EPR steering} as well as
demonstrating entanglement. 

\subsubsection{Generalised Hillery-Zubairy Planar Spin Variance Test}

The results (\ref{Eq.ResultVarSxSyLHS}), (\ref{Eq.ResultNLHS}) and
(\ref{Eq.SzLHSModel}) show that for Category 2 states where the LHS
occurs in sub-system $B$
\begin{align}
  \left\langle \Delta S_{x}^{2}\right\rangle +\left\langle \Delta S_{y}^{2}\right\rangle &-\frac{1}{4}\left\langle N\right\rangle +\frac{1}{2}\left\langle S_{z}\right\rangle
 \nonumber\\
  &\geq\left\langle N_{A}\otimes N_{B}\right\rangle +\frac{1}{2}\left\langle 1_{A}\otimes N_{B}\right\rangle \,,
 \nonumber\\
  &\geq0 \,. \label{Eq.GenPlanarSpinVarCat2States}
\end{align}
The details are set out in Appendix \ref{Appendix - EPR Sterering Othe Approach}.

This provides a generalisation of the \emph{Hillery-Zubairy planar
spin variance} test \cite{Hillery06a} for \emph{EPR steering. }In
the case we see that if 
\begin{equation}
\left\langle \Delta\widehat{S}_{x}^{2}\right\rangle +\left\langle \Delta\widehat{S}_{y}^{2}\right\rangle -\frac{1}{4}\left\langle \widehat{N}\right\rangle +\frac{1}{2}\left\langle \widehat{S}_{z}\right\rangle <0 \,, \label{Eq.EPRSteeringTest}
\end{equation}
then the state is not in Category 2. If sub-system $A$ involves
the LHS then $+\frac{1}{2}\langle \widehat{S}_{z}\rangle $
is replaced by $-\frac{1}{2}\langle \widehat{S}_{z}\rangle $.
Since $+\frac{1}{2}\langle \widehat{N}\rangle \geq\langle \widehat{S}_{z}\rangle \geq-\frac{1}{2}\langle \widehat{N}\rangle $
then $\frac{1}{2}\langle \widehat{N}\rangle \geq\frac{1}{4}\langle \widehat{N}\rangle +\frac{1}{2}\langle \widehat{S}_{z}\rangle \geq0$,
so as $\langle \Delta\widehat{S}_{x}^{2}\rangle +\langle \Delta\widehat{S}_{y}^{2}\rangle -\frac{1}{4}\langle \widehat{N}\rangle +\frac{1}{2}\langle \widehat{S}_{z}\rangle =\langle \Delta\widehat{S}_{x}^{2}\rangle +\langle \Delta\widehat{S}_{y}^{2}\rangle -\frac{1}{2}\langle \widehat{N}\rangle + (\frac{1}{4}\langle \widehat{N}\rangle +\frac{1}{2}\langle \widehat{S}_{z}\rangle )$
and we have just shown that $(\frac{1}{4}\langle \widehat{N}\rangle +\frac{1}{2}\langle \widehat{S}_{z}\rangle )$
is never negative, then if (\ref{Eq.EPRSteeringTest}) is satisfied
then the Hillary-Zubairy planar spin variance test in (\ref{Eq.HiilarySpinVarianceTest})
must also apply, showing (see Ref.\ \cite{Dalton16b} for details)
that the state cannot be in Category 1. The latter test is of course
itself sufficient to demonstrate EPR steering. Since $0\leq\frac{1}{4}\langle \widehat{N}\rangle -\frac{1}{2}\langle \widehat{S}_{z}\rangle \leq$
$\frac{1}{2}\langle \widehat{N}\rangle $ it is of course
harder to find states where $\langle \Delta\widehat{S}_{x}^{2}\rangle +\langle \Delta\widehat{S}_{y}^{2}\rangle <\frac{1}{4}\langle \widehat{N}\rangle -\frac{1}{2}\langle \widehat{S}_{z}\rangle $
to show EPR steering than merely being less than $\frac{1}{2}\langle \widehat{N}\rangle $,
as would also show EPR steering. The generalised Hillery-Zubairy planar
spin variance test (\ref{Eq.EPRSteeringTest}) for EPR steering is
a more difficult test to satisfy than the Hillery-Zubairy test. In
the generalised form (\ref{Eq.EPRSteeringTest}) the EPR steering
test now allows for \emph{asymmetry} ($\langle \widehat{S}_{z}\rangle \neq0$).

The generalised Hillery-Zubairy EPR steering test in~(\ref{Eq.EPRSteeringTest}) can also be written
as 
\begin{equation}
E_{GHZ}=\frac{\left\langle \Delta\widehat{S}_{x}^{2}\right\rangle +\left\langle \Delta\widehat{S}_{y}^{2}\right\rangle }{\frac{1}{2}\left\langle \widehat{N}\right\rangle }<\frac{\left\langle \widehat{N}_{A}\right\rangle }{\left\langle \widehat{N}\right\rangle }\label{Eq.EPRSteerTestB}
\end{equation}
after substituting $\langle \widehat{N}\rangle =\langle \widehat{N}_{A}\rangle +\langle \widehat{N}_{B}\rangle $
and $\langle \widehat{S}_{z}\rangle =(\langle \widehat{N}_{B}\rangle -\langle \widehat{N}_{A}\rangle )/2$,
which is consistent with the result $E_{HZ}<1/2$ 
previously obtained by He et al.\ in
Ref.\ \cite{He12a}
for $\langle \widehat{S}_{z}\rangle =0$. This form of
the test also shows that the EPR steering test in (\ref{Eq.HZTestB})
is satisfied, since the right side is always less than unity because
$\langle \widehat{N}_{A}\rangle \leq\langle \widehat{N}\rangle$.
Note that for EPR steering to apply, it is not \emph{necessary} that
(\ref{Eq.EPRSteerTestB}) applies, since (\ref{Eq.HZTestB}) is \emph{sufficient}
to demonstrate EPR steering. Combining both tests we see that if \emph{either}
$(E_{HZ}<1$ $ and $ $E_{GHZ}<\langle \widehat{N}_{A}\rangle /\langle \widehat{N}\rangle )$ \emph{or}
$(E_{HZ}<1)$ then the state cannot be either Category 1 or Category
2, and hence is EPR steerable.

The tests in (\ref{Eq.EPRSteeringTest}) and (\ref{Eq.EPRSteerTestB}) also follow from the strong correlation condition obtained by Cavalcanti et.\ al \cite{Cavalcanti11a} - set out here as Eq.~(\ref{Eq.InequalCat2}) (see Appendices \ref{Appendix- Correlation Tests for EPR Steering} and \ref{Appendix - Correlation Ineq and Spin Operators}). The derivation of the test (\ref{Eq.EPRSteeringTest}) in terms of spin operators starting from the strong correlation condition (\ref{Eq.InequalCat2}) is set out in Appendix \ref{SubSubSection - Spin Opr Ineqall Cats 1,,2, 3}. The test given in (\ref{Eq.EPRSteerTestB}) was first stated in Ref.\ \cite{Rosales18a}, again starting from the strong correlation condition in Ref.\ \cite{Cavalcanti11a}, and then expressing the latter inequality in terms of spin operators - as derived here in Appendix \ref{SubSubSection - Spin Opr Ineqall Cats 1,,2, 3}.

These two planar spin variance test are involved in discussing the
so-called \emph{depth of EPR steering} in two mode BECs \cite{Rosales18a},
which specifies the number of particles involved in the component
of the density operator which is responsible for EPR steering effects.

\subsection{Two Mode Quadrature Squeezing Test for EPR Steering}

\label{SubSection - Two Mode Quadrature Squeezing Test for EPR Steering}

\subsubsection{Mean Values for Two Mode Quadratures $X_{\theta}(\pm)$ and $P_{\theta}(\pm)$
- Category 2 States}

We now consider the mean value for two mode quadrature observables
for the Category 2 states. For example in the case of the quadratures
$X_{\theta}(\pm)$
\begin{align}
\left\langle X_{\theta}(\pm)\right\rangle =\frac{1}{\sqrt{2}}\sum\limits _{\lambda}&P(\lambda|c)
\nonumber\\ & \times 
[ \left\langle x_{A}(\lambda)\right\rangle \cos\theta+\left\langle p_{A}(\lambda)\right\rangle \sin\theta
\nonumber\\
&\quad\pm\left\langle x_{B}(\lambda)\right\rangle _{Q}\cos\theta\pm\left\langle p_{B}(\lambda)\right\rangle _{Q}\sin\theta ], \nonumber\\
\end{align}
using Eq.\ (\ref{Eq.TwoModeQuadObserv}). A similar result is found
for $P_{\theta}(\pm)$. We then use the previous results (\ref{Eq.MeanSubSystB})
for sub-system $B$ to find 
\begin{eqnarray}
\left\langle X_{\theta}(\pm)\right\rangle  & = & \frac{1}{\sqrt{2}}\sum\limits _{\lambda}P(\lambda|c)\left(\left\langle x_{A}(\lambda)\right\rangle \cos\theta+\left\langle p_{A}(\lambda)\right\rangle \sin\theta\right),\nonumber \\
\left\langle P_{\theta}(\pm)\right\rangle  & = & \frac{1}{\sqrt{2}}\sum\limits _{\lambda}
 P(\lambda|c)\left(-\left\langle x_{A}(\lambda)\right\rangle \sin\theta
+\left\langle p_{A}(\lambda)\right\rangle \cos\theta\right).\nonumber\\ &&\label{Eq.MeanTwoModeQuads}
\end{eqnarray}

\subsubsection{Variances for Two Mode Quadratures - Category 2 States}

Using (\ref{Eq.MeanValueJointMeastsQThyLHVT}) and the LHVT expression
for $X_{\theta}(\pm)^{2}$ obtained from the equivalent of Eq.\ (\ref{Eq.TwoModeQuadSquare})
for classical observables we have for Category 2 states, 
\begin{eqnarray}
\left\langle X_{\theta}(\pm)^{2}\right\rangle  & = & \frac{1}{2}\sum\limits _{\lambda}P(\lambda|c)\left[\left\langle x_{A}^{2}(\lambda)\right\rangle \cos^{2}\theta
\right. \nonumber\\ && \left .\qquad
+\left\langle U_{A}(\lambda)\right\rangle 2\sin\theta\cos\theta+\left\langle p_{A}^{2}(\lambda)\right\rangle \sin^{2}\theta\right]\nonumber \\
 &  & +\frac{1}{2}\sum\limits _{\lambda}P(\lambda|c)\left(\left\langle N_{B}(\lambda)\right\rangle _{Q}+\frac{1}{2}\right),\label{Eq.MeanTwoModeQuadSquare}
\end{eqnarray}
where we have used the previous results (\ref{Eq.MeanSubSystB}) and
(\ref{Eq.MeanUBLHSModel}) for sub-system $B$ to eliminate terms
involving $\left\langle x_{B}(\lambda)\right\rangle _{Q}$, $\left\langle p_{B}(\lambda)\right\rangle _{Q}$
and $\left\langle U_{B}(\lambda)\right\rangle _{Q}$ and the results
(\ref{Eq.ResultXBSquareQuant}) and (\ref{Eq.ResultPBSquareQuant})
for $\left\langle x_{B}^{2}(\lambda)\right\rangle _{Q}$ and $\left\langle p_{B}^{2}(\lambda)\right\rangle _{Q}$
to simplify the last term.

We next use the LHVT - quantum theory equivalences (\ref{Eq.MeanValueSingleMeastQThyLHVT})\ to
replace (\ref{Eq.MeanTwoModeQuads}) and (\ref{Eq.MeanTwoModeQuadSquare})
by their quantum forms. Quantum forms for the variances are then obtained.
Finally we use the result from SubSection \ref{SubSection - Quantum Theory}
the \emph{reduced density operator} for sub-system $A$ satisfies
the local particle number SSR to obtain expressions for $\left\langle x_{A}\right\rangle $,
$\left\langle p_{A}\right\rangle $, $\left\langle x_{A}^{2}\right\rangle $,
$\left\langle p_{A}^{2}\right\rangle $ and $\left\langle U_{A}\right\rangle $
to give the following results for the variances $\left\langle \Delta X_{\theta}(\pm)^{2}\right\rangle $
and $\left\langle \Delta P_{\theta}(\pm)^{2}\right\rangle $ for Category
2 states (see Eq.\ (\ref{Eq.TwoModeQuadVarCat2States})): 
\begin{eqnarray}
\left\langle \Delta X_{\theta}(\pm)^{2}\right\rangle  & = & \frac{1}{2}\left\langle N\right\rangle +\frac{1}{2}\geq\frac{1}{2},\nonumber \\
\left\langle \Delta P_{\theta}(\pm)^{2}\right\rangle  & = & \frac{1}{2}\left\langle N\right\rangle +\frac{1}{2}\geq\frac{1}{2}.\label{Eq.FinalVarQuadCat2}
\end{eqnarray}
Details are given in Appendix \ref{Appendix - Variances of Two Mode Quadratures - Cat 2}.
The same results apply for Category 1 (separable) states (see Appendix
L in Ref.\ \cite{Dalton16b}). 

\subsubsection{Two Mode Quadrature Squeezing Test}

We have shown for Category 2 states (see Eq.\ (\ref{Eq.FinalVarQuadCat2}))
that $\left\langle \Delta X_{\theta}(\pm)^{2}\right\rangle =\left\langle \Delta P_{\theta}(\pm)^{2}\right\rangle =\frac{1}{2}\left\langle N\right\rangle +\frac{1}{2},$ and
the right side is never less than one half. The same result applied
for Category 1 states. Hence it follows that if 
\begin{equation}
  \left\langle \Delta\widehat{X}_{\theta}(\pm)^{2}\right\rangle <\frac{1}{2}\quad\text{or}\quad\left\langle \Delta\widehat{P}_{\theta}(\pm)^{2}\right\rangle <\frac{1}{2},\label{Eq.TwoModeQuadSqueezeTest}
\end{equation}
which is the condition for \emph{squeezing} in either of the \emph{two
mode quadrature} observables $X_{\theta}(\pm)$ or $P_{\theta}(\pm)$,
then the state is not in Categories 1 or 2. Due to the Heisenberg
uncertainty principle $\langle \Delta\widehat{X}_{\theta}(\pm)^{2}\rangle \langle \Delta\widehat{P}_{\theta}(\pm)^{2}\rangle \geq1/4$
only one of the pair of quadrature operators is squeezed. Thus \emph{two
mode quadrature squeezing} as in (\ref{Eq.TwoModeQuadSqueezeTest})
provides a sufficiency test for \emph{EPR steering}.

Experiments in two mode BEC by \cite{Gross11a,Peise15a}
have found two mode quadrature squeezing in $S_{z}$. These experiments
therefore demonstrate EPR steering, which was identified in these
papers.

\subsection{Two Mode Binomial State}

\label{SubSection - Two Mode Binomial State}

The two mode \emph{binomial} state given by 
\begin{equation}
\left\vert \Phi\right\rangle =\frac{\left[ (\widehat{a}^{\dag}+\widehat{b}^{\dag})/\sqrt{2}\right]^{N}}{\sqrt{N!}}\left\vert 0\right\rangle \label{Eq.BinomialState}
\end{equation}
provides for a simple illustration of some of the EPR steering tests.
Results for mean values and variances of the spin operators $\widehat{S}_{x}$,
$\widehat{S}_{y}$, $\widehat{S}_{z}$ and number operators $\widehat{N}_{A}$,
$\widehat{N}_{B}$, $\widehat{N}$ are as follows:
\begin{align}
\left\langle \widehat{N}\right\rangle  &=  N  ,
&\left\langle \widehat{N}_{A}\right\rangle &=\frac{N}{2}, &\left\langle \widehat{N}_{B}\right\rangle &=\frac{N}{2},\nonumber \\
\left\langle \widehat{S}_{x}\right\rangle  &=  \frac{N}{2},&\left\langle \widehat{S}_{y}\right\rangle &=0,&\left\langle \widehat{S}_{z}\right\rangle &=0,\nonumber \\
\left\langle \Delta\widehat{S}_{x}^{2}\right\rangle  &=  0,&\left\langle \Delta\widehat{S}_{y}^{2}\right\rangle &=\frac{N}{4},&\left\langle \Delta\widehat{S}_{z}^{2}\right\rangle &=\frac{N}{4},\label{Eq.MeamsVariancesBinomState}
\end{align}
(see Ref \cite{Dalton16b} for details).
From these results we see that:
\begin{eqnarray}
\left\langle \widehat{S}_{x}\right\rangle  & \neq & 0,\nonumber \\
\left\langle \Delta\widehat{S}_{y}^{2}\right\rangle -\frac{1}{2}\left|\left\langle \widehat{S}_{x}\right\rangle \right| & = & 0 \,,
\nonumber\\ 
\left\langle \Delta\widehat{S}_{z}^{2}\right\rangle -\frac{1}{2}\left|\left\langle \widehat{S}_{x}\right\rangle \right|&=&0\,,\nonumber \\
E_{HZ} & = & \frac{1}{2}<1 \,,
\nonumber\\ 
E_{GHZ}&=&\frac{1}{2}=\frac{\left\langle \widehat{N}_{A}\right\rangle }{\left\langle \widehat{N}\right\rangle } \,.\label{Eq.EPRTests}
\end{eqnarray}
Hence the Bloch vector test and the Hillery-Zubairy planar spin variance
test both predict EPR steering, though neither the spin squeezing
test or the generalised Hillery-Zubairy planar spin variance test
does this. Nevertheless, EPR steering does occur for this state, since
we only require one of the tests to be positive. That the state is
steerable in the EPR sense may be seen if the measurables for the
two modes are the number operators $\widehat{N}_{A}$, $\widehat{N}_{B}$.
The measurement of $\widehat{N}_{A}$ leading to the outcome $n_{A}$
changes the quantum state to be the number state $ (\widehat{a}^{\dag})^{n_{A}}(\widehat{b}^{\dag})^{N-n_{A}}
 \left\vert 0\right\rangle /(\sqrt{n_{A}!}\sqrt{(N-n_{A})!})$,
so that measurement of $\widehat{N}_{B}$ must lead to the outcome
$N-n_{A}$ in accordance with EPR steering.

\section{Summary and Conclusion}

\label{Section - Summary and Conclusion}

Tests for EPR steering (EPR entanglement) based on violation of the LHS
model have been examined for two mode systems of identical massive bosons,
such as occur in BECs. Such tests were obtained based on whether the Bloch
vector is in the $xy$ plane (Bloch vector test) and on whether there is spin
squeezing in any of the spin components $S_{x}$, $S_{y}$ or $S_{z}$ (spin
squeezing test). Experiments that have been carried out on two mode BEC 
\cite{Gross10a,Riedel10a,Maussang10a,Gross11a,Egorov11a,Peise15a} 
have demonstrated EPR steering in such two mode
systems. The Hillery planar spin variance test based on the sum of variances
in $S_{x}$ and $S_{y}$ also demonstrates EPR steering. In addition, two mode
quadrature squeezing also provides a test for EPR steering. A 
generalised Hillery-Zubairy planar spin variance test for EPR steering was
found, involving the sum of variances in $S_{x}$\ and $S_{y}$, but now
containing a different multiple of the mean value for $N$ along with a term
involving the mean value for $S_{z}$. This allows for asymmetry and is a
stronger version of the Hillery planar spin variance test. Correlation tests
based on the mean value of $\left\langle a^{\dag }b\right\rangle$ have also
been obtained by others \cite{Cavalcanti11a}, and these are equivalent to
some of the tests based on the spin operators. No EPR steering test based on
the difference between the variances of the number difference and number sum
was found. We note that some of the tests (Bloch vector, spin squeezing, two
mode quadrature squeezing) were based on applying the super-selection rules
for the total particle number as well as that for the local particle number
for the sub-system LHS. However, since the stronger correlation inequalities
from which they can also be derived do not depend on the SSR (see 
Section~\ref{SubSubSection - Sttronger Correlation Inequalities}) the
Hillery-Zubairy planar spin variance test and its generalisation involving
the mean value for $S_{z}$ do not depend on these rules.

The treatment involved considering two possible classification schemes for
the quantum states of bipartite composite systems. In the first (Quantum
Theory Classification Scheme) the states are classified as being either
quantum separable or quantum entangled. In the second (Local Hidden Variable
Theory Classification Scheme) the states are initially classified as being
Bell local or Bell non-local. The Bell non-local states are quantum
entangled and EPR steerable - these are listed as Category 4 states.
However, the Bell local states can be divided up into three categories
depending on whether both, one or neither of the sub-system single
measurement probability is given by a quantum theory expression involving a
sub-system density operator. The Category 1 states (both) are the same as
the quantum separable states and are non-entangled, LHS states and
non-steered. The Category 2 states (one) are quantum entangled LHS states
(LHS) and are non-steerable. The Category 3 (neither) states are quantum
entangled and EPR steerable. A detailed study of how observables are
treated in terms of quantum theory and local hidden variable theories was
also carried out, including how the two approaches are related and how to
replace quantum operators for observables with classical entities. For
systems involving identical bosons the mode annihilation, creation operators
are replaced by quadrature amplitudes. Certain auxiliary observables also
needed to be introduced.

In a later paper we will consider tests for \emph{Bell non-locality} that
can be applied when the measurable quantities for the two sub-systems have a
range of outcomes other than the more limited $+1,-1$ outcomes considered by
Clauser et al.\ \cite{Clauser69a}.

\section*{Acknowledgements}

The authors thank S.\ Barnett, E.\ Cavalcanti, M.\ Hall, S.\ Jevtic, L.\
Rosales-Zarate, K.\ Rzazewski, T.\ Rudolph, R.\ Y.\ Teh, J.\ A.\ Vaccaro, V.\
Vedral and H.\ M.\ Wiseman for helpful discussions. BJD thanks E.\ Hinds for the
hospitality of the Centre for Cold Matter, Imperial College, London during
this work. MDR acknowledges support from the Australian Research Council via
Discovery Project Grant DP140104584. BMG acknowledges support from the UK
EPSRC via grant EP/M013294/1. This research has been supported by the
Australian Research Council Discovery Project Grants schemes under Grant
DP180102470.

\appendix

\section{Review of Hidden Variable Theory and Quantum States}

\label{Apppendix - Review of Hidden Variable Theory and Local Hidden States}

\subsection{Origin of hidden variable theory}

Local hidden variable theory has its origins in papers by Einstein,
Schr\"{o}dinger, Bell and Werner (\cite{Einstein35a,Schrodinger35a,Schrodinger35b,Bell65a,Werner89a}). Einstein
suggested that quantum theory, though correctly predicting the probabilities
for measurement outcomes was nevertheless an \emph{incomplete} theory
- in that the probabilistic measurement outcomes predicted in quantum
theory could just be the \emph{statistical} outcome of an underlying
\emph{deterministic} theory, where the possible measured outcomes
for all observables \emph{always} have specific values irrespective
of whether an \emph{actual} measurement has taken place. Hence possible
outcomes for observable quantities (such as position and momentum)
could always be regarded as elements of \emph{reality} independent
of measurement The EPR paradox is based on this assumption and involved
an entangled state for two \emph{well-separated} and no longer interacting
distinguishable particles, which had well-defined values for the position
\emph{difference} and the momentum \emph{sum}. Because of these correlations,
the choice of measuring the position (or the momentum) for the first
particle would instantly determine the outcome for the position (or
the momentum) of the second particle - a feature we now refer to as
\emph{steering} - but which Einstein called ``spooky
action at a distance'' because it conflicted with causality
(since no signal would have had time to travel between the two particles).
The paradox is that by measuring (for example) the position for the
first particle, we then know the position for the second particle
\emph{without} doing a measurement, so by then measuring the momentum
for the second particle a joint precise measurement of \emph{both}
the position and momentum for the second particle would have occurred
- which evidently conflicts with the Heisenberg uncertainty principle.
Bohm \cite{Bohm51a} described a similar paradox to EPR, but now involving
a system consisting of two spin $1/2$ particles in a singlet state,
and where the observables were spin components with \emph{quantised}
measured outcomes rather than the \emph{continuous} outcomes that
applied to EPR. The Schr\"{o}dinger cat paradox \cite{Schrodinger35b}
is another example, but now involving a \emph{macroscopic} sub-system
(the cat) in an entangled state with a \emph{microscopic} sub-system
(the two state radioactive atom). From the Einstein concept of reality,
the cat must be \emph{either} alive \emph{or }dead even \emph{before}
the box is opened to see what is the case. However, from the Copenhagen
interpretation of quantum theory (see \cite{Copenhagen} for a discussion),
the values for observables do not have a presence in reality \emph{until}
measurement takes place. Hence from the Copenhagen viewpoint the cat
is \emph{neither} dead \emph{nor} alive \emph{until} the box is opened.
Similarly, in the EPR experiment the second particle does not have
a position (or momentum) until the observable is measured. Reality
thus \emph{emerges} as the result of measurement. Thus from the Copenhagen
perspective of what constitutes reality, there are no paradoxes in
either the EPR or Schr\"{o}dinger cat scenarios.

Einstein believed that an underlying \emph{realist} theory could be
found, based on what are now referred to as \emph{hidden variables}
- which would specify the real or underlying state of the system.
Thus, quantum theory is not wrong, it is merely \emph{incomplete}.
However, it was not until 1965 before a quantitative general form
for \emph{local} \emph{hidden variable theory} was proposed by Bell
\cite{Bell65a}. This was relevant for the EPR paradox and could be
tested in experiments. In its simplest form, the key idea is that
hidden variables are specified probabilistically when the state for
the composite system is prepared, and these would determine the \emph{actual}
values for \emph{all} the sub-system observables even after the
sub-systems have separated - and even if the observables were\emph{\
incompatible} with \emph{simultaneous} precise measurements according
to quantum theory (such as two different spin components). In the
EPR\ experiment the hidden variables would specify \emph{both} the
position and momentum for each distinguishable particle. More elaborate
versions of local hidden variable theory only require the hidden variables
to determine the \emph{probabilities} of measurement outcomes for
each of the separate sub-systems, with the overall expressions for
the joint sub-system measurement outcomes then being obtained in accordance
with classical probability theory (see \cite{Wiseman07a,Jevtic15a,Dalton16a} and Section \ref{Section - Classes of Quantum States}
for details). Quantum states for composite systems that could be described
by local hidden variable theory are referred to as \emph{Bell local}.
Quantum states for composite systems that could be described by local
hidden variable theory were such that certain inequalities would apply
involving the mean values of products for the results of measuring
pairs of observables for the two sub-systems - the \emph{Bell inequalities}
\cite{Bell65a,Bell71a}. States for which a local hidden
variable theory does not apply (and hence do not satisfy Bell inequalities)
are the \emph{Bell non-local} states. Based on the entangled singlet
state of two spin $1/2$ particles Clauser et al.\ \cite{Clauser69a}
proposed an experiment that could demonstrate a violation of a Bell
inequality. This showed that local hidden variable theory could not
account for an experiment which was explained by quantum theory. Subsequent
experimental work violating Bell inequalities confirmed that there
are other quantum states for which a local hidden variable theory
does \emph{not} apply, and where quantum theory was needed to explain
the results (see Brunner et al.\ \cite{Brunner14a} for a recent review).
Numerous loopholes preventing LHVT being ruled out were shown not
to apply. However, the existence of \emph{some} quantum states (such
as the two qubit singlet \emph{Bell states} \cite{Barnett09a}) for
which the Bell inequalities are \emph{not} obeyed and where the results
were confirmed experimentally to agree with quantum theory, is itself
sufficient to show that Einstein's hope that an underlying reality
represented by a local hidden variable theory could \emph{always}
underpin quantum theory \emph{cannot} be realised.

In spite of this, there has been continued interest in determining
the circumstances in which the ideas of Einstein, Bell and others
could not be applied - that the predictions of quantum theory are
correct, and the experimental results could not be explained by a
hidden variable theory. However experience has shown that finding
Bell inequality violations is not easy. Such research is important
because it enables the regimes in which quantum theory \emph{must}
be applied to be better understood - for example, what states for
\emph{macroscopic} systems are Bell non-local? And even for states
that are Bell local, which of them exhibit the feature of EPR steering?
Although not ruling out local hidden variable theory, EPR steering
is itself a strange effect in terms of Einstein's viewpoint on reality,
so it is of interest to identify circumstances where it occurs. For
this research program \emph{bipartite} systems are often studied due
to their relative simplicity, and the simplest of these would just
involve \emph{two modes}. Since its origins HVT has been focused
on the \emph{probabilistic} predictions of quantum theory. However,
it should be noted that no \emph{unique} form for HVT has been found
that satisfies the constraint of agreeing with \emph{every} feature
of quantum theory, even for states and measurement choices where \emph{some}
of the predictions agree. As well as being probabilistic, such features
include the \emph{quantisation} for measured outcomes of certain observables
(such as angular momentum components), Heisenberg \emph{uncertainty
principle} requirements for the variances of pairs of incompatible
observables (such as position and momentum), the presence in quantum
theory of observables with \emph{non-classical} counterparts (such
as parity), the existence of a \emph{classical regimes} in quantum
theory - as well as general effects such as \emph{quantum interference}.
Although it may be possible to find versions of HVT that account for
some of these general quantum features, testing whether HVT can account
more generally for quantum results is best done via the study of phenomena
for which the predictions of HVT and quantum theory are unambiguously
different, and cannot be made to agree via minor changes to the details
in HVT. It is here that the role of measurements such as \emph{Bell
tests} are particularly important, since \emph{Bell inequality violations}
rule out \emph{all} versions of at least \emph{local }HVT (though
not excluding \emph{non-local} forms of HVT where the hidden variables
do not determine probabilities for the sub-systems separately). As
we will see, \emph{spin squeezing} for two mode systems implies EPR
steering, and hence at least ruling out some forms of LHVT - namely
those involving Category 1 and Category 2 LHVT states (see below).

\subsection{Categories of Quantum States - Overview}

It was recognised \cite{Werner89a} that \emph{all} separable states
could be described by hidden variable theory (and hence are Bell local)
and hence a state had to be \emph{entangled} to be Bell non-local.
However, Werner \cite{Werner89a} showed that \emph{some} entangled
states could also be described by hidden variable theory - and hence
not violate a Bell inequality. The relationship between the classification
of states into separable or entangled on one hand, and a classification
into Bell local and Bell non-local states on the other hand is therefore
not a simple one. This issue will be discussed in detail in Section
\ref{Section - Classes of Quantum States}. In addition to Bell locality
or non\textendash locality, there is the question of which categories
of states demonstrate the feature of steering
\cite{Einstein35a,Schrodinger35a,Schrodinger35b}, in which a choice
of measurement on one sub-system can be used to instantly affect the
outcomes for possible measurements on the other sub-system - even
if it they are well separated. For separable states, both sub-system
states are specified by quantum density operators which are determined
probabilistically in the preparation process. These are examples of
the general concept of \emph{local hidden states} (see
\cite{Wiseman07a,Jones07a,Cavalcanti09a,Jevtic15a,Dalton16a})
- which are sub-system quantum states whose density operator is specified
by hidden variables. Steerability requires the absence of local hidden
states. The physical reason for this is described in
\cite{Wiseman07a,Jones07a,Cavalcanti09a}, but for completeness this
is set out in Appendix \ref{Appendix - EPR Steering}.

In the work by Wiseman et al.\
\cite{Wiseman07a,Jones07a,Cavalcanti09a}
states for bipartite systems defined in terms
of local hidden variable theory were first categorised by whether
they are Bell local or Bell non-local. Within the states that are
Bell local a more detailed categorisation was made based on a hierarchy
of \emph{non-disjoint} sub-sets - firstly by whether they are EPR
steerable or not, and then secondly for EPR non-steerable states by
whether they are separable or not. In the present paper we apply the
concept of local hidden quantum states (whose density operators are
determined from the hidden variables) that were introduced by Wiseman
et al.\ to propose a \emph{different} categorisation of the Bell local
states into three sub-sets which are \emph{disjoint}. These are related
to the hierarchy of non-disjoint sub-sets introduced by Wiseman et
al.. The disjoint sub-sets of states are defined by whether two, one
or none of the sub-system hidden variable probabilities is \emph{also}
obtained from a local hidden quantum state. \emph{Category 1} states
involve two hidden states, and this Bell local sub-set is the same
as the separable states. These are non-steerable. \emph{Category 2}
states involve only one hidden state and for this Bell local sub-set
the states are entangled, though non-steerable. \emph{Category 3}
states do not involve any hidden state, and these Bell local states
are both entangled and steerable. We will also designate the states
that are Bell non-local as \emph{Category 4} states, and these states
are both entangled and steerable. The categorisation of the quantum
states both in terms of entanglement versus separability and alternatively
Bell locality versus Bell non-locality is summarised in Fig.~1.

It is of some interest to devise tests for which specific category
a quantum state falls into in the context of \emph{bipartite} systems
of \emph{identical massive bosons}, such as occur in Bose-Einstein
condensates for cold bosonic atomic gases. We treat the simplest situation
where each sub-system involves just a \emph{single} mode. For these
systems, both the \emph{symmetrisation principle} and the \emph{super-selection
rule} for particle number must be applied. The focus of this paper
is on whether the quantum state is \emph{EPR\ steerable} - which
means showing that it is not a Category 1 or a Category 2 state. In
previous work tests have been obtained (see \cite{Dalton16b} for
details of a range of tests found by various authors) for showing
that a state is \emph{entangled}, which therefore rules them out from
being in Category 1. Hence we only need to consider tests for showing
that the state is also not in Category 2. Based on local hidden variable
theory, predictions can be made for Category 2 states involving the
mean values and variances for measurement outcomes. For observables
associated with the sub-system for which there is a local hidden state,
quantum expressions may be applied.

\section{Basic Measurement Probabilities for Bipartite Systems}

\label{Appendix A - Basic Measurement Probabilities}

This paper deals with measurements on \emph{bipartite} composite
quantum systems, where we have two \emph{distinguishable} sub-systems
$A$ and $B$ which are each associated with measurable physical
\emph{observables} $\Omega_{A}$ and $\Omega_{B}$ for which possible
\emph{outcomes} are denoted $\alpha$ and $\beta$. The composite
system exists in various \emph{quantum} \emph{states}, whose \emph{preparation}
is symbolised by $c$. Quantum theory has the key feature that such
measurements the occurrence of particular outcomes are specified by
\emph{probabilities} rather than being \emph{deterministic}, and the
basic quantity of interest is the \emph{joint probability} $P(\alpha,\beta|\Omega_{A},\Omega_{B},c)$
for measurement of \emph{any} pair of sub-system \emph{observables}
$\Omega_{A}$ and $\Omega_{B}$ to obtain \emph{any} of their possible
\emph{outcomes} $\alpha$ and $\beta$ when the \emph{preparation}
process is $c$. As the sub-systems are distinct \emph{simultaneous
precise} \emph{measurement} outcomes apply for the pairs of observables
$\Omega_{A}$ and $\Omega_{B}$ in both quantum and hidden variable
theory (in the latter case the observables are classical variables
and not Hermitian operators). The probability $P(\alpha,\beta|\Omega_{A},\Omega_{B},c)$
is of course \emph{real} and \emph{positive} and its sum for all outcomes
for both $\Omega_{A}$ and $\Omega_{B}$ is equal to unity. The sum
of the joint probability over the possible outcomes $\alpha$ for
measuring $\Omega_{A}$ \emph{defines} the \emph{single probability}
$P(\beta|\Omega_{B},c)$ for measuring $\Omega_{B}$ with outcome
$\beta$, \emph{irrespective} of the outcome for measuring $\Omega_{A}$.
A similar definition applies for the single probability $P(\alpha|\Omega_{A},c)$
for measuring $\Omega_{A}$ with outcome $\alpha$, irrespective of
the outcome for measuring $\Omega_{B}$. Thus:
\begin{align}
\sum\limits _{\alpha,\beta}P(\alpha,\beta|\Omega_{A},\Omega_{B},c) & =  1 , \label{Eq.SumJointProballOutcomes}\\
P(\beta|\Omega_{B},c) & =  \sum\limits _{\alpha}P(\alpha,\beta|\Omega_{A},\Omega_{B},c) , \label{Eq.SingleProbMeasureB}\\
P(\alpha|\Omega_{A},c) & =  \sum\limits _{\beta}P(\alpha,\beta|\Omega_{A},\Omega_{B},c) . \label{Eq.SingleProbMeastA}
\end{align}
The single probabilities also satisfy the expected \emph{probability
sum rules} 
\begin{align}
\sum\limits _{\beta}P(\beta|\Omega_{B},c)&=1,&\sum\limits _{\alpha}P(\alpha|\Omega_{A},c)&=1,\label{Eq.SingleProbSumRules}
\end{align}
which follow from (\ref{Eq.SumJointProballOutcomes}).

From the joint measurement probability $P(\alpha, \beta | \Omega_{A}, \Omega_{B}, c)$
and the single measurement probabilities $P(\alpha|\Omega_{A},c)$ and
$P(\beta|\Omega_{B},c)$, we can introduce \emph{conditional probabilities}
$P(\beta|\Omega_{B}||\alpha,\Omega_{A},c)$ and $P(\alpha|\Omega_{A}||\beta,\Omega_{B},c)$.
Here $P(\beta|\Omega_{B}||\alpha,\Omega_{A},c)$ is the probability
that measurement of the observable $\Omega_{B}$\ yields the outcome
$\beta$\ given that measurement of the observable $\Omega_{A}$\ yields
the outcome $\alpha$. This (and the corresponding expression for
$P(\alpha|\Omega_{A}| |\beta,$ $\Omega_{B},c)$) is given by \emph{Bayes'
theorem} as 
\begin{eqnarray}
P(\beta|\Omega_{B}||\alpha,\Omega_{A},c) & = & \frac{P(\alpha,\beta|\Omega_{A},\Omega_{B},c)}{P(\alpha|\Omega_{A},c)},\nonumber \\
P(\alpha|\Omega_{A}||\beta,\Omega_{B},c) & = & \frac{P(\alpha,\beta|\Omega_{A},\Omega_{B},c)}{P(\beta|\Omega_{B},c)}.\label{Eq.LHVTConditionalProb}
\end{eqnarray}
All these expressions apply irrespective of whether the joint and
single measurement probabilities are obtained from \emph{quantum theory}
or \emph{local hidden variable theory} formulae.

\section{Mean Values and Variances - General Features}

\label{Appendix - Mean Values and Variances}

\subsection{Mean Values and Variances - Quantum Models}

In a fully quantum treatment, any\ observable represented by a Hermitian
operator $\widehat{\Omega}$\ - whose measured outcomes are its eigenvalues
$\theta$, can be written as $\widehat{\Omega}=\sum_{\theta}\theta\,\widehat{\Pi}_{\theta}$ in
terms of its projectors $\widehat{\Pi}_{\theta}$ and we can determine
the probability $P(\widehat{\Omega},\theta)$ for the outcome
$\theta$ via $P(\widehat{\Omega},\theta)=\Tr(\widehat{\Pi}_{\theta}\,\widehat{\rho})$ -
where $\widehat{\rho}$\ is the density operator that specifies the
quantum state. Hence the mean value of the measured outcomes can be
defined and then determined as follows
\begin{eqnarray}
\left\langle \widehat{\Omega}\right\rangle _{Q} & = & \sum_{\theta}\theta\,P(\widehat{\Omega},\theta),\label{Eq.MeanGeneralQuantum}\\
 & = & \Tr(\widehat{\Omega}\,\widehat{\rho}) . \label{Eq.TraceResultMean}
\end{eqnarray}

We can also extend the concept of the mean value for measured outcomes
to the case of a non-Hermitian operator $\widehat{\Omega}$ - which
although it does not correspond to an observable can be written in
the form $\widehat{\Omega}=\widehat{\Omega}_{1}+i\widehat{\Omega}_{2}$,
where both $\widehat{\Omega}_{1}$\ and $\widehat{\Omega}_{2}$\ are
each observable Hermitian operators, not necessarily commuting. We
simple define the mean for $\widehat{\Omega}$\ via 
\begin{eqnarray}
\left\langle \widehat{\Omega}\right\rangle  & \equiv & \left\langle \widehat{\Omega}_{1}\right\rangle +i\left\langle \widehat{\Omega}_{2}\right\rangle \nonumber \\
 & = & \Tr\left[ (\widehat{\Omega}_{1}+i\widehat{\Omega}_{2})\widehat{\rho}\right] , \label{Eq.MeanNonHermOpr}
\end{eqnarray}
where $\langle \widehat{\Omega}_{1}\rangle $\ and $\langle \widehat{\Omega}_{2}\rangle $ are
defined as in (\ref{Eq.MeanGeneralQuantum}), and we see that the
result is given by the trace process. This definition and result can
be applied to provide a meaning for the quantum mean values of operators
such as an annihilation operator $\widehat{a}$ $=\frac{1}{\sqrt{2}}(\widehat{x}_{A}+i\widehat{p}_{A})$ -
which can be written in terms of \emph{quadrature operators} or
a transition operator $\widehat{b}^{\dag}\widehat{a}=\widehat{S}_{x}+i\widehat{S}_{y}$ -
which can be expressed in terms of \emph{spin operators}.
The latter case applies for considering \emph{correlation tests}.
If $\widehat{\Omega}$ can be written as the sum of products of
Hermitian sub-system operators $\widehat{\Omega}_{A}$\ and $\widehat{\Omega}_{B}$\
the last expression can be used to evaluate the mean value based on
the quantum probability distributions for measurements of each $\widehat{\Omega}_{A}$\ and
$\widehat{\Omega}_{B}$.

Note that in expressing $\langle \widehat{\Omega}\rangle $
in terms of $\langle \widehat{\Omega}_{1}\rangle $ and $\langle \widehat{\Omega}_{2}\rangle $
we are considering the results of two \emph{independent} sets of measurements,
one set for $\widehat{\Omega}_{1}$ and the other for $\widehat{\Omega}_{2}$.
We do not imply that there is a joint probability $P(\omega_{1},\omega_{2}|\Omega_{1},\Omega_{2},c)$
for simultaneous outcomes $\omega_{1},\omega_{2}$ of a combined measurement
of $\Omega_{1},\Omega_{2}$ following preparation $c$. We only require
\emph{single} measurement probabilities $P(\omega_{1}|\Omega_{1},c)$
and $P(\omega_{2}|\Omega_{2},c)$ to exist in order to define the
mean values via $\langle \widehat{\Omega}_{1}\rangle =\sum_{\omega_{1}}\omega_{1}\,P(\omega_{1}|\Omega_{1},c)$,
which corresponds to the set of measurements on $\widehat{\Omega}_{1}$
\emph{alone}. In von-Neumann's proof that hidden variable theories
were \emph{inconsistent} with quantum theory, he had evidently used
the equivalent of $\langle \widehat{\Omega}\rangle =\sum_{\omega_{1}}\sum_{\omega_{2}}(\omega_{1}+i\omega_{2})\,P(\omega_{1},\omega_{2}|\Omega_{1},\Omega_{2},c)$
based on \emph{one} set of measurements, whereas we just use $\langle \widehat{\Omega}\rangle =\sum_{\omega_{1}}(\omega_{1})\,P(\omega_{1}|\Omega_{1},c)+i\sum_{\omega_{2}}(\omega_{2})\,P(\omega_{2}|\Omega_{2},c)$,
which rests on two independent sets of measurements.

In the case of quantum separable states the \emph{mean values} for
jointly measuring $\Omega_{A}$\ in sub-system $A$\ and $\Omega_{B}$\ in
sub-system $B$\ for preparation $\rho$\ would be given by
\begin{equation}
\left\langle \Omega_{A}\Omega_{B}\right\rangle =\sum\limits _{R}P_{R}\,\left\langle \Omega_{A}\right\rangle _{R}\,\left\langle \Omega_{B}\right\rangle _{R} , 
\label{Eq.MeanValQSS}
\end{equation}
where $\left\langle \Omega_{A}\right\rangle _{R} = \dsum_{\alpha} \alpha P_{Q}(\alpha|\Omega_{A},\rho,R)=\Tr(\widehat{\Omega}_{A} \widehat{\rho}_{R}^{A})$ and
$\left\langle \Omega_{B}(\lambda)\right\rangle _{Q} = \sum_{\beta}  \beta P_{Q}(\beta|\Omega_{B},\rho,R)=\Tr(\widehat{\Omega}_{B} \widehat{\rho}_{R}^{B})$ are
the mean values for measurement outcomes for $\Omega_{A}$\ and $\Omega_{B}$.
For the quantum separable state the mean value for \emph{any} sum
of products of sub-system operators which is Hermitian overall would
be given by 
\begin{equation}
\left\langle \sum\limits _{i}\widehat{\Omega}_{Ai}\widehat{\Omega}_{Bi}\right\rangle =\sum\limits _{R}P_{R}\sum\limits _{i}\left\langle \widehat{\Omega}_{Ai}\right\rangle _{R}\,\left\langle \widehat{\Omega}_{Bi}\right\rangle _{R} , \label{Eq.MeanSumProducts}
\end{equation}
where $\langle \widehat{\Omega}_{Ai}\rangle _{R}=\Tr(\widehat{\Omega}_{Ai}\widehat{\rho}_{R}^{A})$
and $\langle \widehat{\Omega}_{Bi}\rangle _{R}=\Tr(\widehat{\Omega}_{Bi}\widehat{\rho}_{R}^{B})$
are quantum mean values, since we can always write $\widehat{\Omega}_{Ai}=\widehat{\Omega}_{Ai}^{(1)}+i\widehat{\Omega}_{Ai}^{(2)}$
where both $\widehat{\Omega}_{Ai}^{(1)}$ and $\widehat{\Omega}_{Ai}^{(2)}$ are
Hermitian and can be regarded as observables. So with $\widehat{\Omega}_{Ai}\widehat{\Omega}_{Bi}=\widehat{\Omega}_{Ai}^{(1)}\widehat{\Omega}_{Bi}^{(1)}-\widehat{\Omega}_{Ai}^{(2)}\widehat{\Omega}_{Bi}^{(2)}+i(\widehat{\Omega}_{Ai}^{(1)}\widehat{\Omega}_{Bi}^{(2)}-\widehat{\Omega}_{Ai}^{(2)}\widehat{\Omega}_{Bi}^{(1)})$
which is of the form $\widehat{\Omega}_{1}+i\widehat{\Omega}_{2}$,
where both $\widehat{\Omega}_{1}$ and $\widehat{\Omega}_{2}$ are
each observable Hermitian operators (the $A$ and $B$ operators commute),
we can then invoke the probability distributions for the $\widehat{\Omega}_{Ai}^{(1)}$,
$\widehat{\Omega}_{Bi}^{(1)}$, $\widehat{\Omega}_{Ai}^{(2)}$ and
$\widehat{\Omega}_{Bi}^{(2)}$ to derive the expression for the mean
value of $\widehat{\Omega}_{Ai}\widehat{\Omega}_{Bi}$ by also using
(\ref{Eq.MeanNonHermOpr}). So (\ref{Eq.MeanSumProducts}) applies
even if quantum operators $\widehat{\Omega}_{Ai}$ and $\widehat{\Omega}_{Bi}$
do not represent observables.

\emph{Variances} can be obtained based on considering the
mean values of the square of $\widehat{\Omega}$. For an observable
represented by a Hermitian operator $\widehat{\Omega}$ the variance
is defined by the mean of the squared variation of outcomes from the
mean and equal to the difference between the mean of $\widehat{\Omega}^{2}$
and the square of the mean of $\widehat{\Omega}$:
\begin{eqnarray}
\left\langle \Delta\widehat{\Omega}^{2}\right\rangle _{Q} & = & \sum\limits _{\theta}\left(\theta-\left\langle \widehat{\Omega}\right\rangle _{Q}\right)^{2}P(\widehat{\Omega},\theta) 
, \nonumber \\
 & = & \left\langle \widehat{\Omega}^{2}\right\rangle _{Q}-\left\langle \widehat{\Omega}\right\rangle _{Q}^{2} . \label{Eq.VarianceQuantum}
\end{eqnarray}

In the case of a \emph{mixed} state (such as
the QSS)
\begin{equation}
\widehat{\rho}=\sum_{R}P_{R}\,\widehat{\rho}_{R}\label{Eq.MixedState}
\end{equation}
the \emph{mean} for a Hermitian operator $\widehat{\Omega}$
is the average of means for separate components
\begin{equation}
\left\langle \widehat{\Omega}\right\rangle =\sum_{R}P_{R}\,\left\langle \widehat{\Omega}\right\rangle _{R} , \label{Eq.MeanResult}
\end{equation}
where $\langle \widehat{\Omega}\rangle _{R}=\Tr(\widehat{\rho}_{R}\widehat{\Omega})$.
The variance for a Hermitian operator $\widehat{\Omega}$ in a mixed
state is always never less than the the average of the variances for
the separate components (see \cite{Hoffmann03a}) 
\begin{equation}
\left\langle \Delta\widehat{\Omega}\,^{2}\right\rangle \geq\sum_{R}P_{R}\,\left\langle \Delta\widehat{\Omega}_{R}{}^{2}\right\rangle _{R} , \label{Eq.VarianceResult}
\end{equation}
where $\langle \Delta\widehat{\Omega}^{2}\rangle =\Tr(\widehat{\rho}\Delta\widehat{\Omega}^{2})$ with
$\Delta\widehat{\Omega}=\widehat{\Omega}-\langle \widehat{\Omega}\rangle $, and
$\langle \Delta\widehat{\Omega}^{2}\rangle _{R}=\Tr(\widehat{\rho}_{R}\Delta\widehat{\Omega}_{R}^{2})$ with
$\Delta\widehat{\Omega}_{R}=\widehat{\Omega}-\langle \widehat{\Omega}\rangle _{R}$.
To prove this result we have using (\ref{Eq.MeanResult}) both for
$\widehat{\Omega}$ and $\widehat{\Omega}^{2}$,
\begin{eqnarray}
\left\langle \Delta \widehat{\Omega }\,^{2}\right\rangle &=&\left\langle 
\widehat{\Omega }\,^{2}\right\rangle -\left\langle \widehat{\Omega }%
\right\rangle ^{2}  , \nonumber \\
&=&\sum_{R}P_{R}\,\left( \left\langle \widehat{\Omega }{}^{2}\right\rangle
_{R}-\left\langle \widehat{\Omega }\right\rangle _{R}^{2}\right)  \nonumber
\\
&&+\sum_{R}P_{R}\,\left\langle \widehat{\Omega }\right\rangle
_{R}^{2}-\left( \sum_{R}P_{R}\,\left\langle \widehat{\Omega }\right\rangle
_{R}\right) ^{2} ,  \nonumber \\
&=&\sum_{R}P_{R}\,\left\langle \Delta \widehat{\Omega }{}_{R}^{2}\right%
\rangle _{R}+\sum_{R}P_{R}\,\left\langle \widehat{\Omega }\right\rangle
_{R}^{2}  \nonumber \\
&&-\left( \sum_{R}P_{R}\,\left|\left\langle \widehat{\Omega }\right\rangle
_{R}\right|\right) ^{2}   .
\end{eqnarray}%
The variance result (\ref{Eq.VarianceResult}) follows because the
sum of the last two terms is always $\geq0$ using the result (135) in
Appendix E of Ref \cite{Dalton16a}, with $C_{R}=\langle \widehat{\Omega}\rangle _{R}^{2}$,
and $\sqrt{C_{R}} = |\langle \widehat{\Omega}\rangle_{R}|$,
which are real and positive.

In considering the means and variances in the context of LHVT several
difficult issues need to be dealt with. Firstly, in a LHV the observables
are basically considered as classical c-numbers, but given that the
predictions from quantum theory are accepted as being correct these
classical observables must correspond to underlying quantum Hermitian
operators - especially as when a local hidden state occurs where the
probabilities $P_{Q}(\beta|\Omega_{B},c,\lambda)$ for sub-system
$B$ are also to be given by quantum formulae. Also, there are several
entanglement tests involving spin components, these are represented
by the spin operators $\widehat{S}_{x}=(\widehat{b}^{\dag}\widehat{a}+\widehat{a}^{\dag}\widehat{b})/2$,
$\widehat{S}_{y}=(\widehat{b}^{\dag}\widehat{a}-\widehat{a}^{\dag}\widehat{b})/2i$\ and
$\widehat{S}_{z}=(\widehat{b}^{\dag}\widehat{b}-\widehat{a}^{\dag}\widehat{a})/2$,
where $\widehat{a}$\ and $\widehat{b}$ are mode annihilation operators.
The tests also involve the total number operator $\widehat{N}=(\widehat{b}^{\dag}\widehat{b}+\widehat{a}^{\dag}\widehat{a})$.
All these operators are Hermitian and represent observable quantities
applying for the overall two mode system. We may also consider number
operators for the two modal sub-systems defined by $\widehat{N}_{A}=\widehat{a}^{\dag}\widehat{a}$
and $\widehat{N}_{B}=\widehat{b}^{\dag}\widehat{b}$, which again
are Hermitian and represent observable quantities for each sub-system.
The question then arises: How do you define the spin components and
the boson number when the observables are supposed to be non-quantum?
Secondly, when considering entanglement tests involving spin
components, both sub-system $A$ and $B$ involve mode annihilation
operators - which are non-Hermitian and not themselves associated
with measurable observables. What meaning can we give to LHVT probabilities
$P(\alpha|\Omega_{A},c,\lambda)$ and associated mean values 
$\left\langle \Omega_{A}(\lambda)\right\rangle =\sum_{\alpha}\alpha\,P(\alpha|\Omega_{A},c,\lambda)$ for
sub-system $A$\ when during the discussion of spin squeezing tests
we consider situations where $\Omega_{A}$ corresponds to a mode
annihilation or creation operator? Do we need to consider non-local
HVT\ probabilities $P(\alpha_{1},\alpha_{2}|\Omega_{A1},\Omega_{A2},c,\lambda)$\ associated
with the outcomes of measuring \emph{two} observables
$\Omega_{A1},\Omega_{A2}$ for sub-system $A$ when the hidden
variables are $\lambda$ and which may correspond to quantum operators
that do not commute? What happens when we need to consider a product
such as $\Omega_{A1}\Omega_{A2}\Omega_{B1}\Omega_{B2}$\ such as
may occur when we are considering expressions for variances? Would
this mean that for products of sub-system observables we should 
use the expression
\begin{align}
\left\langle \Omega_{A1}\Omega_{A2}\Omega_{B1}\Omega_{B2}\right\rangle =&\sum\limits _{\lambda}P(\lambda|c)
\nonumber \\
& \times
\left\langle \Omega_{A1}\Omega_{A2}(\lambda)\right\rangle \,\left\langle \Omega_{B1}\Omega_{B2}(\lambda)\right\rangle _{Q} \,,
\nonumber \\&
\end{align}
where 
\begin{align}
\left\langle \Omega_{A1}\Omega_{A2}(\lambda)\right\rangle 
&=\sum\limits _{\alpha_{1},\alpha_{2}} \alpha\,_{1}\alpha_{2}\,P(\alpha_{1},\alpha_{2}|\Omega_{A1},\Omega_{A2},c,\lambda)
\nonumber \\
\left\langle \Omega_{B1}\Omega_{B2}(\lambda)\right\rangle _{Q}
&=\sum\limits _{\beta_{1},\beta_{2}} \beta\,_{1}\beta_{2}\,P_{Q}(\beta_{1},\beta_{2}|\Omega_{B1},\Omega_{B2},c,\lambda)
\,,
\nonumber \\
\end{align}
to determine the mean values?
But what meaning is there to the quantum expression when the corresponding
operators $\widehat{\Omega}_{B1},\widehat{\Omega}_{B2}$ do not
commute?

None of these questions arose in considering whether spin squeezing
is a test for standard quantum entanglement, since no hidden variables
are involved nor are issues of the existence of probabilities for
measurement of individual sub-system operators that may become involved
in the evaluation. However, when non-quantum LHVT expressions for
measurement probabilities are involved, the analogous results to those
for quantum mean values need further consideration. Until these
issues are resolved we cannot begin to modify the operator based proof
regarding the consequences for spin variances and means for LHVT state.
The proof would involve expressions giving meaningful interpretations
to the mean values of what would appear to be non-physical quantities
such as mode annihilation and creation operators for sub-system $A$.

\subsection{General Results for Mean and Variance in LHVT}

Before dealing with the above issues it is useful to prove some results
for mean values and variances in general HVT that are analogous to
similar results in quantum theory. We now consider the measurement
of an observable $\Omega$ with outcomes $\omega$ for a preparation
process $c$. The probability $P(\omega|\Omega,c)$ for this outcome
can be written in LHV as 
\begin{equation}
P(\omega|\Omega,c)=\sum P(\lambda|c)\,P(\omega|\Omega,c,\lambda),\label{Eq.MeastProbHVT}
\end{equation}
where $\lambda$ are the hidden variables and $P(\lambda|c)$ is the
probability for preparation process $c$ that the hidden variables
are $\lambda$ and $\,P(\omega|\Omega,c,\lambda)$ is the probability
of outcome $\omega$ for measurement of $\Omega$ when the hidden
variables are $\lambda$.

The \emph{mean value} for measurement outcomes for observable $\Omega$
will then be given by
\begin{eqnarray}
\left\langle \Omega\right\rangle  & = & \sum\limits _{\omega}\omega\,P(\omega|\Omega,c)\label{Eq.MeanHVTDefn}\\
 & = & \sum\limits _{\lambda}P(\lambda|c)\,\left\langle \Omega(\lambda)\right\rangle , \label{Eq.MeanHVTResult}\\
\left\langle \Omega(\lambda)\right\rangle  & = & \sum\limits _{\omega}\omega\,P(\omega|\Omega,c,\lambda) , \label{Eq.MeanHVTVariableLambda}
\end{eqnarray}
where the first equation is the definition and the second equation
shows that the mean value is given by weighting the mean value $\,\left\langle \Omega(\lambda)\right\rangle $
that would apply if the hidden variables are $\lambda$, by the probability
$P(\lambda|c)$ for these hidden variables when the preparation is
$c$. The result (\ref{Eq.MeanHVTResult}) is similar to the quantum
result for the mixed state $\widehat{\rho}=\sum_{R}P_{R}\widehat{\rho}_{R}$
where $\langle \widehat{\Omega}\rangle =\sum P_{R}\langle \widehat{\Omega}\rangle _{R}$ and
$\langle \widehat{\Omega}\rangle _{R}=\Tr(\widehat{\Omega}\,\widehat{\rho}_{R})$.
The result for the mean value of a \emph{function} $F(\Omega)$ would
be
\begin{eqnarray}
\left\langle F(\Omega)\right\rangle  & = & \sum\limits _{\lambda}P(\lambda|c)\,\left\langle F(\Omega)_{\lambda}\right\rangle , \nonumber \\
\left\langle F(\Omega)_{\lambda}\right\rangle  & = & \sum\limits _{\omega}F(\omega)\,P(\omega|\Omega,c,\lambda) . \label{Eq.MeanHHVTFn}
\end{eqnarray}

In the case of \emph{two} observables $\Omega$ and
$\Lambda$ with outcomes $\omega$ and $\mu$, the mean value for a
\emph{function} $F(\Omega,\Lambda)$ when the preparation process is
$c$, would be
\begin{eqnarray}
\left\langle F(\Omega,\Lambda)\right\rangle  & = & \sum\limits _{\lambda}P(\lambda|c)\,\left\langle F(\Omega,\Lambda)_{\lambda}\right\rangle , \nonumber \\
\left\langle F(\Omega,\Lambda)_{\lambda}\right\rangle  & = & \sum\limits _{\omega\mu}F(\omega,\mu)\,P(\omega,\mu|\Omega,\Lambda,c,\lambda) . \label{Eq.LHVTFunction}
\end{eqnarray}
This result will be useful when we consider steering tests.

The \emph{variance} for measurement outcomes for observable $\Omega$
will then be given by
\begin{eqnarray}
\left\langle \Delta\Omega^{2}\right\rangle  & = & \sum\limits _{\omega}(\omega\,-\left\langle \Omega\right\rangle )^{2}P(\omega|\Omega,c),\label{Eq.VarHVTDefn}\\
 & = & \sum\limits _{\omega}\bigl(\omega^{2}\,-2\omega\left\langle \Omega\right\rangle +\left\langle \Omega\right\rangle ^{2}\bigr)P(\omega|\Omega,c),\nonumber \\
 & = & \left\langle \Omega^{2}\right\rangle -\left\langle \Omega\right\rangle ^{2},\label{Eq.VarHVTResult}\\
\left\langle \Omega^{2}\right\rangle  & = & \sum\limits _{\omega}\omega^{2}\,P(\omega|\Omega,c),\label{Eq.MeanSquareHVTDefn}
\end{eqnarray}
where the first equation is the definition and the third equation
shows that the variance is given by the difference between the mean
of the squared observable and the square of the mean, as in standard
statistics. Here we have used $\sum_{\omega}P(\Omega|\omega,c)=1$
and (\ref{Eq.MeanHVTDefn}). We can then write
\begin{eqnarray}
\left\langle \Omega^{2}\right\rangle  & = & \sum\limits _{\lambda}P(\lambda|c)\,\left\langle \Omega^{2}(\lambda)\right\rangle,\label{Eq.ResultMeanSquaredObservableHVT}\\
\left\langle \Omega^{2}(\lambda)\right\rangle  & = & \sum\limits _{\omega}\omega^{2}\,P(\omega|\Omega,\lambda,c)\,,\label{Eq.MeanSquareObservParticularLambdaDefn}
\end{eqnarray}
where the second line gives the definition for the mean of the square
of the observable when the hidden variables are $\lambda$ and the
first line expresses the mean of the square of the observable in terms
of an average over this quantity.

We then have
\begin{align}
\left\langle \Delta\Omega^{2}\right\rangle  & =  \sum\limits _{\lambda}P(\lambda|c)\,\left\langle \Omega^{2}(\lambda)\right\rangle - \biggl(\sum\limits _{\lambda}P(\lambda|c)\,\left\langle \Omega(\lambda)\right\rangle \biggr)^{2},\nonumber \\
 & \geq  \sum\limits _{\lambda}P(\lambda|c)\,\bigl(\left\langle \Omega^{2}(\lambda)\right\rangle -\left\langle \Omega(\lambda)\right\rangle ^{2}\bigr)
 \nonumber \\
 & \quad
+\sum\limits _{\lambda}P(\lambda|c)\left\langle \Omega(\lambda)\right\rangle ^{2}- \biggl(\sum\limits _{\lambda}P(\lambda|c) |\left\langle \Omega(\lambda)\right\rangle |\biggr)^{2},\nonumber \\
 & \geq  \sum\limits _{\lambda}P(\lambda|c) \bigl(\left\langle \Omega^{2}(\lambda)\right\rangle -\left\langle \Omega(\lambda)\right\rangle ^{2}\bigr),\label{Eq.InequalityVarHVT}
\end{align}
which establishes an important inequality. The second line follows
from the modulus of a sum being less than the sum of the moduli, and
the last line follows from the Cauchy inequality $\sum_{R}P_{R}C_{R}\geq(\sum_{R}P_{R}\sqrt{C_{R}})^{2}$
with $\sqrt{C_{R}} = |\langle \Omega(\lambda)\rangle |$.
But we also have
\begin{eqnarray}
\left\langle \Delta\Omega^{2}(\lambda)\right\rangle  & = & \sum\limits _{\omega}(\omega-\left\langle \Omega(\lambda)\right\rangle )^{2}\,P(\omega|\Omega,c,\lambda)\label{Eq.VarHVTParticularLambdaDefn}\\
 & = & \sum\limits _{\omega}\omega^{2}\,P(\omega|\Omega,c,\lambda)-\left\langle \Omega(\lambda)\right\rangle ^{2}\nonumber \\
 & = & \left\langle \Omega^{2}(\lambda)\right\rangle -\left\langle \Omega(\lambda)\right\rangle ^{2}\label{Eq.VarHVTParticularLambdaResult}
\end{eqnarray}
showing that when the hidden variable is $\lambda$ the variance for
measured outcomes of observable $\Omega$ is equal to the difference
between the mean value for measured outcomes of the square of the
observable and the square of the mean value (as expected).

We finally have the inequality
\begin{equation}
\left\langle \Delta\Omega^{2}\right\rangle \geq\sum_{\lambda}P(\lambda|c)\,\left\langle \Delta\Omega^{2}(\lambda)\right\rangle . \label{Eq.VarianceInequalityHVT}
\end{equation}
This result may be compared to the quantum theory result $\langle \Delta\widehat{\Omega}^{2}\rangle \geq\sum_{R}P_{R}\langle \Delta\widehat{\Omega}^{2}\rangle _{R}$.

Finally, we consider mean values in general HVT for complex combinations
of observables $\Omega_{1}$ and $\Omega_{2}$, which have measured
outcomes $\omega_{1}$ and $\omega_{2}$. We can easily show that
\begin{equation}
\left\langle (\Omega_{1}+i\Omega_{2})\right\rangle =\left\langle \Omega_{1}\right\rangle +i\left\langle \Omega_{2}\right\rangle, \label{Eq.MeanComplexCombHVTDefn}
\end{equation}
where in HVT we have
\begin{align}
\left\langle \Omega _{1}\right\rangle &=\sum\limits_{\lambda }P(\lambda
|c)\,\sum\limits_{\omega _{1}}\omega _{1}\,P(\omega _{1},\omega _{2}|\Omega
_{1},\Omega _{2},c,\lambda )  ,\nonumber \\
\left\langle \Omega _{2}\right\rangle &=\sum\limits_{\lambda }P(\lambda
|c)\,\sum\limits_{\omega _{2}}\omega _{2}\,P(\omega _{1},\omega _{2}|\Omega
_{1},\Omega _{2},c,\lambda )  ,\label{Eq.MeanHVTTwoObserv}
\end{align}
since the fundamental probability $P(\omega _{1},\omega _{2}|\Omega
_{1},\Omega _{2},c,\lambda )$
always exists 
in a LHV, \emph{even if} in quantum theory the corresponding operators $\widehat{\Omega}_{1}$
and $\widehat{\Omega}_{2}$ do \emph{not} commute. This is an important
feature to recognise about LHV. The result (\ref{Eq.MeanComplexCombHVTDefn})
may be compared to the quantum result (\ref{Eq.MeanNonHermOpr}).
Thus, we see that many results in HVT are analogous to the results
in quantum theory.

With these results now established we can see that for Category 2
states the \emph{mean values} for jointly measuring $\Omega_{A}$
in sub-system $A$ and $\Omega_{B}$ in sub-system $B$ for preparation
$c$ would be given by
\begin{equation}
\left\langle \Omega_{A}\otimes\Omega_{B}\right\rangle =\sum\limits _{\lambda}P(\lambda|c)\left\langle \Omega_{A}(\lambda)\right\rangle \,\left\langle \Omega_{B}(\lambda)\right\rangle _{Q}\label{Eq.MeanValues}
\end{equation}
where $\left\langle \Omega_{A}(\lambda)\right\rangle $ $=\sum_{\alpha}$
$\alpha\,P(\alpha|\Omega_{A},c,\lambda)$ and $\left\langle \Omega_{B}(\lambda)\right\rangle _{Q}$
$=\sum_{\beta}$ $\beta\,P_{Q}(\beta|\Omega_{B},c,\lambda)=\Tr(\widehat{\Omega}_{B}\,\widehat{\rho}_{\lambda}^{B})$
are the definitions of the mean values for measurement outcomes for
$\Omega_{A}$ and $\Omega_{B}$. The latter is also determined from
quantum theory; the former is not. \emph{Variances} can be obtained
based on considering the mean values of the squares of $\Omega_{A}$
and $\Omega_{B}$. The similarities and differences between the Category
2 states and the quantum Separable (Category 1) states expressions
(\ref{Eq.MeanValues}) and (\ref{Eq.MeanValQSS}) should be noted.

\subsection{Links between Quantum Theory and LHVT}

We will also need to consider the mean values for observables which
in quantum theory are given by the \emph{sum} of \emph{products} of
sub-system Hermitian operators, where the operators for each sub-system
do not necessarily commute - $[\widehat{\Omega}_{A1},\widehat{\Omega}_{A2}]\neq0$
etc.. The links between quantum theory and LHVT for these cases are
set out here. Thus for 
\begin{equation}
\widehat{\Omega}=\widehat{\Omega}_{A1}\otimes\widehat{\Omega}_{B1}+\widehat{\Omega}_{A2}\otimes\widehat{\Omega}_{B2}\label{Eq.SumProdQOprs}
\end{equation}
the mean value will be given in \emph{quantum theory} by 
\begin{eqnarray}
\left\langle \widehat{\Omega}\right\rangle  & = & \left\langle \widehat{\Omega}_{A1}\otimes\widehat{\Omega}_{B1}\right\rangle +\left\langle \widehat{\Omega}_{A2}\otimes\widehat{\Omega}_{B2}\right\rangle, \nonumber \\
 & = & \Tr(\widehat{\Omega}_{A1}\otimes\widehat{\Omega}_{B1})\widehat{\rho}\;+\;\Tr(\widehat{\Omega}_{A2}\otimes\widehat{\Omega}_{B2})\widehat{\rho},\nonumber \\
                                            & = & \sum\limits _{\alpha_{1}\beta_{1}}\alpha_{1}\beta_{1}\,P(\alpha_{1},\beta_{1}|\Omega_{A1},\Omega_{B1},c)
\nonumber\\ &&
+\sum\limits_{\alpha_{2}\beta2}\alpha_{2}\beta_{2}\,P(\alpha_{2},\beta_{2}|\Omega_{A2},\Omega_{B2},c),\nonumber \\
\label{Eq.MeanSumProdQThy}
\end{eqnarray}
where 
\begin{align}
P(\alpha_{1},\beta_{1}|\Omega_{A1},\Omega_{B1},c)&=\Tr(\widehat{\Pi}_{\alpha1}\otimes\widehat{\Pi}_{\beta1})\widehat{\rho}
\nonumber\\  P(\alpha_{2},\beta_{2}|\Omega_{A2},\Omega_{B2},c)&=\Tr(\widehat{\Pi}_{\alpha2}\otimes\widehat{\Pi}_{\beta2})\widehat{\rho}. \label{Eq.JointProbsQThy}
\end{align}

In \emph{LHVT} the corresponding observable is 
\begin{equation}
\Omega=\Omega_{A1}\otimes\Omega_{B1}+\Omega_{A2}\otimes\Omega_{B2}\label{Eq.SumProdLHVTObserv}
\end{equation}
and for \emph{Bell local} states, the mean value of $\Omega$ is given
by
\begin{eqnarray}
\left\langle \Omega\right\rangle  & = & \left\langle \Omega_{A1}\otimes\Omega_{B1}\right\rangle 
+\left\langle \Omega_{A2}\otimes\Omega_{B2}\right\rangle, \nonumber \\
 & = & \sum\limits _{\lambda}P(\lambda|c)\,\left\langle \Omega_{A1}(\lambda)\right\rangle \,\left\langle \Omega_{B1}(\lambda)\right\rangle 
\nonumber \\ &&
+\sum\limits _{\lambda}P(\lambda|c)\,\left\langle \Omega_{A2}(\lambda)\right\rangle \,\left\langle \Omega_{B2}(\lambda)\right\rangle, \nonumber \\
 & = & \sum\limits _{\alpha_{1}\beta_{1}}\alpha_{1}\beta_{1}\,P(\alpha_{1},\beta_{1}|\Omega_{A1},\Omega_{B1},c)
\nonumber \\ &&
+\sum\limits _{\alpha_{2}\beta2}\alpha_{2}\beta_{2}\,P(\alpha_{2},\beta_{2}|\Omega_{A2},\Omega_{B2},c),\nonumber \\
\label{Eq.MeanSumProdLHVT}
\end{eqnarray}
where in LHVT
\begin{eqnarray}
P(\alpha_{1},\beta_{1}|\Omega_{A1},\Omega_{B1},c) & = & \sum\limits _{\lambda}P(\lambda|c)
\nonumber \\ &&
\times
P(\alpha_{1}|\Omega_{A1},c,\lambda)P(\beta_{1}|\Omega_{B1},c,\lambda),\nonumber \\
P(\alpha_{2},\beta_{2}|\Omega_{A2},\Omega_{B2},c) & = & \sum\limits _{\lambda}P(\lambda|c)
\nonumber \\ &&
\times
P(\alpha_{2}|\Omega_{A2},c,\lambda)P(\beta_{2}|\Omega_{B2},c,\lambda),\label{Eq.JointProbsLHVT}
\nonumber \\ &&
\end{eqnarray}
We will use these expressions (\ref{Eq.MeanSumProdQThy}) and (\ref{Eq.MeanSumProdLHVT})
to interconvert between quantum theory and LHVT when the latter applies.

To determine these mean values \emph{experimentally}, two sets of
joint measurements for $\widehat{\Omega}_{A1},\widehat{\Omega}_{B1}$
and \emph{then} $\widehat{\Omega}_{A2},\widehat{\Omega}_{B2}$ (or
the classical observables $\Omega_{A1},\Omega_{B1}$ and then $\Omega_{A2},\Omega_{B2}$)
would be required, unless a technique exists for measuring the outcomes
for $\widehat{\Omega}$ (or $\Omega$) directly.

\section{Classical Observables and Quadrature Amplitudes}

\label{Appendix - Classical Observables and Quadratures}

For the square of the spin components $S_{x}^{2}$ and $S_{y}^{2}$
we have 
\begin{eqnarray}
S_{x}^{2} & = & \frac{1}{4}(x_{A}^{2}x_{B}^{2}+p_{A}^{2}p_{B}^{2})+\frac{1}{2}(U_{A}U_{B}-V_{A}V_{B}),\label{Eq.SquareSx}\\
S_{y}^{2} & = & \frac{1}{4}(p_{A}^{2}x_{B}^{2}+x_{A}^{2}p_{B}^{2})-\frac{1}{2}(U_{A}U_{B}+V_{A}V_{B}),\label{Eq,SquareSy}
\end{eqnarray}
and the square of $X_{\theta}(\pm)$ is given by
\begin{eqnarray}
X_{\theta}(\pm)^{2} & = & \frac{1}{2}\left( x_{A}^{2}\cos^{2}\theta+p_{A}^{2}\sin^{2}\theta+2U_{A}\sin\theta\cos\theta\right) \nonumber \\
 &  & +\frac{1}{2}\left( x_{B}^{2}\cos^{2}\theta+p_{B}^{2}\sin^{2}\theta+2U_{B}\sin\theta\cos\theta\right) 
\nonumber \\
 &  & \pm\bigl(
x_{A}x_{B}\cos^{2}\theta+p_{A}p_{B}\sin^{2}\theta
\nonumber \\ && ~~~
+x_{A}p_{B}\sin\theta\cos\theta+p_{A}x_{B}\sin\theta\cos\theta\bigr).
\nonumber \\
\label{Eq.TwoModeQuadObservSquare}
\end{eqnarray}

\section{Werner States}

\label{Appendix - Werner States}

As examples of the three categories of Bell local states we may consider
the states introduced by Werner \cite{Werner89a} as $U\otimes U$
invariant states ($(\widehat{U}\otimes\widehat{U})\,\widehat{\rho}_{W}\,(\widehat{U}^{\dag}\otimes\widehat{U}^{\dag})=\widehat{\rho}_{W}$,
where $\widehat{U}$ is any \emph{unitary} operator) for two $d$
dimensional sub-systems. Depending on the parameter $\eta$ (or $\phi$)
the Werner states, may be separable or entangled. They may also be
Bell local in one of the three categories described above, or they
may be Bell non-local. The density operator for the \emph{Werner states}
is given by 
\begin{eqnarray}
\widehat{\rho}_{W} & = & (d^{3}-d)^{-1}\left[(d-\phi)\,\widehat{1}+(d\,\phi-1)\,\widehat{V}\right]\nonumber \\
 & = & \left(\frac{(d-1+\eta)}{(d-1)}\right)\,\frac{\widehat{1}}{d^{2}}-\left(\frac{\eta}{(d-1)}\right)\,\frac{\widehat{V}}{d},\label{Eq.WernerStates}
\end{eqnarray}
where $\widehat{1}$ is the \emph{unit} operator and $\widehat{V}$
is the \emph{flip} operator defined as 
$\widehat{V}(\left\vert \psi\right\rangle \otimes\left\vert \chi\right\rangle ) = \left\vert \chi\right\rangle \otimes\left\vert \psi\right\rangle $.
The two expressions are interconvertable with $\phi=(1-(d+1)\eta)/d$.
For a positive density operator we have $-1\leq\phi\leq+1$. Werner
has shown that if $\eta<1/(d+1)$ (or $\phi>0$) the state $\widehat{\rho}_{W}$
is separable, but for $\eta>1/(d+1)$ (or $\phi<0$) the state is entangled.
Thus Werner states with $\eta<1/(d+1)$ or $\phi>0$ are separable.
Wiseman et al.\ \cite{Wiseman07a} considered the above categories for
such Werner states and determined the parameter boundaries for the
various categories. These results are shown in Fig.~2 (taken from
Fig.~1a in Ref \cite{Wiseman07a}), where the parameter regimes for
the various categories of quantum states are explained.

\begin{figure}[htb]
  \centering
  \includegraphics[width=\columnwidth]{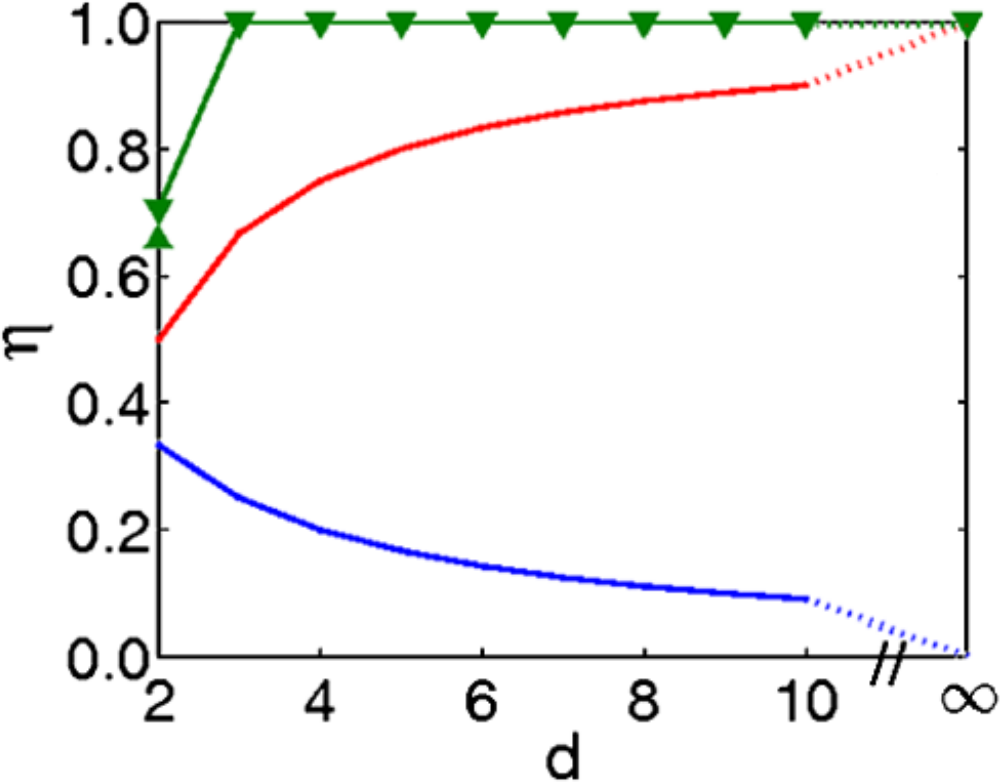}
\caption[]{Parameter $\eta$ (see text) boundaries for Werner States. The blue
line corresponds to $\eta=1/(d+1)$, the red line to $\eta=(1-d^{-1})$
and the green line to $\eta=1$ for $d\geq3$. For $\eta$ below blue
line the states are Category 1 - separable states. These states are
also Bell local, LHS and non-steerable. For $\eta$ between blue line
and red line the states are Category 2. These states are also Bell
local, non-steerable and entangled. For $\eta$ between red line and
green line the states are Category 3 - Bell local, steerable and entangled
(EPR entangled). For $\eta$ above green line the states are Category
4 - Bell non-local, steerable and entangled. This is only possible
for $d=2$. Figure taken from Wiseman et al.\ Ref.\ \protect\cite{Wiseman07a}.}
\end{figure}
\begin{center}
\end{center}

\section{Idea of EPR Steering}

\label{Appendix - EPR Steering}

In this Appendix we consider for reasons of completeness the physical
idea behind EPR steering, as presented in the papers
\cite{Wiseman07a,Jones07a,Cavalcanti09a}.

We can derive expressions within LHV theory for the \emph{conditional
probabilities} defined in (\ref{Eq.LHVTConditionalProb}). These expressions
apply for all three Bell local categories considered here. We will
focus on LHS states, which in terms of our LHVCS may be either in
Category 1 or Category 2. We will initially consider the latter.

In the case of \emph{Category 2} states (which are \emph{LHS states})
we obtain from (\ref{Eq.CategoryTwoStates}) and (\ref{Eq.LHVTConditionalProb})
\begin{equation}
P(\beta|\Omega_{B}||\alpha,\Omega_{A},c)=\frac{\sum_{\lambda}P(\alpha|\Omega_{A},c,\lambda)\,\Tr_{B}((\widehat{\Pi}_{\beta}^{B})\widehat{\rho}^{B}(\lambda))}{\sum_{\lambda}P(\alpha|\Omega_{A},c,\lambda)\,P(\lambda|c)}\label{Eq.CondProbLHSStates}
\end{equation}
using (\ref{Eq.LHVTSingleProb}) and (\ref{Eq.SingleProbLHVQuantumSubSys}).

It is also important to realise that these LHS model states are still
related to an overall quantum state, but one which is \emph{non-separable}
since we cannot derive the density operator (\ref{Eq.QuantSepState})\ for
separable states from Category 2 expression (\ref{Eq.CategoryTwoStates})
for the joint probability. For Category 2 LHS states, $P(\alpha|\Omega_{A},c,\lambda)$
is \emph{not} given by a quantum expression. However, as in
\cite{Jones07a,Cavalcanti09a} we can relate the quantities in the LHS model
(\ref{Eq.CategoryTwoStates}) to a density operator for sub-system
$B$ that is \emph{conditional} on the results for measurements on
sub-system $A$.

From (\ref{Eq.QuantumSingleProb}) the \emph{quantum theory} result
for the probability that measurement of observable $\Omega_{A}$ results
in outcome $\alpha$ is given by
\begin{equation}
P(\alpha|\Omega_{A},\rho)=\Tr((\widehat{\Pi}_{\alpha}^{A}\otimes\widehat{1}^{B})\,\widehat{\rho}),\label{Eq.GeneralQuantumSingleProb}
\end{equation}
where $\widehat{\rho}$ is the density operator for the overall quantum
state (the preparation symbol $c$ is left out for simplicity). In
the Copenhagen interpretation of quantum theory the \emph{normalised}
state that is produced as a \emph{result} of this measurement is the
\emph{conditional state} 
\begin{eqnarray}
\widehat{\rho}_{cond}(\alpha|\Omega_{A},\rho)&=&(\widehat{\Pi}_{\alpha}^{A}\otimes\widehat{1}^{B})\,\widehat{\rho}\,(\widehat{\Pi}_{\alpha}^{A}\otimes\widehat{1}^{B})/P(\alpha|\Omega_{A},\rho).
\nonumber\\ &&
\label{Eq.CondQState}
\end{eqnarray}
This state has a trace of unity, as required. To confirm that $\widehat{\rho}_{cond}(\alpha|\Omega_{A},\rho)$
\emph{does} lead to the correct quantum expression for
the \emph{conditional probability} $P(\beta|\Omega_{B}||\alpha|\Omega_{A},\rho)$
(i.e.\ that measurement of $\Omega_{B}$ in sub-system $B$ will result
in outcome $\beta$ \emph{given} that measurement of $\Omega_{A}$
resulted in outcome $\alpha$ based on the quantum state
$\widehat{\rho}$),
we calculate the probability of that measurement of $\Omega_{B}$
in sub-system $B$ which will result in outcome $\beta$ for the quantum
state $\widehat{\rho}_{cond}(\alpha|\Omega_{A},\rho)$. 
This is given by 
\begin{align}
P(\beta|\Omega_{B},\rho_{cond}) & =  \Tr((\widehat{1}^{A}\otimes\widehat{\Pi}_{\beta}^{B})\,\widehat{\rho}_{cond}(\alpha|\Omega_{A},\rho)),\nonumber \\
 & =  \Tr((\widehat{\Pi}_{\alpha}^{A}\otimes\widehat{\Pi}_{\beta}^{B})
\widehat{\rho}(\widehat{\Pi}_{\alpha}^{A}\otimes\widehat{1}^{B}))
/P(\alpha|\Omega_{A},\rho),\nonumber \\
 & =  \Tr\left((\widehat{\Pi}_{\alpha}^{A}\otimes\widehat{\Pi}_{\beta}^{B})\,\widehat{\rho}\right)
/P(\alpha|\Omega_{A},\rho),\nonumber \\
 & =  P(\alpha,\beta|\Omega_{A},\Omega_{B},\rho)/P(\alpha|\Omega_{A},\rho),\nonumber \\
 & =  P(\beta|\Omega_{B}\,||\alpha|\Omega_{A},\rho),\label{Eq.CondProb}
\end{align}
using the cyclic properties of the trace and $(\widehat{\Pi}_{\alpha}^{A})^{2}=\widehat{\Pi}_{\alpha}^{A}$,
with the last line (see (\ref{Eq.LHVTConditionalProb})) following
from \emph{Bayes' theorem}. This confirms the status of $\widehat{\rho}_{cond}(\alpha|\Omega_{A},\rho)$.

The \emph{physical} concept of \emph{steering} has been discussed
in several papers, including
\cite{Wiseman07a,Jones07a}
and \cite{Cavalcanti09a} and was originally introduced by Schr\"{o}dinger
\cite{Schrodinger35a} following the important EPR paper \cite{Einstein35a}.
The key idea is that when a measurement of $\Omega_{A}$\ is made
on sub-system $A$ resulting in outcome $\alpha$ (the bipartite quantum
state prepared being $\rho$ ) this results in both the overall quantum
state changing to a new conditioned state $\widehat{\rho}_{cond}(\alpha|\Omega_{A},\rho)$\ (given
in Eq.\ (\ref{Eq.CondQState})) and hence the \emph{post-measurement}
state describing sub-system $B$ changing to
\begin{equation}
\widehat{\rho}_{cond}(\alpha|\Omega_{A},\rho)^{B}=\Tr_{A}(\widehat{\rho}_{cond}(\alpha|\Omega_{A},\rho))\label{Eq.DefnCondDensOprSubB}
\end{equation}
from its \emph{pre-measurement} state $\widehat{\rho}^{B}=\Tr_{A}(\widehat{\rho})$
given by the \emph{reduced density operator} (Eq.\ \ref{Eq.ReducedDensOprs}).
This strange quantum effect allows for an experiment carried out on
sub-system $A$ to instantly change (or ``\emph{steer}'')
the quantum state for sub-system $B$ into a new quantum state, even
when the two sub-systems are localised in well-separated spatial regions
and the experimenter on $A$ may have no direct access to sub-system
$B$. For those who accept the Copenhagen interpretation of quantum
theory there is nothing really strange involved. Quantum states merely
specify all that can be known about the physical state (and no distinction
between ``physical state'' and ``quantum
state'' is made), so as the measurement of $\Omega_{A}$
has led to a particular outcome $\alpha$ our knowledge about the
state has changed, and hence the quantum state for both the overall
system and its sub-systems should change accordingly. Using quantum
theory we can obtain an explicit formula for $\widehat{\rho}_{cond}(\alpha|\Omega_{A},\rho)^{B}$
and this is
\begin{align}
\widehat{\rho}_{cond}(\alpha|\Omega_{A},\rho)^{B}=&\sum\limits _{\beta l,\gamma n}\left\vert B\beta l\right\rangle \left\langle B\gamma n\right\vert 
\nonumber \\ 
&\times\sum\limits _{i}\,\rho_{A\alpha i,B\beta l::A\alpha i,B\gamma n} \,,
\label{Eq.CondDensOprSubB}
\end{align}
where the original density operator $\rho$ is expressed in terms
of orthonormal basis states $\left\vert A\alpha i\right\rangle \otimes\left\vert B\beta n\right\rangle $
that are eigenstates for $\widehat{\Omega}_{A}$ and $\widehat{\Omega}_{B}$,
with $i=1,2,..,d_{\alpha}$ and $n=1,2,..,d_{\beta}$ allowing for
degeneracy.

We can also show that the sum of the conditional density operators
$\widehat{\rho}_{cond}(\alpha|\Omega_{A},\rho)^{B}$ each weighted
by the probability $P(\alpha|\Omega_{A},\rho)$ for the measurement
outcome $\alpha$ for $\Omega_{A}$ gives the reduced density operator
$\widehat{\rho}^{B}$ associated with the original state $\rho$.
This result is not surprising, since carrying out the measurement
of any choice of $\Omega_{A}$ and then discarding the results would
be described by reduced density operator:
\begin{equation}
\sum\limits _{\alpha}P(\alpha|\Omega_{A},\rho)\,\widehat{\rho}_{cond}(\alpha|\Omega_{A},\rho)^{B}=\widehat{\rho}^{B}=\Tr_{A}\widehat{\rho}.\label{Eq.WeightedSumCondDensOprB}
\end{equation}
The proofs of (\ref{Eq.CondDensOprSubB}) and (\ref{Eq.WeightedSumCondDensOprB})
are straightforward.

Thus, we have seen how according to quantum theory the quantum state
describing sub-system $B$ changes as a result of measuring $\Omega_{A}$
on sub-system $A$\ and obtaining outcome $\alpha$. Furthermore,
we have obtained quantum theory expressions (\ref{Eq.CondProb}) for
the conditional probability $P(\beta|\Omega_{B},\,\rho_{cond})$ for
measurement of $\Omega_{B}$ on sub-system $B$\ and obtaining outcome
$\beta$ when measurement of $\Omega_{A}$ on sub-system $A$ resulted
in outcome $\alpha$ and (\ref{Eq.CondDensOprSubB}) for the quantum
state describing sub-system $B$. The question then is: Although quantum
theory gives the correct results for the conditional probability $P(\beta|\Omega_{B},\,\rho_{cond})$,
can the same results \emph{also} be explained in a local hidden variable
theory?

Following the operational definition for steering in Refs.\ \cite{Wiseman07a,Jones07a} and \cite{Cavalcanti09a}, the quantum state $\rho$
is only considered to be \emph{EPR steerable} when the conditional
probability $P(\beta|\Omega_{B}||\alpha,\Omega_{A},c)$ can \emph{not}
be explained via a local hidden variable theory. For the LHS\ cases
of Category 1 and Category 2 states we will see that a LHV theory
explanation applies. We consider what expression for a density operator
for sub-system $B$ would give the LHS result for the conditional
probability $P(\beta|\Omega_{B}||\alpha,\Omega_{A},c)$ for measurement
of $\Omega_{B\text{ }}$to have outcome $\beta$, given that measurement
of $\Omega_{A\text{ }}$has outcome $\alpha$ and the preparation
process is $c$. In the case of Category 2 states we use Eqs.\ (\ref{Eq.CategoryTwoStates})
and (\ref{Eq.SingleProbLHVQuantumSubSys}) in conjunction with (\ref{Eq.LHVTConditionalProb})
and (\ref{Eq.LHVTSingleProb}) to find 
\begin{align}
P(\beta|\Omega_{B}&||\alpha,\Omega_{A},c) = 
\nonumber \\ 
&\frac{\sum\limits _{\lambda}P(\alpha|\Omega_{A},c,\lambda)\,\Tr_{B}((\widehat{\Pi}_{\beta}^{B})\widehat{\rho}^{B}(\lambda))\,P(\lambda|c)\,}{\sum_{\lambda}P(\alpha|\Omega_{A},c,\lambda)\,P(\lambda|c)\,}.
\nonumber \\ &
\end{align}
We then \emph{define} a \emph{new} normalised quantum state 
for sub-system $B$,
$\widehat{\rho}_{cond}^{B}(\alpha|\Omega_{A},c)$,
by the expression
\begin{eqnarray}
\widehat{\rho}_{cond}^{B}(\alpha|\Omega_{A},c) & = & \frac{\sum\limits _{\lambda}P(\alpha|\Omega_{A},c,\lambda)\,\widehat{\rho}^{B}(\lambda)\,P(\lambda|c)\,}{\Tr_{B}\left(\sum\limits _{\lambda}P(\alpha|\Omega_{A},c,\lambda)\,\widehat{\rho}^{B}(\lambda)\,P(\lambda|c)\,\right)},\nonumber \\
 & = & \frac{\sum\limits _{\lambda}P(\alpha|\Omega_{A},c,\lambda)\,\widehat{\rho}^{B}(\lambda)\,P(\lambda|c)\,}{\left(\sum\limits _{\lambda}P(\alpha|\Omega_{A},c,\lambda)\,P(\lambda|c)\,\right)}.\label{Eq.CondQuantStateLHSModel}
\end{eqnarray}
It is to be noted that this state for sub-system $B$ involves local
HVT and not quantum expressions for the measurement probabilities
$P(\alpha|\Omega_{A},c,\lambda)$ for sub-system $A$. We then see
from (\ref{Eq.QuantumSingleProb}) that for \emph{this state} the
probability for measurement of $\Omega_{B\text{ }}$to have outcome
$\beta$ is given by 
\begin{align}
\Tr_{B}(\widehat{\Pi}_{\beta}^{B}\,&\widehat{\rho}_{cond}^{B}(\alpha|\Omega_{A},c)) 
\nonumber\\
& =  \frac{\sum\limits _{\lambda}P(\alpha|\Omega_{A},c,\lambda)\,\Tr_{B}((\widehat{\Pi}_{\beta}^{B})\widehat{\rho}^{B}(\lambda))\,P(\lambda|c)\,}{\sum\limits _{\lambda}P(\alpha|\Omega_{A},c,\lambda)\,P(\lambda|c)},\nonumber \\
 & =  P(\beta|\Omega_{B}||\alpha,\Omega_{A},c),
\end{align}
which is the same as (\ref{Eq.CondProbLHSStates}) obtained for the
Category 2 states (which are LHS states). Thus the sub-system $B$
quantum state (\ref{Eq.CondQuantStateLHSModel}) has been constructed
purely from the Category 2 LHS\ model probabilities $P(\alpha|\Omega_{A},c,\lambda)\,$and
$P(\lambda|c)$, together with the LHS model quantum state $\widehat{\rho}^{B}(\lambda)$
- which is a possible quantum state for sub-system $B$ based on hidden
variables $\lambda$. The sub-system $B$ quantum state $\widehat{\rho}_{cond}^{B}(\alpha|\Omega_{A},c)$
in (\ref{Eq.CondQuantStateLHSModel})\ determines the correct probability
for measurement of $\Omega_{B\text{ }}$to have outcome $\beta$.
The same analysis would apply to the LHS states in Category 1, the
only difference being that $P(\alpha|\Omega_{A},c,\lambda)$ would
be replaced by $P_{Q}(\alpha|\Omega_{A},c,\lambda)$ in terms of our
notation. So in both of these cases there could be a hidden state
$\widehat{\rho}^{B}(\lambda)$ associated with hidden variables that
could explain (along with suitable choices for $P(\alpha|\Omega_{A},c,\lambda)\,$and
$P(\lambda|c)$) the measurements on sub-system $B$. The treatment
however does not apply to the quantum states in Category 3, where
the LHV model in Eq.\ (\ref{Eq.CategoryThreeStates}) does \emph{not}
include a quantum state $\widehat{\rho}^{B}(\lambda)$ for sub-system
$B$. Hence, the conditional probability $P(\beta|\Omega_{B}||\alpha,\Omega_{A},c)$
\emph{can} be explained via the \emph{LHS model} for both Category
1 and Category 2 states, showing that the Category 1 and
Category 2 quantum states are \emph{non-steerable}. However, the Category
3 states are \emph{EPR steerable}.

\section{Spin Variances: EPR Steering Test}

\label{Appendix - EPR Sterering Othe Approach}

The EPR steering test in (\ref{Eq.EPRSteeringTest}) can be obtained
from the results in Sections \ref{SubSection - Spin Squeezing Test for EPR Steering}
and \ref{SubSection - Planar Spin Variance Tests for EPR Steering} by using
(\ref{Eq.ResultVarSxSyLHS}), (\ref{Eq.ModeNumbersObserv}) and (\ref{Eq.SzLHSModel}).
We find using LHVT\ that for Category 2 states
\begin{align}
\left\langle \Delta S_{x}^{2}\right\rangle +\left\langle \Delta S_{y}^{2}\right\rangle &-\frac{1}{4}\left\langle N\right\rangle +\frac{1}{2}\left\langle S_{z}\right\rangle 
\nonumber\\&
\geq\left\langle N_{A}\otimes N_{B}\right\rangle +\frac{1}{2}\left\langle 1_{A}\otimes N_{B}\right\rangle,
\nonumber\\&
\geq0.
\end{align}
Details are:
\begin{align}
\left\langle \Delta S_{x}^{2}\right\rangle &+\left\langle \Delta S_{y}^{2}\right\rangle -\frac{1}{4}\left\langle N\right\rangle +\frac{1}{2}\left\langle S_{z}\right\rangle  
\nonumber\\
 \geq&  \left\langle N_{A}\otimes N_{B}\right\rangle +\frac{1}{2}\left\langle 1_{A}\otimes N_{B}\right\rangle +\frac{1}{2}\left\langle N_{A}\otimes1_{B}\right\rangle \nonumber\\
  & -\frac{1}{4}\left\langle 1_{A}\otimes N_{B}\right\rangle -\frac{1}{4}\left\langle N_{A}\otimes1_{B}\right\rangle \nonumber\\
  & +\frac{1}{4}\left\langle 1_{A}\otimes N_{B}\right\rangle -\frac{1}{4}\left\langle N_{A}\otimes1_{B}\right\rangle, \nonumber\\
  \geq &\left\langle N_{A}\otimes N_{B}\right\rangle +\frac{1}{2}\left\langle 1_{A}\otimes N_{B}\right\rangle, \nonumber\\
  \geq &0.
\end{align}

As LHVT is required to predict the same result as quantum theory we
have 
\begin{align}
\left\langle \Delta\widehat{S}_{x}^{2}\right\rangle +\left\langle \Delta\widehat{S}_{y}^{2}\right\rangle &-\frac{1}{4}\left\langle \widehat{N}\right\rangle +\frac{1}{2}\left\langle \widehat{S}_{z}\right\rangle   
\nonumber\\
\geq & \left\langle \widehat{N}_{A}\otimes\widehat{N}_{B}\right\rangle +\frac{1}{2}\left\langle \widehat{1}_{A}\otimes\widehat{N}_{B}\right\rangle, \nonumber \\
  \geq & 0,\label{Eq.Result2}
\end{align}
since both $\langle \widehat{N}_{A}\otimes\widehat{N}_{B}\rangle $
and $\langle \widehat{1}_{A}\otimes\widehat{N}_{B}\rangle $
are positive quantities. In this form it shows that if $\langle \Delta\widehat{S}_{x}^{2}\rangle + \langle \Delta\widehat{S}_{y}^{2}\rangle - \frac{1}{4}\langle \widehat{N}\rangle + \frac{1}{2}\langle \widehat{S}_{z}\rangle <0$
then the state cannot be Category 2. This result is also obtained
by Appendices \ref{Appendix- Correlation Tests for EPR Steering} and \ref{Appendix - Correlation Ineq and Spin Operators}. 

\section{Variances of Two Mode Quadratures - Category 2 States}

\label{Appendix - Variances of Two Mode Quadratures - Cat 2}

Using the LHVT expressions (\ref{Eq.TwoModeQuadObserv},\ref{Eq.TwoModeQuadObservSquare})
for $X_{\theta}(\pm)$ and $X_{\theta}(\pm)^{2}$ together with the
results (\ref{Eq.MeanSubSystB}) and (\ref{Eq.MeanUBLHSModel}) for
$\left\langle x_{B}\right\rangle $, $\left\langle p_{B}\right\rangle $
and $\left\langle U_{B}\right\rangle $, together with $U_{A}=\frac{1}{2}(x_{A}p_{A}+p_{A}x_{A})$,
we find for Category 2 states the mean values of the two mode quadratures
and their square are given by 
\begin{align}
\left\langle X_{\theta}(\pm)\right\rangle   = &
 \frac{1}{\sqrt{2}}\left(\left\langle x_{A}\right\rangle \cos\theta+\left\langle p_{A}\right\rangle \sin\theta\right),\label{Eq.MeanTwoModeQuadQThy}\\
\left\langle X_{\theta}(\pm)\right\rangle ^{2}  = & \frac{1}{2}\big( \left\langle x_{A}\right\rangle ^{2}\cos^{2}\theta+\left\langle p_{A}\right\rangle ^{2}\sin^{2}\theta
\nonumber\\ &
~~~+2\left\langle x_{A}\right\rangle \left\langle p_{A}\right\rangle \sin\theta\cos\theta\big),\nonumber \\
\left\langle X_{\theta}(\pm)^{2}\right\rangle   = & \frac{1}{2}\big(\left\langle x_{A}^{2}\right\rangle \cos^{2}\theta+\left\langle (x_{A}p_{A}+p_{A}x_{A})\right\rangle \sin\theta\cos\theta
\nonumber\\
& ~~~+\left\langle p_{A}^{2}\right\rangle \sin^{2}\theta\big)\nonumber \\
   & +\frac{1}{2}\left(\left\langle N_{B}\right\rangle +\frac{1}{2}\right).
\label{Eq.MeanSquareTwoModeQuadQThy}
\end{align}

The variance for Category 2 states is then given by the LHVT expression
\begin{align}
  \left\langle \Delta X_{\theta}(\pm)^{2}\right\rangle 
 & =  \frac{1}{2}\big\langle \left(\Delta x_{A}\cos\theta+\Delta p_{A}\sin\theta\right)
\nonumber\\& \qquad
\times
\left(\Delta x_{A}\cos\theta+\Delta p_{A}\sin\theta\right)\big\rangle 
\nonumber\\ & \quad
+\frac{1}{2}\left(\left\langle N_{B}\right\rangle +\frac{1}{2}\right),\nonumber \\
    \left\langle \Delta P_{\theta}(\pm)^{2}\right\rangle 
 &=  \frac{1}{2}\big\langle \left(-\Delta x_{A}\sin\theta+\Delta p_{A}\cos\theta\right)
\nonumber\\ & \qquad \times
\left(-\Delta x_{A}\sin\theta+\Delta p_{A}\cos\theta\right)\big\rangle 
\nonumber\\ & \quad
+\frac{1}{2}\left(\left\langle N_{B}\right\rangle +\frac{1}{2}\right),
\label{Eq.VarianceTwoModeQuadQThy}
\end{align}
where $\Delta x_{A}=x_{A}-\left\langle x_{A}\right\rangle $ and $\Delta p_{A}=p_{A}-\left\langle p_{A}\right\rangle $.
The expression for $\left\langle \Delta P_{\theta}(\pm)^{2}\right\rangle $
is obtained using $P_{\theta}(\pm)=X_{\theta+\pi/2}(\pm)$.

As LHVT underlies quantum theory then we also have for the quantum
theory treatment of Category 2 states
\begin{align}
 \left\langle \Delta\widehat{X}_{\theta}(\pm)^{2}\right\rangle 
 & =  \frac{1}{2}\big\langle \left(\Delta\widehat{x}_{A}\cos\theta+\Delta\widehat{p}_{A}\sin\theta\right)
\nonumber\\ & \qquad \times 
\left(\Delta\widehat{x}_{A}\cos\theta+\Delta\widehat{p}_{A}\sin\theta\right)\big\rangle 
\nonumber\\ & \quad
+\frac{1}{2}\left(\left\langle \widehat{1}_{A}\otimes\widehat{N}_{B}\right\rangle +\frac{1}{2}\right),
\end{align}
where now $\Delta\widehat{x}_{A}=\widehat{x}_{A}-\left\langle \widehat{x}_{A}\right\rangle $,
$\Delta\widehat{p}_{A}=\widehat{p}_{A}-\left\langle \widehat{p}_{A}\right\rangle $.
However, we can make use of the SSR to simplify these expressions further.
As shown in SubSection \ref{SubSection - Quantum Theory} the reduced density
operator for sub-system $A$ satisfies the local particle number SSR. This is
the case even though the reduced density operator depends on the full
density matrix for both sub-systems, unlike that for a local hidden state.
Consequently 
\begin{align}
\left\langle \widehat{x}_{A}\right\rangle &=\Tr_{A}(\widehat{x}_{A}\widehat{\rho}^{A})=0,
& \left\langle \widehat{p}_{A}\right\rangle &=\Tr_{A}(\widehat{p}_{A}\widehat{\rho}^{A})=0,
\nonumber\\ &&&
\label{Eq.MeanQuadSysA}
\end{align}
using the same arguments as for $\left\langle x_{B}(\lambda)\right\rangle _{Q}$
and $\left\langle p_{B}(\lambda)\right\rangle _{Q}$ in Eq.\ (\ref{Eq.MeanSubSystB}).
Furthermore, the same steps as for $\left\langle x_{B}^{2}(\lambda)\right\rangle _{Q}$,
$\left\langle p_{B}^{2}(\lambda)\right\rangle _{Q}$ and $\left\langle U_{B}(\lambda)\right\rangle _{Q}$
lead to 
\begin{align}
\left\langle \widehat{x}_{A}^{2}\right\rangle  &=  \left\langle \widehat{N}_{A}\right\rangle +\frac{1}{2} ,
& \left\langle \widehat{p}_{A}^{2}\right\rangle &=\left\langle \widehat{N}_{A}\right\rangle +\frac{1}{2},\nonumber \\
\left\langle \widehat{U}_{A}\right\rangle  & =  0, &&\label{Eq.MeanSquaresQuadSysA}
\end{align}
(see SubSubSection \ref{SuSubSection- Evaluation of Expressions Cat 2}).
Using these results we then find that
\begin{align}
\left\langle \Delta\widehat{X}_{\theta}(\pm)^{2}\right\rangle  & =  \frac{1}{2}\left(\left\langle \widehat{N}_{A}\otimes\widehat{1}_{B}\right\rangle +\frac{1}{2}\right)
\nonumber\\ & \quad
+\frac{1}{2}\left(\left\langle \widehat{1}_{A}\otimes\widehat{N}_{B}\right\rangle +\frac{1}{2}\right),\nonumber \\
 & =  \frac{1}{2}\left\langle \widehat{N}\right\rangle +\frac{1}{2},\nonumber \\
\left\langle \Delta\widehat{P}_{\theta}(\pm)^{2}\right\rangle  & =  \frac{1}{2}\left\langle \widehat{N}\right\rangle +\frac{1}{2}.
\label{Eq.TwoModeQuadVarCat2States}
\end{align}
(The calculation for $\langle \Delta\widehat{P}_{\theta}(\pm)^{2}\rangle $
is trivial, as $\widehat{P}_{\theta}(\pm)=\widehat{X}_{\theta+\pi/2}(\pm)$).
Exactly the same results apply for Category 1 (separable) states (see
Appendix L in Ref.\ \cite{Dalton16b}). 

\section{Correlation Tests for EPR Steering}

\label{Appendix- Correlation Tests for EPR Steering}

The paper by Cavalcanti et al.\ \cite{Cavalcanti11a} derives certain
inequalities for $|\langle \widehat{a}^{\dag}\widehat{b}\rangle |^{2}$
for Category 1 and Category 2 states which lead to \emph{strong correlation}
tests for EPR steering. We will show here that these inequalities
lead to more useful tests in terms of spin operators for quantum entanglement
and EPR steering. These inequalities are set out here in Eqs.\ (\ref{Eq.IneqCat1})
and (\ref{Eq.InequalCat2}) for Category 1 and Category 2 states respectively.
The inequality in Eq.\ (\ref{Eq.IneqCat1}) has also been previously
obtained for separable states by Hillery and Zubairy \cite{Hillery06a}.
They two inequalities correspond to Eqs.\ (15) and (14) in Ref.\ \cite{Cavalcanti11a}
where there are $N=2$ sub-systems (``sites''),
with Eq.\ (15) applying when both sub-systems are associated with a
LHS ($T=2$ - two ``trusted sites'') and
Eq.\ (14) when only one sub-system has a LHS ($T=1$ - one ``trusted
site''). The inequalities obtained by Cavalcanti et al
\cite{Cavalcanti11a} were based on their general expression in Eq.
(4) for the LHV theory joint measurement probability, for which Eqs.
(\ref{Eq.CategoryOneStates}) and (\ref{Eq.CategoryTwoStates}) for
Category 1 and Category 2 states are special cases. Hence these inequalities
would apply for the present paper. For completeness however, rather
than just quoting the inequalities in Ref.\ \cite{Cavalcanti11a} we
will also derive them here using the approach set out in the present
paper. A further inequality for $|\langle \widehat{a}^{\dag}\widehat{b}\rangle |^{2}$
will also be derived that would apply to Category 3 states.

For Category 1 states the result gives a strong correlation test and
the Hillery-Zubairy \cite{Hillery06a} test for quantum entanglement,
whilst for Category 2 states the result gives a strong correlation
test plus a generalised Hillery-Zubairy test for EPR steering, originally
set out in He et al.\ \cite{He12a} for the case where $\langle \widehat{S}_{z}\rangle =0$.
The new test allows for $\langle \widehat{S}_{z}\rangle \neq0$.
For Category 3 states no useful test for Bell non-locality occurs.

\subsection{General Correlation Inequality for $|\langle \widehat{a}^{\dag}\widehat{b}\rangle |^{2}$:
Bell Local States}

Using Eqs.\ (\ref{Eq.PositionMtmOprs},\ref{Eq.SpinCompts}) to
introduce quadrature operators and spin operators, the quantity $\widehat{a}^{\dag}\widehat{b}$
can be written as 
\begin{eqnarray}
\widehat{a}^{\dag}\widehat{b} & = & \frac{1}{2}(\widehat{x}_{A}-i\widehat{p}_{A})(\widehat{x}_{B}+i\widehat{p}_{B})\nonumber \\
 & = & \widehat{S}_{x}-i\widehat{S}_{y}\label{Eq.ADaggB}
\end{eqnarray}
so that the LHVT quantity $\left\langle a^{\dag}b\right\rangle $
becomes 
\begin{equation}
\left\langle a^{\dag}b\right\rangle =\frac{1}{2}\left(\left\langle x_{A}x_{B}\right\rangle+\left\langle p_{A}p_{B}\right\rangle +i\left(\left\langle x_{A}p_{B}\right\rangle -\left\langle p_{A}x_{B}\right\rangle \right)\right).\label{Eq.HVTExpressionMeanAdaggB}
\end{equation}
Then introducing the LHVT expression
\begin{multline*}
\left\langle a^{\dag} b\right\rangle =\frac{1}{2}\sum_{\lambda}P(\lambda|c)\left(\left\langle x_{A}(\lambda)\right\rangle -i\left\langle p_{A}(\lambda)\right\rangle \right)
\\ \times
\left(\left\langle x_{B}(\lambda)\right\rangle +i\left\langle p_{B}(\lambda)\right\rangle \right),
\end{multline*}
and
\begin{multline*}
\left|\left\langle a^{\dag} b\right\rangle \right|\leq\frac{1}{2}\sum_{\lambda}P(\lambda|c)|\left(\left\langle x_{A}(\lambda)\right\rangle -i\left\langle p_{A}(\lambda)\right\rangle \right)|
\\ \times
|\left(\left\langle x_{B}(\lambda)\right\rangle +i\left\langle p_{B}(\lambda)\right\rangle \right)|
\end{multline*}
with 
$|\left(\left\langle x_{A}(\lambda)\right\rangle -i\left\langle p_{A}(\lambda)\right\rangle \right)|
=\sqrt{\left\langle x_{A}(\lambda)\right\rangle ^{2}+\left\langle p_{A}(\lambda)\right\rangle ^{2}}$
etc., we then find that 
\begin{multline}
|\left\langle a^{\dag}b\right\rangle |^{2}\;\leq\frac{1}{4}
\Biggl(
\sum_{\lambda}P(\lambda|c)\sqrt{\left\langle x_{A}(\lambda)\right\rangle ^{2}+\left\langle p_{A}(\lambda)\right\rangle ^{2}}
\\  \times
\sqrt{\left\langle x_{B}(\lambda)\right\rangle ^{2}+\left\langle p_{B}(\lambda)\right\rangle ^{2}}\Biggr)^{2}.
\end{multline}
Using the inequality (\ref{Eq.CauchyInequality}) with 
\begin{multline*}
C(\lambda)=\left(\left\langle x_{A}(\lambda)\right\rangle ^{2}+\left\langle p_{A}(\lambda)\right\rangle ^{2}\right)
\\ \times
\left(\left\langle x_{B}(\lambda)\right\rangle ^{2}+\left\langle p_{B}(\lambda)\right\rangle ^{2}\right)\geq0,
\end{multline*}
we then have the key inequality
\begin{multline}
\left|\left\langle a^{\dag}b\right\rangle \right|^{2}\;\leq\frac{1}{4}\sum_{\lambda}P(\lambda|c)\,\left(\left\langle x_{A}(\lambda)\right\rangle ^{2}+\left\langle p_{A}(\lambda)\right\rangle ^{2}\right)
\\ \times
\left(\left\langle x_{B}(\lambda)\right\rangle ^{2}+\left\langle p_{B}(\lambda)\right\rangle ^{2}\right)\label{Eq.KeyInequality}
\end{multline}
that would follow from the approach in Ref.\ \cite{Cavalcanti11a}.
Again, as LHVT underlies quantum theory we can use 
(\ref{Eq.ModeNumbersObserv}),
(\ref{Eq.MeanNA}),
(\ref{Eq.MeanValueJointMeastsQThyLHVT}), 
and (\ref{Eq.MeanValueSingleMeastQThyLHVT}) to write this inequality
for \emph{all} Bell local states in terms of quantum operators as
\begin{align}
  \left|\left\langle \widehat{a}^{\dag}\widehat{b}\right\rangle \right|^{2}
   & \leq  \left\langle (\widehat{N}_{A}+\widehat{V}_{A})\otimes(\widehat{N}_{B}+\widehat{V}_{B})\right\rangle, \nonumber \\
 & =  \left\langle \widehat{N}_{A}\otimes \widehat{N}_{B}\right\rangle +\frac{1}{2}\left\langle \widehat{1}_{A}\otimes \widehat{N}_{B}\right\rangle 
\nonumber\\ & \quad
+\frac{1}{2}\left\langle \widehat{N}_{A}\otimes\widehat{1}_{B}\right\rangle +\frac{1}{4}.\label{Eq.LHVTGenInequality}
\end{align}

\subsection{Stronger Correlation Inequalities for Bell Local States}

\label{SubSubSection - Sttronger Correlation Inequalities}

Stronger inequalities can now be derived for the quantities $\langle x_{A}(\lambda)\rangle ^{2}+\langle p_{A}(\lambda)\rangle ^{2}$
and $\langle x_{B}(\lambda)\rangle ^{2}+\langle p_{B}(\lambda)\rangle ^{2}$
in the cases of Categories 1, 2 and 3 states. This leads to some outcomes
different to (\ref{Eq.LHVTGenInequality}).

Even if the sub-system $C$ does \emph{not} involve a local hidden
state $\widehat{\rho}_{\lambda}^{C}$ then we can always use the inequality
(\ref{Eq.ClassMeanSquareResult}) to give $\langle x_{C}(\lambda)\rangle ^{2}\leq\langle x_{C}^{2}(\lambda)\rangle $
and $\langle p_{C}(\lambda)\rangle ^{2}\leq\langle p_{C}^{2}(\lambda)\rangle $.
This is equivalent to the variances of $x_{C}$ and $p_{C}$ being
non-negative. Thus
\begin{equation}
\left\langle x_{C}\left(\lambda\right)\right\rangle ^{2}+\left\langle p_{C}\left(\lambda\right)\right\rangle ^{2}\leq\left\langle x_{C}^{2}(\lambda)\right\rangle +\left\langle p_{C}^{2}(\lambda)\right\rangle. \label{Eq.IneqWithoutLHS}
\end{equation}

On the other hand, if the sub-system $C$ \emph{does} involve a
local hidden state $\widehat{\rho}_{\lambda}^{C}$ then we can obtain
a \emph{stronger inequality} via quantum theory. For any real $\eta$
the quantity $\left\langle \left(\Delta\widehat{x}_{C}-i\eta\Delta\widehat{p}_{C}\right)\left(\Delta\widehat{x}_{C}+i\eta\Delta\widehat{p}_{C}\right)\right\rangle _{\lambda}=\Tr[\left(\Delta\widehat{x}_{C}-i\eta\Delta\widehat{p}_{C}\right)\left(\Delta\widehat{x}_{C}+i\eta\Delta\widehat{p}_{C}\right)\widehat{\rho}_{\lambda}^{C}]\geq0$,
where $\Delta\widehat{x}_{C}=\widehat{x}_{C}-\left\langle \widehat{x}_{C}\right\rangle _{\lambda}$,
$\Delta\widehat{p}_{C}=\widehat{p}_{C}-\left\langle \widehat{p}_{C}\right\rangle _{\lambda}$.
Thus for all $\eta$ we have $\left\langle \Delta\widehat{x}_{C}^{2}\right\rangle _{\lambda}-\eta+\eta^{2}\left\langle \Delta\widehat{p}_{C}^{2}\right\rangle _{\lambda}\geq0$
using $[\widehat{x}_{C},\widehat{p}_{C}]=i$. Putting $\eta=1$ gives
the inequality $\left\langle \Delta\widehat{x}_{C}^{2}\right\rangle _{\lambda}+\left\langle \Delta\widehat{p}_{C}^{2}\right\rangle _{\lambda}-1\geq0$,
which can be written as $\left\langle \widehat{x}_{C}\right\rangle _{\lambda}^{2}+\left\langle \widehat{p}_{C}\right\rangle _{\lambda}^{2}\leq\left\langle \widehat{x}_{C}^{2}\right\rangle _{\lambda}+\left\langle \widehat{p}_{C}^{2}\right\rangle _{\lambda}-1$.
In terms of LHVT notation this inequality is
\begin{equation}
\left\langle x_{C}\left(\lambda\right)\right\rangle ^{2}+\left\langle p_{C}\left(\lambda\right)\right\rangle ^{2}\leq\left\langle x_{C}^{2}(\lambda)\right\rangle +\left\langle p_{C}^{2}(\lambda)\right\rangle -1.\label{Eq.IneqForLHS}
\end{equation}

For Category 1 states both sub-systems involve a local hidden state,
so the key inequality (\ref{Eq.KeyInequality}) gives
\begin{multline}
|\left\langle a^{\dag}b\right\rangle |^{2}\;\leq\frac{1}{4}\sum_{\lambda}P(\lambda|c)\,\left(\left\langle x_{A}^{2}(\lambda)\right\rangle +\left\langle p_{A}^{2}(\lambda)\right\rangle -1\right)
\\ \times 
\left(\left\langle x_{B}^{2}(\lambda)\right\rangle +\left\langle p_{B}^{2}(\lambda)\right\rangle -1\right).\label{Eq.KeyInequalityCat1}
\end{multline}
Using (\ref{Eq.MeanValueJointMeastsQThyLHVT}), (\ref{Eq.MeanValueSingleMeastQThyLHVT}),
(\ref{Eq.ModeNumbOprs}) and (\ref{Eq.AuxOprsV}) we can then convert
these inequalities to quantum expressions involving number operators,
$\widehat{N}_{C}=\widehat{c}^{\dag}\widehat{c}$ (where $C=A,B$):
\begin{align}
\left|\left\langle \widehat{a}^{\dag}\widehat{b}\right\rangle \right|^{2}
&\leq   \left\langle \left(\widehat{N}_{A}+\widehat{V}_{A}-\widehat{1}_{A}/2\right)\otimes\left(\widehat{N}_{B}+\widehat{V}_{B}-\widehat{1}_{B}/2\right)\right\rangle, \nonumber \\
 & =  \left\langle \widehat{N}_{A}\otimes\widehat{N}_{B}\right\rangle. \label{Eq.IneqCat1}
\end{align}

For Category 2 states with sub-system $B$ involving a local hidden
state $\widehat{\rho}_{\lambda}^{B}$, the key inequality (\ref{Eq.KeyInequality})
gives
\begin{multline}
\left|\left\langle a^{\dag}b\right\rangle \right|^{2}\;\leq\frac{1}{4}\sum_{\lambda}P(\lambda|c)\,\left(\left\langle x_{A}^{2}(\lambda)\right\rangle +\left\langle p_{A}^{2}(\lambda)\right\rangle \right)
\\  \times
\left(\left\langle x_{B}^{2}(\lambda)\right\rangle +\left\langle p_{B}^{2}(\lambda)\right\rangle -1\right).\label{Eq.KeyInequalityCat2}
\end{multline}
Similarly to the Category 1 case we then find that for Category 2
states (with $B$ involving the local hidden state)
\begin{eqnarray}
\left|\left\langle \widehat{a}^{\dag}\widehat{b}\right\rangle \right|^{2}\; & \leq & \left\langle \left(\widehat{N}_{A}+\widehat{V}_{A}\right)\otimes\left(\widehat{N}_{B}+\widehat{V}_{B}-\frac{1}{2}\widehat{1}_{B}\right)\right\rangle \nonumber \\
 & = & \left\langle \left(\widehat{N}_{A}+\frac{1}{2}\widehat{1}_{A}\right)\otimes\widehat{N}_{B}\right\rangle. \label{Eq.InequalCat2}
\end{eqnarray}

For Category 3 states with neither sub-system involving a local hidden
state, the key inequality (\ref{Eq.KeyInequality}) gives
\begin{multline}
\left|\left\langle a^{\dag}b\right\rangle \right|^{2}\;\leq\frac{1}{4}\sum_{\lambda}P(\lambda|c)\,\left(\left\langle x_{A}^{2}(\lambda)\right\rangle +\left\langle p_{A}^{2}(\lambda)\right\rangle \right)
\\ \times
\left(\left\langle x_{B}^{2}(\lambda)\right\rangle +\left\langle p_{B}^{2}(\lambda)\right\rangle \right).\label{Eq.KeyInequalityCat3}
\end{multline}
In the case of the Category 3 states we then have
\begin{eqnarray}
\left|\left\langle \widehat{a}^{\dag}\widehat{b}\right\rangle \right|^{2}\; & \leq & \left\langle \left(\widehat{N}_{A}+\widehat{V}_{A}\right)\otimes\left(\widehat{N}_{B}+\widehat{V}_{B}\right)\right\rangle, \nonumber \\
 & = & \left\langle \left(\widehat{N}_{A}+\frac{1}{2}\widehat{1}_{A}\right)\otimes\left(\widehat{N}_{B}+\frac{1}{2}\widehat{1}_{B}\right)\right\rangle,
\nonumber \\ &&\label{Eq.IneqalCat3}
\end{eqnarray}
where we note that $\widehat{N}_{A}+\frac{1}{2}\widehat{1}_{A}=\widehat{a}^{\dag}\widehat{a}+\frac{1}{2}=(\widehat{a}\widehat{a}^{\dag}+\widehat{a}^{\dag}\widehat{a})/2$.
This result is the same as the general result (\ref{Eq.LHVTGenInequality})
found for all Bell local states. Note also that this derivation of
Eqs.\ (\ref{Eq.IneqCat1},\ref{Eq.InequalCat2}) and (\ref{Eq.IneqalCat3})
did not make use of the SSR. Only the presence or absence of a local
hidden state was invoked, and whether the LHS satisfied the SSR was
not used.

As will be seen in the next Section, all these inequalities (\ref{Eq.IneqCat1}),
(\ref{Eq.InequalCat2}) and (\ref{Eq.IneqalCat3}) can be expressed
in terms of spin operator variances.

\subsection{Correlations as Spin Operator Inequalities: Bell Local States}

\label{SubSubSection - Spin Opr Ineqall Cats 1,,2, 3}

The inequalities (\ref{Eq.IneqCat1},\ref{Eq.InequalCat2}) and
(\ref{Eq.IneqalCat3}) derived above can be put into a more useful
form involving \emph{spin operators} - whose mean values and variances
can be measured. From (\ref{Eq.ADaggB}) we have (see also Ref.\ \cite{Dalton16b})
\begin{align}
\left|\left\langle \widehat{a}^{\dag}\widehat{b}\right\rangle \right|^{2} & =  \left\langle \widehat{S}_{x}\right\rangle ^{2}+\left\langle \widehat{S}_{y}\right\rangle ^{2},&&\nonumber \\
\widehat{N}_{A} & =  \frac{1}{2}\widehat{N}-\widehat{S}_{z}, 
&\widehat{N}_{B}&=\frac{1}{2}\widehat{N}+\widehat{S}_{z},\nonumber \\
\widehat{S}_{x}^{2}+\widehat{S}_{y}^{2}+\widehat{S}_{z}^{2} & =  \frac{\widehat{N}}{2}(\frac{\widehat{N}}{2}+1).&&\label{Eq.SpinResults}
\end{align}
Then we find, after some straightforward calculations and introducing the
variances $\langle \Delta\widehat{S}_{x}^{2}\rangle =\langle \widehat{S}_{x}^{2}\rangle -\langle \widehat{S}_{x}\rangle ^{2}$
etc., the following results 
for Category 1, 2 and 3 states:
\begin{align}
\left\langle \Delta\widehat{S}_{x}^{2}\right\rangle +\left\langle \Delta\widehat{S}_{y}^{2}\right\rangle -\frac{1}{2}\left\langle \widehat{N}\right\rangle  & \geq  0
\nonumber\\ &
\text{Category 1 States}\label{Eq.IneqSpinOprsCat1}\\
\left\langle \Delta\widehat{S}_{x}^{2}\right\rangle +\left\langle \Delta\widehat{S}_{y}^{2}\right\rangle -\frac{1}{4}\left\langle \widehat{N}\right\rangle +\frac{1}{2}\left\langle \widehat{S}_{z}\right\rangle  
& \geq  0
\nonumber\\ & 
\text{Category 2 States}
\label{Eq.IneqSpinOprsCat2}\\
\left\langle \Delta\widehat{S}_{x}^{2}\right\rangle +\left\langle \Delta\widehat{S}_{y}^{2}\right\rangle +\frac{1}{4} & \geq  0
\nonumber\\ &
\text{Category 3 States}
\label{Eq.InequalSpinOprsCat3}
\end{align}
Details are given in Appendix \ref{Appendix - Correlation Ineq and Spin Operators}.
For Category 2 states with $A$\ involving the LHS then the left
side would have involved $-\frac{1}{2}\langle \widehat{S}_{z}\rangle $.

The inequality (\ref{Eq.IneqSpinOprsCat2}) for Category 2 states
was obtained more directly \emph{without} using the strong correlation
inequalities in Sections \ref{SubSection - Spin Squeezing Test for EPR Steering},
\ref{SubSection - Planar Spin Variance Tests for EPR Steering} -
see Eqs.\ (\ref{Eq.ResultVarSxSyLHS},\ref{Eq.ResultNLHS})
and (\ref{Eq.SzLHSModel}).
Details were given in Appendix \ref{Appendix - EPR Sterering Othe Approach}.
The inequality (\ref{Eq.IneqSpinOprsCat1}) for Category 1 states
was also derived in Refs.\ \cite{Hillery06a} and \cite{Dalton16b}.

We note in passing that Eq.\ (\ref{Eq.InequalSpinOprsCat3}) does not
lead to a test for Bell non-locality. From the Heisenberg Uncertainty
Principle this inequality applies for \emph{all} quantum states. Hence
the inequalities (\ref{Eq.IneqalCat3}) or (\ref{Eq.InequalSpinOprsCat3})
do \emph{not} provide a test for Bell non-locality.

\subsection{Weak Correlation Test}

The quantum operator $\widehat{a}^{\dag}\widehat{b}$ is not an observable,
but from the definitions for the spin operator we can write $\widehat{a}^{\dag}\widehat{b}=\widehat{S}_{x}-i\widehat{S}_{y}$.
We have interpreted $a^{\dag}b$ to be $S_{x}-iS_{y}$, where now $S_{x}$
and $S_{y}$ are observables whose mean values are definable in a
LHV theory.

From (\ref{Eq.MeanComplexCombHVTDefn}) and (\ref{Eq.MeanSxSy}) we
see that for Category 2 (and Category 1) states
\begin{eqnarray}
\left\langle a^{\dag}b\right\rangle  & = & \left\langle S_{x}\right\rangle -i\left\langle S_{y}\right\rangle, \nonumber \\
 & = & 0,\label{Eq.ResultMeanAdagBLHS}
\end{eqnarray}
so that 
\begin{equation}
|\left\langle a^{\dag}b\right\rangle |^{2}=\left\langle S_{x}\right\rangle ^{2}+\left\langle S_{y}\right\rangle ^{2}=0
\end{equation}
\textbf{f}or quantum states in Category 2 (or Category 1). This means
that if 
\begin{equation}
\left|\left\langle \widehat{a}^{\dag}\widehat{b}\right\rangle \right|^{2}>0,\label{Eq.WeakCorrelTestEPRSteer}
\end{equation}
the state cannot be either Category 1 or Category 2. This constitutes
a so-called \emph{weak correlation} test for EPR steering. However
because $|\langle \widehat{a}^{\dag}\widehat{b}\rangle |^{2}=\langle \widehat{S}_{x}\rangle ^{2}+\langle \widehat{S}_{y}\rangle ^{2}$
this test is really just \emph{equivalent} to the Bloch vector test.
So no useful test for either quantum entanglement or EPR steering
involving $\langle \widehat{S}_{x}\rangle ^{2}+\langle \widehat{S}_{y}\rangle ^{2}$\ and
$\langle \widehat{N}_{A}\otimes\widehat{N}_{B}\rangle $
is established at this point. However (see Section \ref{SubSection - Tests for EPR - Strong Correlation})
it was shown that related tests can be obtained both for quantum entanglement
and EPR steering. 

\subsection{Strong Correlation Test}

\label{SubSection - Tests for EPR - Strong Correlation}

Hillery and Zubairy \cite{Hillery06a} showed that for separable states
(Category 1 states) that $|\langle \widehat{a}^{\dag}\widehat{b}\rangle |^{2}\leq\langle \widehat{a}^{\dag}\widehat{a} \widehat{b}^{\dag}\widehat{b}\rangle =\langle \widehat{N}_{A}\otimes\widehat{N}_{B}\rangle $.
This result is also obtained here in Eq.\ (\ref{Eq.IneqCat1}). The
proof of this result was valid irrespective of whether the sub-system
states $\widehat{\rho}_{R}^{A}$ and $\widehat{\rho}_{R}^{B}$ were
local particle number SSR compliant or not (see Ref.\ \cite{Dalton16b}
for details). The quantum result
\begin{eqnarray}
\left|\left\langle \widehat{a}^{\dag}\widehat{b}\right\rangle \right|^{2} & = & \left\langle \widehat{S}_{x}\right\rangle ^{2}+\left\langle \widehat{S}_{y}\right\rangle ^{2},\nonumber \\
 & > & \left\langle \widehat{N}_{A}\otimes\widehat{N}_{B}\right\rangle, \label{Eq.CorrelTestLHS}
\end{eqnarray}
is a \emph{strong correlation} test for quantum \emph{entanglement}.
Hence as the numbers of bosons $N_{A}$ and $N_{B}$ are observables
in the LHV model (and therefore the mean $\left\langle N_{A}\otimes N_{B}\right\rangle $
can be defined) we see that for Category 1 states the LHVT result
\begin{equation}
\left|\left\langle a^{\dag}b\right\rangle \right|^{2}\leq\left\langle N_{A}\otimes N_{B}\right\rangle \label{Eq.CorrelInequalLHSModel}
\end{equation}
applies. Thus if 
\begin{equation}
\left|\left\langle \widehat{a}^{\dag}\widehat{b}\right\rangle \right|^{2}>\left\langle \widehat{N}_{A}\otimes\widehat{N}_{B}\right\rangle, \label{Eq.StrongCorrelTestEntangle}
\end{equation}
we have a strong correlation test for entanglement. However, there
is a \emph{different} strong correlation test for EPR steering that
applies - and which is harder to satisfy.

In the case of Category 2 states from the inequality in Eq.\ (\ref{Eq.InequalCat2})
we see that if 
\begin{equation}
\left|\left\langle \widehat{a}^{\dag}\widehat{b}\right\rangle \right|^{2}>\left\langle \widehat{N}_{A}\otimes\widehat{N}_{B}\right\rangle +\frac{1}{2}\left\langle \widehat{1}_{A}\otimes\widehat{N}_{B}\right\rangle, \label{Eq.StrongCorrelTestEPR}
\end{equation}
the state cannot be in Category 2 (nor in Category 1) so it must be
\emph{EPR steerable}. Thus the inequality (\ref{Eq.StrongCorrelTestEPR})
is a \emph{strong correlation} test for \emph{EPR steering}. Note
that the condition is harder to satisfy than the strong correlation
test (\ref{Eq.CorrelTestLHS}) for entanglement since $\langle \widehat{1}_{A}\otimes\widehat{N}_{B}\rangle $
is positive, but obviously if (\ref{Eq.StrongCorrelTestEPR}) is satisfied
the state is entangled as well as being EPR steerable. If $A$\ involved
the LHS then the right side would have been $\langle \widehat{N}_{A}\otimes(\widehat{N}_{B}+\frac{1}{2}\widehat{1}_{B})\rangle $.

However, as these tests are just \emph{equivalent} to the
Hillery-Zubairy planar spin variance test and the generalised
Hillery-Zubairy planar spin variance test, no additional test has
been obtained.

\section{Correlation Inequalities and Spin Operators}

\label{Appendix - Correlation Ineq and Spin Operators}

The inequalities (\ref{Eq.IneqCat1}), (\ref{Eq.InequalCat2}) and
(\ref{Eq.IneqalCat3}) derived above can be put into a more useful
form involving \emph{spin operators} - whose mean values and variances
can be measured. We use the definitions of the spin operators in Section
\ref{SubSection - Quadrature Amplitudes} (see also Ref.\ \cite{Dalton16b})
\begin{eqnarray}
|\left\langle \widehat{a}^{\dag}\widehat{b}\right\rangle |^{2} & = & \left\langle \widehat{S}_{x}\right\rangle ^{2}+\left\langle \widehat{S}_{y}\right\rangle ^{2},\nonumber \\
\widehat{N}_{A} & = & \frac{1}{2}\widehat{N}-\widehat{S}_{z},\qquad\widehat{N}_{B}=\frac{1}{2}\widehat{N}+\widehat{S}_{z},\nonumber \\
\widehat{S}_{x}^{2}+\widehat{S}_{y}^{2}+\widehat{S}_{z}^{2} & = & \frac{\widehat{N}}{2}(\frac{\widehat{N}}{2}+1) .
\end{eqnarray}
We see that 
\begin{align*}
\left\langle \Delta \widehat{S}_{x}^{2}\right\rangle +\left\langle \Delta 
\widehat{S}_{y}^{2}\right\rangle  &=\frac{1}{4}\left\langle (\widehat{N}%
_{A}+\widehat{N}_{B})^{2}\right\rangle +\frac{1}{2}\left\langle \widehat{N}%
_{A}+\widehat{N}_{B}\right\rangle   \nonumber \\
&\quad-\left|\left\langle \widehat{a}^{\dag }\widehat{b}\right\rangle \right|^{2}-\frac{1}{4%
}\left\langle (\widehat{N}_{B}-\widehat{N}_{A})^{2}\right\rangle,   \nonumber
\\
\end{align*}
\begin{align*}
\left\langle \Delta \widehat{S}_{x}^{2}\right\rangle +\left\langle \Delta 
\widehat{S}_{y}^{2}\right\rangle  &\geq \left\langle \widehat{N}_{A}\otimes 
\widehat{N}_{B}\right\rangle +\frac{1}{2}\left\langle \widehat{N}_{A}\otimes 
\widehat{1}_{B}\right\rangle   \nonumber \\
&\quad+\frac{1}{2}\left\langle \widehat{1}_{A}\otimes \widehat{N}%
_{B}\right\rangle -\left\langle \widehat{N}_{A}\otimes \widehat{N}%
_{B}\right\rangle,   \nonumber \\
&\geq \frac{1}{2}\left\langle \widehat{N}_{A}\otimes \widehat{1}%
_{B}\right\rangle +\frac{1}{2}\left\langle \widehat{1}_{A}\otimes \widehat{N}%
_{B}\right\rangle,   \nonumber \\
&\qquad \qquad \qquad\qquad ~~ \text{Cat 1 States}  \nonumber \\
\end{align*}
\begin{align*}
\left\langle \Delta \widehat{S}_{x}^{2}\right\rangle +\left\langle \Delta 
\widehat{S}_{y}^{2}\right\rangle  &\geq \left\langle \widehat{N}_{A}\otimes 
\widehat{N}_{B}\right\rangle +\frac{1}{2}\left\langle \widehat{N}_{A}\otimes 
\widehat{1}_{B}\right\rangle   \nonumber \\
&\quad+\frac{1}{2}\left\langle \widehat{1}_{A}\otimes \widehat{N}%
_{B}\right\rangle   \nonumber \\
&\quad-\left\langle \left( \widehat{N}_{A}+\frac{1}{2}\widehat{1}_{A}\right)
\otimes \widehat{N}_{B}\right\rangle,   \nonumber \\
&\geq \frac{1}{2}\left\langle \widehat{N}_{A}\otimes \widehat{1}%
_{B}\right\rangle, \qquad \text{Cat 2 States}  \nonumber \\
\end{align*}
\begin{align}
\left\langle \Delta \widehat{S}_{x}^{2}\right\rangle +\left\langle \Delta 
\widehat{S}_{y}^{2}\right\rangle  &\geq \left\langle \widehat{N}_{A}\otimes 
\widehat{N}_{B}\right\rangle +\frac{1}{2}\left\langle \widehat{N}_{A}\otimes 
\widehat{1}_{B}\right\rangle   \nonumber \\
&\quad+\frac{1}{2}\left\langle \widehat{1}_{A}\otimes \widehat{N}%
_{B}\right\rangle   \nonumber \\
&\quad-\left\langle \left( \widehat{N}_{A}+\frac{1}{2}\widehat{1}_{A}\right)
\otimes \left( \widehat{N}_{B}+\frac{1}{2}\widehat{1}_{B}\right)
\right\rangle,   \nonumber \\
&\geq -\frac{1}{4}. \qquad\qquad\quad \text{Cat 3 States}
\end{align}
So we have:
\begin{align}
\left\langle \Delta\widehat{S}_{x}^{2}\right\rangle +\left\langle \Delta\widehat{S}_{y}^{2}\right\rangle -\frac{1}{2}\left\langle \widehat{N}\right\rangle  & \geq  0 \,,
\nonumber \\ & 
\qquad \text{Cat 1 States}\nonumber \\
\left\langle \Delta\widehat{S}_{x}^{2}\right\rangle +\left\langle \Delta\widehat{S}_{y}^{2}\right\rangle -\frac{1}{4}\left\langle \widehat{N}\right\rangle +\frac{1}{2}\left\langle \widehat{S}_{z}\right\rangle  
& \geq  0 \,,
\nonumber \\ & 
\qquad \text{Cat 2 States}\nonumber \\
\left\langle \Delta\widehat{S}_{x}^{2}\right\rangle +\left\langle \Delta\widehat{S}_{y}^{2}\right\rangle +\frac{1}{4} 
& \geq  0 \,.
\nonumber \\ & 
\qquad \text{Cat 3 States}\nonumber \\
& 
\end{align}


\end{document}